\documentclass[12pt]{article}
\input{epsf}
\newcommand{\postscriptscale}[2]
{\setlength{\epsfxsize}{#2\hsize} \centerline{\epsfbox{#1}}}

\makeatletter
\def\Ddots{\mathinner{\mkern1mu\raise\p@
\vbox{\kern7\p@\hbox{.}}\mkern2mu
\raise4\p@\hbox{.}\mkern2mu\raise7\p@\hbox{.}\mkern1mu}}
\makeatother
\font\bm=cmmib10 at 10pt
\font\bms=cmmib10 at 7pt \textfont9=\bm \scriptfont9=\bms
\usepackage{amsfonts}
\usepackage{multirow}
\mathchardef\balpha= "790B
\mathchardef\bbeta= "790C
\mathchardef\bTheta= "7902
\mathchardef\bzeta= "7910
\mathchardef\bOmega= "790A
\mathchardef\bGamma= "7900
\mathchardef\bDelta= "7901
\mathchardef\bPhi= "7908
\mathchardef\bphi= "791E
\mathchardef\bomega= "7921
\mathchardef\bxi= "7918
\mathchardef\bet= "7911
\mathchardef\brho= "791A
\mathchardef\btau= "791C
\mathchardef\bmu= "7916
\mathchardef\bvarpi= "7924

\def \lvec{(\kern-.26em(}
\def\pmb#1{\setbox0=\hbox{#1}
\kern-.025em\copy0\kern-\wd0
\kern.05em\copy0\kern-\wd0
\kern-.025em\raise.0433em\box0 }
\mathchardef\btheta= "7912
\usepackage{amsmath}
\usepackage{amssymb}
\usepackage{authblk}
\usepackage{url}
\providecommand{\keywords}[1]{\textbf{\textit{Keywords:  }} #1}

\begin{document}
\title{2n-Stream Radiative Transfer}
\author[1]{ W. A. van Wijngaarden}
\author[2] {W. Happer}
\affil[1]{Department of Physics and Astronomy, York University, Canada}
\affil[2]{Department of Physics, Princeton University, USA}
\renewcommand\Affilfont{\itshape\small}
\date{\today}
\maketitle

\begin{abstract}
\noindent We use $2n= 2,4,6,\ldots$ `` streams" of axially symmetric radiation to solve the equation of transfer for a layered medium. This is a generalization of Schuster’s classic 2-stream model. As is well known, using only the first  $2n$ Legendre polynomials to describe the angular dependence of radiation reduces the equation of transfer to a first order differential equation in a space of $2n$ dimensions.  It is convenient to characterize the radiation as $2n$ stream intensities $I(\mu_i)$,  propagating at angles close to the zenith angles $\theta_i=\cos^{-1}\mu_i$. The $2n$ Gauss-Legendre  cosines $\mu_i$ are defined by $P_{2n}(\mu_i)=0$, where $P_{2n}(\mu)$ is the Legendre polynomial of degree $2n$.  We show how to efficiently and accurately solve the equation of transfer with  vector and matrix methods analogous to those used to solve Schr\"odinger's equation of quantum mechanics.  To model strong forward scattering, like that of visible light by Earth's clouds, we have introduced a new family of phase functions $\varpi^{\{p\}}(\mu)$. These give the maximum possible forward scattering, $\varpi^{\{p\}}(\mu=1)=p(p+1)$, for a phase function constructed from the first $2p=2,4,6,\ldots$ Legendre polynomials. We show illustrative examples of radiative-transfer phenomena calculated with this new method.  
\end{abstract}
\keywords{radiative transfer, multiple scattering, absorption, emission, phase functions, equation of transfer, Gaussian quadrature}
\newpage
\section{Introduction}
Radiative transfer in semi-transparent media is controlled by absorption, emission and scattering.  The {\it equation of transfer} which describes these processes has many similarities to the {\it Schr\"odinger equation} of quantum mechanics. In this paper we show how methods of linear-algebra, similar to those used to solve the Schr\"odinger equation, can facilitate the analysis of radiative transfer, especially for strongly anistropic scattering like that of Earth's clouds. 

An instructive history of radiative transfer theory has been given by Mobley\cite{Mobley}.  We will frequently refer to the authoritative review of radiative transfer  in Chandrasekhar's classic book, {\it Radiative Transfer}\cite{Chandrasekhar}. Other useful books on radiative transfer are those of Petty \cite{Petty} and Thomas and Stamnes\cite{Stamnes}.

 \vskip  0.5 in
\section{Equation of transfer}
For axial symmetry we will characterize  monochromatic radiation of spatial frequency $\nu$ at an altitude $z$  above Earth's surface and at time $t$ with  the monochromatic {\it intensity} $I(\nu,\mu,z,t)$, also called the {\it radiance}. One can think of the flux as streams of photons making various angles, $\theta =\cos^{-1}\mu$, with the vertical. $I(\nu,\mu,z,t)d\mu\,d\nu$ is the radiative flux carried by photons with direction cosines between $\mu$ and $\mu+d\mu$ and with spatial frequencies between $\nu$ and $\nu+d\nu$. A representative unit of $I(\nu,\mu,z,t)$ is W m$^{-2}$ cm sr$^{-1}$, or watts per (square meter-wave number-steradian). In \S {\bf 1}(3) Chandrasekhar uses the symbol $I_{\nu}$ to denote our intensity  $I$.
For the remainder of this paper we will discuss only monochromatic radiation and omit the frequency variable $\nu$. 

It will be convenient to specify the altitude $z$ in terms of the optical depth, $\tau$, from the surface
\begin{equation}
\tau =\int_0^{\tau}d\tau' = \int_0^{z}\alpha(z')dz'.  \label{in2}
\end{equation}
The coefficient $\alpha$ gives the attenuation, due to absorption and scattering, of radiation by molecules and particulates. In \S{\bf 3}(22) Chandrasekhar\cite{Chandrasekhar} defines a  mass attenuation coefficient $\kappa_{\nu}$ and a mass density $\rho$. Our attenuation coefficient is $\alpha = \kappa_{\nu}\rho$. Our (\ref{in2}) is the same as \S{\bf 9}(62) of Chandrasekhar\cite{Chandrasekhar}.
For Earth's atmosphere,
the value of $\alpha$ can change rapidly with $z$ because of changes in temperature, pressure and the density of molecules or particulates with which the photons collide. 

Most of this paper is focused on steady-state, time-independent conditions, but we include a time variable $t$ in the first few sections to clarify the physical meanings of basic equations.  It is convenient to use a relative time $\vartheta$, defined in analogy to (\ref{in2}), by
\begin{equation}
\vartheta= c\, \alpha t.\label{in6}
\end{equation}
In (\ref{in6}) we have denoted the speed of light by the symbol $c$. We assume negligible time dependence of $\alpha$ so 
an increment $d\vartheta$ of relative time is related to an increment $dt$ of absolute time by 
\begin{equation}
\partial \vartheta = \alpha\, c\, \partial t.
\label{in8}
\end{equation}
In terms of these dimensionless versions of altitude and time, the intensity, $I(\mu,\tau,\vartheta)$  is a solution of the {\it equation of transfer}\cite{Chandrasekhar,Code},
\begin{equation}
\left(\mu\frac{\partial}{\partial\tau}+\frac{\partial}{\partial \vartheta}+1\right) I(\mu, \tau,\vartheta) =(1-\tilde \omega) B(\tau)+\frac{\tilde\omega}{2}
\int_{-1}^1 d\mu' p(\mu,\mu')I(\mu',\tau,\vartheta).\label{in10}
\end{equation}
We neglect any variation of the intensity in horizontal spatial directions. 
The time-dependent equation of transfer (\ref{in10})  is given  by Code\cite{Code} as his Eq. (4). The steady-state, time-independent version of (\ref{in10}), with $\partial/\partial\vartheta \to 0
$, is given by Chandrasekhar\cite{Chandrasekhar} as \S{\bf 6}(47). He uses the symbol $\mathcal{J}_{\nu}$ to denote both source terms on the right of (\ref{in10}). He writes the emissive part of the source, which is proportional to the Planck intensity $B(\tau)$, as \S{\bf 5}(42)  and the scattering part of the source, that is proportional to the scattering phase, $p(\mu,\mu')$, as  in \S{\bf 5}(41). 

The source terms are characterized by the {\it single-scattering albedo} $\tilde\omega$ and by the {\it scattering phase function} $p(\mu,\mu')$. We assume both are independent of position and  time.  Later, we will show how to combine the effects of many cloud layers, each with its  own optical thickness $\tau_c$ and values of $\tilde\omega$ and $p(\mu,\mu')$.

The single-scattering albedo $\tilde\omega$ is the probability that a photon, after being removed from a stream of radiation by a molecule or cloud-particulate, is elastically scattered into another direction, rather than being inelastically scattered into a photon of a different frequency (Raman scattering) or 
absorbed.  We neglect Raman scattering and assume that a fraction $1-\tilde\omega$ of the attenuated photons is absorbed and converted to atmospheric heat.
Chandrasekhar\cite{Chandrasekhar} uses the symbol $\varpi_0$, the first term in his multipole expansion \S{\bf 3}(33) of the scattering phase function, to denote the single-scattering  albedo $\tilde \omega$ of (\ref{in10}).

The phase function $p(\mu,\mu')$ of (\ref{in10}) is a symmetric and nonnegative function of the direction cosine, $\mu=\cos\theta$, of the scattered radiation and the direction cosine, $\mu'=\cos\theta'$, of incident radiation
\begin{equation}
 p(\mu,\mu')=p(\mu',\mu)\ge 0.
\label{in14}
\end{equation}
The probability for scattering radiation with a direction cosine $\mu'$ to radiation with direction cosines in the interval from $\mu$ to $\mu+d\mu$ is $p(\mu,\mu')\,d\mu /2$.
In keeping with its significance as a conditional probability,  the phase function satisfies the identity
\begin{equation}
\frac{1}{2}\int_{-1}^1 d\mu \,p(\mu,\mu')=1.
\label{in16}
\end{equation}

The monochromatic Planck intensity $B=B(\tau)$ of (\ref{in10})  depends on the local temperature, $T$ of the scattering medium, and on the spatial frequency $\nu$ of the radiation, as described by Planck's formula
\begin{equation}
B=\frac{2h_{\rm P}c^2\nu^3}{e^{\nu c\, h_{\rm P}/(k_{\rm B}T)}-1}.
\label{in18}
\end{equation}
We are using cgs units, where $h_{\rm P}$ is Planck's constant and $k_{\rm B}$ is Boltzmann's constant. We assume that the absolute air temperature, $T=T(z,t)$ may depend on altitude $z$  and time $t$. Then the Planck intensity, $B=B(\tau,\vartheta)$, may depend  on the optical depth $\tau$, and on the relative time $\vartheta$. For future reference, we note that the frequency-integrated Planck intensity is related to the Stefan-Boltzmann flux by
\begin{equation}
\sigma_{\rm SB}T^4=\pi \int_0^{\infty}d\nu B.
\label{in20}
\end{equation}
The Stefan-Boltzman constant has the value $\sigma_{\rm SB}= 5.67 \times 10^{-5}$ erg s$^{-1}$ cm$^{-2}$ K$^{-4}$.

In this analysis, we assume that there is negligible difference between the speed of light in the scattering medium and the speed of light $c$ in vacuum. Some equations need minor modifications for applications in the ocean or other media where the speed of light is substantially smaller than the speed in vacuum.

\subsection{Multipole moments}
There are advantages to describing the intensity $I(\mu,\tau,\vartheta)$ in terms of its multipole moments
\begin{equation}
I_l(\tau,\vartheta)=\frac{1}{2}\int_{-1}^1 d\mu\,P_l(\mu) I(\mu,\tau,\vartheta). \label{mm2}
\end{equation}
Here $P_l(\mu)$ is the Legendre polynomial of  multipolarity $l=0,1,2,\ldots\infty$.  The inverse of 
(\ref{mm2}) is 
\begin{equation}
I(\mu,\tau,\vartheta)=\sum_{l=0}^{\infty}(2l+1)P_l(\mu) I_l(\tau,\vartheta). 
\label{mm4}
\end{equation}
In terms of Legendre polynomials, we can write the phase function 
for randomly oriented scatterers as
\begin{equation}
p(\mu,\mu')= \sum_{l=0}^{\infty}P_l(\mu)(2l+1)p_l\,P_l(\mu').
\label{mm6}
\end{equation}
The phase function (\ref{mm6}) is parameterized by the  multipole transfer coefficients, $p_l$, for $l=0,1,2,\ldots,\infty$.
As implied by (\ref{in10}) and (\ref{mm6}), a single scattering converts a component $I_lP_l(\mu) $ of the intensity to 
\begin{eqnarray}
I_lP_l(\mu) &\to&\frac{1}{2}\int_{-1}^1 d\mu'\, p(\mu,\mu')P_l(\mu')I_l \nonumber\\
&=&
\int_{-1}^1 d\mu'\, \sum_{l'=0}^{\infty}p_l'\,\frac{2l'+1}{2}P_{l'}(\mu)P_{l'}(\mu')P_l(\mu')I_l\nonumber\\
&=&p_l I_lP_l(\mu).
\label{mm8}
\end{eqnarray}
To evaluate the integral of (\ref{mm8})
 we used the orthogonality property of Legendre polynomials
\begin{equation}
\int_{-1}^1 d\mu\frac{2l'+1}{2} P_{l}(\mu)P_{l'}(\mu)
=\delta_{ll'}.
\label{mm10}
\end{equation}
After $n$ successive scatterings, the amplitude will be multiplied by a factor $(p_l)^n$. 
To ensure that a sufficiently large number $n$ of successive scatterings attenuates any initially anisotropic multipole moment of the intensity to zero, the transfer coefficients $p_l$ for nonzero multipoles must satisfy the constraint
\begin{eqnarray}
|p_l| <1,\quad\hbox{for}\quad l=1,2,3,\ldots,\infty.
\label{mm12}
\end{eqnarray}
Any number of scatterings must have no effect on isotropic radiation so the monopole transfer coefficient must be
\begin{eqnarray}
p_0=1.
\label{mm14}
\end{eqnarray}
The $p_l$ must satisfy constraints in addition to (\ref{mm12}) and (\ref{mm14}) to ensure  that the single-scattering phase function is  nonnegative, as required by (\ref{in14}).
Chandrasekhar\cite{Chandrasekhar} uses coefficients $\varpi_l=(2l+1)p_l\,\tilde\omega$ for his expansion 
\S {\bf 3} (33).

Multiplying both sides of (\ref{in10}) on the left by $\frac{1}{2}\int_{-1}^{1} d\mu P_l(\mu)$, and using (\ref{mm2}), (\ref{mm4}), (\ref{mm6}) and (\ref{mm10}), we find the equation of  transfer for intensity multipoles,
\begin{equation}
\sum_{l'=0}^{\infty}\lvec l|\hat \mu|l')\frac{\partial}{\partial\tau}I_{l'}(\tau,\vartheta)+\frac{\partial}{\partial \vartheta}I_l(\tau,\vartheta) =(1-\tilde \omega) B(\tau)\delta_{l0}-(1-\tilde\omega p_l )I_l(\tau,\vartheta).\label{mm16}
\end{equation}
The matrix elements $\hat\mu_{ll'}$ of the direction cosine  $\mu$ are 
\begin{eqnarray}
\lvec l|\hat \mu|l') &=& \int_{-1}^1 d\mu 
\frac{2l+1}{2}\mu P_l(\mu)P_{l'}(\mu)\nonumber\\
&=&\frac{(l)\delta_{l',l-1}+(l+1)\delta_{l',l+1}}{2l+1}.\label{mm18}
\end{eqnarray}
To evaluate (\ref{mm18})  we used the orthogonality relation (\ref{mm10}) and Bonnet's recursion formula  for Legendre polynomials,
\begin{equation}
(2l+1)\mu P_l(\mu) = lP_{l-1}(\mu)+(l+1)P_{l+1}(\mu).\label{mm20}
\end{equation}
\subsection{Conservation of energy}
At the vertical optical depth $\tau$ and relative time $\vartheta,$ the radiant energy density $u(\tau,\vartheta)$ is related to the intensity $I(\mu,\tau)$ by
\begin{eqnarray}
 u(\tau,\vartheta)&=&\frac{2\pi}{c}\int_{-1}^{1}d\mu\, I(\mu,\tau,\vartheta)\nonumber\\
&=&\frac{4\pi}{c}\,\frac{1}{2}\int_{-1}^{1}d\mu P_0(\mu) I(\mu,\tau,\vartheta)\nonumber\\
&=&\frac{4\pi}{c} I_0(\tau,\vartheta).
\label{mn30}
\end{eqnarray}
For azimuthal symmetry the increment of solid angle is $d\Omega = 2\pi\, d\mu$ steradians. This is the source of the factor of $2\pi$ in the first line of (\ref{mn30}). We noted that $P_0(\mu)=1$, and we used (\ref{mm2}) to write the energy density as $4\pi/c$ times the monopole moment $I_0(\tau)$ of the intensity. The energy density $u$ of the radiation is many orders of magnitude smaller than the energy density due to the translation, rotation and vibration of gas molecules. In \S{\bf 2} (19), Chandrasekhar\cite{Chandrasekhar} uses the symbol $J_{\nu}$, which he calls the average intensity, to denote our monopole intensity $I_0$.  

Similarly, the upward monochromatic flux $Z(\tau)$,  often called the  {\it irradiance}, is
\begin{eqnarray}
 Z(\tau,\vartheta)&=&2\pi\int_{-1}^{1}d\mu\, \mu \, I(\mu,\tau,\vartheta)\nonumber\\
&=&4\pi\,\frac{1}{2}\int_{-1}^{1}d\mu P_1(\mu) I(\mu,\tau,\vartheta)\nonumber\\
&=&4\pi I_1(\tau,\vartheta).
\label{mn32}
\end{eqnarray}
Here we noted that $P_1(\mu)=\mu$ and we used (\ref{mm2}) to write
the upward flux  as $4\pi$ times the dipole moment $I_1(\tau,\vartheta)$ of the intensity. In \S {\bf 2} (13) Chandrasekhar\cite{Chandrasekhar} uses the symbol $F_{\nu}$ to denote 4 times our dipole moment $I_1$, that is, $F_{\nu}=4I_1$.

For the special case $l=0$, we can use (\ref{mm18}) to write (\ref{mm16}) as
\begin{equation}
\frac{\partial}{\partial \tau}I_1(\tau,\vartheta)+\frac{\partial}{\partial \vartheta}I_0(\tau,\vartheta)
=(1-\tilde \omega) \left[B(\tau)-I_0(\tau,\vartheta)\right].\label{ce4}
\end{equation}
In terms of the vertical flux $Z$ of (\ref{mn32}), the energy density $u$ of (\ref{mn30}), the increments $\partial \vartheta$ of relative time from (\ref{in8}), and the increment of optical depth $\partial \tau = \alpha\,\partial z$,  we can write (\ref{ce4}) as
\begin{equation}
\frac{\partial}{\partial t}u(z,t)=
-\frac{\partial}{\partial z}Z(z,t)+c\,\alpha (1-\tilde \omega) \left[\frac{4\pi B(z)}{c}-u(z,t)\right].\label{ce6}
\end{equation}
Multiplying the left and right sides of (\ref{ce6}) by the altitude increment $dz$, we can interpret the resulting terms as follows:

\begin{itemize}
\item{$dz[\partial u(z,t)/\partial t]$ is the rate of increase with time of the thermal radiation energy per unit area in an infinitesimal layer between the altitude $z$ and the altitude $z+dz$.}
\item{$dz[-\partial Z(z,t)/\partial z] = Z(z,t)-Z(z+dz, t)$ is the net flux into the infinitesimal layer, the flux $Z(z)$ into the bottom minus the flux $Z(z+dz)$ out of the top.} 
\item{$c\alpha$ is the rate of collisions of photons with with molecules or particulates. }
\item{ $dz\, c\,\alpha (1-\tilde \omega)4\pi B/c$ is the rate of  increase of radiation energy per unit area of the infinitesimal layer due to thermal emission by molecules and particulates.}
\item{ $-dz\, c\,\alpha (1-\tilde \omega) u $ is the rate of decrease of radiation energy per unit area of the infinitesimal layer due to absorption by molecules and particulates.}
\end{itemize}
The last two terms of this list, the rate of increase of radiation energy due to thermal emission and the rate of decrease due to absorption are both proportional to $c\alpha(1-\tilde \omega)$, the absorption probability of photons per unit time.  
In \S{\bf 4}(38) and elsewhere, Chandrasekhar\cite{Chandrasekhar} calls the relation between these terms {\it Kirchhoff's law.} Later, we will discuss a closely related version of Kirchhoff's law for clouds, where the single-scattering albedo $\tilde\omega$ is replaced by a cloud scattering matrix $\mathcal{S}$, and the single-scattering absorption probability $1-\tilde\omega$ is replaced by the   emissivity matrix $\mathcal{E}=\hat 1-\mathcal{S} $ of an isothermal cloud.

Eq. (\ref{ce6}) describes the conservation of energy.  The rate of change of the monochromatic energy density, $u$, is equal to the net monochromatic flux, $-\partial Z/\partial z$, of radiation into a unit volume plus the excess of the rate of  photon emission over that of absorption by molecules and particulates, which is described by the last term of the equation.

The total energy density, $U=\int_0^{\nu} u \,d\nu $ of radiation at all frequencies 
$\nu$ is many orders of magnitude smaller than the thermal energy density of molecules or particulates in the atmosphere.  For example, at sea level, the thermal energy density stored per cubic centimeter of dry air is 
\begin{equation}
U(\hbox{molecules})=N k_{\rm B} c_{v}T =2.78\times 10^6 \hbox{erg cm}^{-3} \label{ce8}
\end{equation}
Here $N=2.69 \times 10^{19}\hbox{ cm}^{-3}$ is the density of air molecules (Loschmidt's number per cubic centimeter), and $k_{\rm B} = 1.38 \times 10^{-16} \hbox{erg K}^{-1}$ is Boltzmann's constant. In units of $k_{\rm B}$, the constant-volume heat capacity per molecule of air is  $c_v = 2.5$. Most of the heat is carried by the diatomic molecules N$_2$ and O$_2$ which have three fully activated translational degrees of freedom and two rotational degrees of freedom.  The atmosphere is not warm enough to unfreeze the vibrational modes, and their contribution to the heat capacity can be neglected.
 A representative absolute temperature, $T= 300$ K, was chosen to evaluate (\ref{ce8}).

The energy density of radiation of all frequencies $\nu$ at the same  temperature is some 11 orders of magnitude smaller than that of the molecules,
\begin{equation}
U(\hbox{radiation})=\int_0^{\infty}d\nu\frac{4\pi B}{c} =\frac{4\sigma_{\rm SB}T^4}{c} =6.12\times 10^{-5} \hbox{erg cm}^{-3}.  \label{ce10}
\end{equation}
The radiant  heat capacity of air is negligible compared to that of its molecules and particulates. But radiation can efficiently transfer energy because the speed of light is so large.
\section{Vector notation}
As is the case for the Schr\"odinger equation in quantum mechanics, Dirac vector
notation simplifies the appearance of the equation of transfer, makes the physical significance clearer, and simplifies numerical analysis with contemporary mathematical software like Matlab.  We  interpret the intensity $I(\mu,\tau,\vartheta)$ of (\ref{in10}) as the projection of an abstract intensity vector $|I(\tau,\vartheta)\}$ onto a direction-cosine basis vector $\langle \mu|$,
\begin{equation}
I(\mu,\tau,\vartheta) = \langle\mu|I(\tau,\vartheta)\}.\label{vc2}
\end{equation}
The intensity $|I(\tau,\vartheta)\}$ at the optical depth $\tau$ is analogous to the abstract wave function $|\psi(t)\rangle$ of a quantum mechanical particle at time $t$.
 Aside from normalization, the basis vector $\langle \mu|$ and its Hermitian conjugate, $|\mu\rangle = \langle \mu|^{\dag}$, are defined by the eigenvalue equation
\begin{equation}
\langle\mu|\hat\mu = \langle\mu|\mu,\quad\hbox{and}\quad \hat\mu|\mu\rangle = \mu|\mu\rangle.\label{vc4}
\end{equation}
The eigenvalues $\mu$ can be any of the real numbers $|\mu|\le 1$.  We normalize the  eigenvectors such that
\begin{equation}
\langle\mu|\mu'\rangle =\delta(\mu-\mu').
\label {vc6}
\end{equation}
Here $\delta(\mu-\mu')$ is a Dirac delta function.
The eigenvectors have the completeness property
\begin{equation}
\int_{-1}^1 d\mu\,|\mu\rangle\langle\mu| = \hat 1.
\label {vc8}
\end{equation}

In like manner,
we will assume that the multipole amplitude $I_l(\tau)$ of (\ref{mm2}) is the projection of the abstract intensity vector $|I(\tau,\vartheta)\}$ onto a left {\it multipole basis vector}, $\lvec l|$,
\begin{equation}
\lvec l|I(\tau,\vartheta)\}=I_l(\tau,\vartheta).
\label {vc10}
\end{equation}
Using (\ref{vc2})  and (\ref{vc8}) in (\ref{vc10})  we find
\begin{equation}
I_l(\tau,\vartheta)=\int_{-1}^1 d\mu\, \lvec l|\mu\rangle\langle\mu|I(\tau,\vartheta)\}=\int_{-1}^1 d\mu\, \lvec l|\mu\rangle I(\mu,\tau,\vartheta).
\label {vc12}
\end{equation}
Comparing (\ref{vc12}) to (\ref{mm2}) we see that the projection of  $\lvec l|$ onto $|\mu\rangle$ must be 
\begin{equation}
 \lvec l|\mu\rangle =\frac{1}{2}P_l(\mu).
\label {vc14}
\end{equation}
In analogy to (\ref{vc6}), 
for each left multipole basis vector, $\lvec l|$, we can choose a right basis vector $|l)$ such that
\begin{equation}
 \lvec l|l')=\delta_{ll'}.
\label {vc16}
\end{equation}
The Kronecker-delta coefficient has the values $\delta_{ll'}=0$ if $l\ne l'$ and $\delta_{ll} = 1$. The multipole basis vectors also satisfy the completeness relation, analogous to (\ref{vc8})
\begin{equation}
 \sum_{l=0}^{\infty}|l)\lvec l|=\hat 1.
\label {vc18}
\end{equation}
Here and subsequently we will use the symbol $\hat 1$ to denote a unit operator, which can be represented by a square unit matrix with 1's along the main diagonal and zeros elsewhere.  In (\ref{vc18}) a square matrix with an infinite number of dimensions is needed to represent $\hat 1$. But in the later parts of this paper, the symbol $\hat 1$ will often mean a $2n\times 2n$ unit matrix, where $n=1,2,3,\ldots$ is the number of ``stream pairs" used to represent the angular distribution of radiation. The context will normally make the dimension of $\hat 1$ clear.
Using (\ref{vc8}) and (\ref{vc14}) with (\ref{vc16}) we find
\begin{eqnarray}
\delta_{l l'}&=&\int_{-1}^{1}d\mu\, \lvec l|\mu\rangle\langle\mu|l)\nonumber\\
&=&\int_{-1}^{1}d\mu\, \frac{1}{2}P_l(\mu)\langle\mu|l').
\label {vc20}
\end{eqnarray}
Comparing (\ref{vc20}) with (\ref{mm10}),  the orthogonality property of Legendre polynomials, we see that the projection of $|l')$ onto $\langle \mu|$ must be
\begin{equation}
\langle\mu|l') = (2l+1)P_{l'}(\mu).
\label {vc22}
\end{equation}
As indicated in (\ref{vc14}) and (\ref{vc22}), we will often find it convenient to use left (row or bra) basis vectors like $\lvec l|$ that are not simple Hermitian conjugates of right (column or ket) basis vectors like $|l)$.  We use the symbols $\lvec\cdots|$ and $|\cdots)$ to denote such basis vectors. They are analogous to the reciprocal lattice vectors that are often used in crystallography\cite {reciprocal}.
\subsection{Matrix representations}
For multipole space, we can use (\ref{mm18}) to represent the direction-cosine operator, $\hat\mu$ by the matrix
\begin{equation}
\lvec l|\hat\mu|l')
	=\left[\begin{array}{lllllll} 0&1/1&0&0&0&0&\cdots \\ 1/3&0&2/3&0&0&0&\cdots \\ 
 0&2/5&0&3/5&0&0&\cdots \\   0&0&3/7&0&4/7&0&\cdots \\ 
0&0&0&4/9&0&5/9&\cdots \\ 0&0&0&0&5/11&0&\cdots \\ 
 \vdots&\vdots &\vdots&\vdots&\vdots&\vdots&\ddots\\ \end{array}\right]
\label{me3}
\end{equation}
The indices of the rows and columns have the values $l=0,1,2,\dots\infty$ and $l'=0,1,2,\ldots,\infty$.
From (\ref{vc4}) and (\ref{vc6}) we see that the matrix elements of the direction-cosine operator $\hat\mu$ in continuous $\mu$-space are diagonal,
\begin{equation}
\langle \mu|\hat\mu|\mu'\rangle = \mu\,\delta(\mu-\mu').
\label {me4}
\end{equation}

 In continuous $\mu$-space the matrix elements of the scattering operator $\hat p$ are
\begin{equation}
\langle\mu|\hat p|\mu'\rangle = p(\mu,\mu').
\label {me8}
\end{equation}
Using (\ref{mm6}), (\ref{mm10}),  (\ref{vc14}),  and (\ref{vc22})  with (\ref{me8}) we see that  $\hat p$ is diagonal in multipole space
\begin{eqnarray}
\lvec l|\hat p|l')&=&\int_{-1}^1 d\mu\int_{-1}^1 d\mu'\lvec l|\mu\rangle \langle\mu|\hat p|\mu'\rangle \langle\mu'|l')\nonumber\\
&=&\frac{2l'+1}{2}\int_{-1}^1 d\mu\int_{-1}^1 d\mu' P_l(\mu) 
\sum_{l''} p_{l''}(2l''+1)P_{l''}(\mu)P_{l''}(\mu')P_{l'}(\mu')\nonumber\\
&=&\frac{2l'+1}{2}
\sum_{l''} p_{l''}(2l''+1)\frac{2^2}{(2l''+1)^2}\delta_{l\, l''}\delta_{l''\,l'}\nonumber\\
&=&2 p_l\delta_{l\,l'}
\label {me10}
\end{eqnarray}
or
\begin{equation}
\hat p=2\sum_{l=0}^{\infty}p_l|l)\lvec l|.
\label {me12}
\end{equation}
The diagonality of $\hat p$ comes from the assumed random orientation of scattering particles. The eigenvalues of $\hat p$ are $2p_l$.

For example, in Section \ref{pfb} we show that  the phase matrix $\lvec l|\hat p|l')$ which describes the maximum backward-scattering phase function that can be constructed from the first six Legendre polynomials is
\begin{equation}
\lvec l|\hat p|l')
=2\left[\begin{array}{rrrrrrrr} 1&0&0&0&0&0&0&\cdots \\ 0&-5/7&0&0&0&0&0&\cdots \\ 
 0&0&4/7&0&0&0&0&\cdots \\   0&0&0&-8/21&0&0&0&\cdots\\ 0&0&0&0&5/21&0&0&\cdots \\ 0&0&0&0&0&-25/231&0&\cdots \\ 
0&0&0&0&0&0&0&\cdots\\ 
 \vdots&\vdots &\vdots&\vdots&\vdots&\vdots&\vdots&\ddots\\ \end{array}\right].
\label{mei13}
\end{equation}
All other elements of the $\infty\times\infty$ matrix (\ref{mei13}) are zero. The six, non-zero diagonal elements, $p_0, p_1, p_2, p_3,p_4, p_5$, alternate in sign. Using  the six values of $p_l$ from (\ref{mei13}) in (\ref{mm6}) one finds that the forward scattering phase function vanishes, $p(1,1)=\sum_l p_l(2l+1)=0$. The backward scattering phase function is $p(-1,1)=\sum_l(-1)^l  p_l(2l+1)= 12$.

The left and right eigenvectors of the direction-cosine operator $\hat\mu$ are also left and right eigenvectors of the direction-secant operator $\hat\varsigma=\hat\mu^{-1}$,
\begin{equation}
\langle\mu|\hat\varsigma = \langle\mu|\varsigma,\quad\hbox{and}\quad \hat\varsigma|\mu\rangle = \varsigma|\mu\rangle,\quad\hbox{where}\quad \varsigma = \frac{1}{\mu}.\label{me14}
\end{equation}
In continuous $\mu$-space, the matrix elements of the direction-secant operator $\hat\varsigma$ are simply
\begin{equation}
\langle \mu|\hat\varsigma|\mu'\rangle = \frac{1}{\mu} \delta(\mu-\mu').
\label {me16}
\end{equation}
Like $\hat\mu$ of (\ref{me3}), the matrix elements of $\hat\varsigma$ in multipole space are ratios of integers. The elements in the first six rows and columns are
\begin{eqnarray}
\lvec l|\hat\varsigma|l')
&=&\int_{-1}^{1}d\mu\frac{2l'+1}{2\mu}P_l(\mu)P_{l'}(\mu)\nonumber\\
&=&\left[\begin{array}{rrrrrrr} 0&3&0&-14/3&0&88/15&\cdots \\ 1&0&0&0&0&0&\cdots \\ 
 0&0&0&7/3&0&-44/15&\cdots \\   -2/3&0&5/3&0&0&0&\cdots\\ 0&0&0&0&0&11/5&\cdots \\ 8/15&0&-4/3&0&9/5&0&\cdots \\ 
 \vdots&\vdots &\vdots&\vdots&\vdots&\vdots&\ddots\\ \end{array}\right]
\label{me17}
\end{eqnarray}

We define the operator for reflections of a basis vector $|\mu\rangle $, with eigenvalue $\mu$, to a basis vector $|-\mu\rangle$, with an equal and opposite eigenvalue, $-\mu$ by
\begin{equation}
\hat r|\mu\rangle = |-\mu\rangle.
\label {me18}
\end{equation}
The matrix elements of the reflection operator in $\mu$-space are
\begin{equation}
\langle \mu |\hat r|\mu'\rangle = \delta(\mu+\mu').
\label {me20}
\end{equation}
The reflection operator  $\hat r$ is analogous to the parity operator for a quantum mechanical wave function.
For multipole space one can use (\ref{vc8}), (\ref{vc22}) and (\ref{me18}) to show that 
\begin{equation}
\hat r|l)=(-1)^l |l\rangle,
\label {me22}
\end{equation}
From (\ref{me22}) we find the matrix elements 
\begin{equation}
\lvec l|\hat r|l')=(-1)^l \delta_{l\,l'}.
\label {me24}
\end{equation}
In $l$-space, the reflection operator $\hat r$ is diagonal with the elements alternating from 1 to -1 along the main diagonal.

The reflection operator is its own inverse
\begin{equation}
\hat r \hat r =\hat 1.
\label {me26}
\end{equation}
The direction-cosine operator $\hat\mu$ and direction-secant operator $\varsigma$ are odd under reflections 
\begin{equation}
\hat r\hat\mu \hat r =-\hat\mu,\quad\hbox{and}\quad\hat r\hat\varsigma\hat r = -\hat\varsigma.
\label {me28}
\end{equation}
The scattering-phase operator $\hat p$ is even under reflections
\begin{equation} \hat r \hat p\hat r =\hat p.
\label {me30}
\end{equation}
\subsection{Vector equation of transfer}
In the notation discussed above, the steady-state version of the equation of transfer (\ref{in10}) becomes 
\begin{equation}
\left(\hat\mu\frac{\partial}{\partial\tau}+\frac{\partial}{\partial\vartheta}+\hat 1\right) |I \}=(1-\tilde \omega) |B\}+\frac{\tilde\omega}{2}\hat p|I\}.\label{vet2}
\end{equation}
To simplify equations in this section, we will not explicitly show the dependence of variables  on optical depth $\tau$ or relative time $\vartheta$. It will be understood that $|I\}=|I(\tau,\vartheta)\}$, $|B\}=|B(\tau,\vartheta)\}$, etc.
In (\ref{vet2}) the Planck-intensity vector,
\begin{equation}
|B\}=|0)B,\label{vet4}
\end{equation}
describes isotropic thermal emission by greenhouse molecules or cloud particulates. Here $|0)$ is the right basis vector, $|l)$,  of (\ref{vc22}) for $l=0$.

From (\ref{vet2}) we see that it is natural to 
introduce an {\it efficiency matrix}
\begin{equation}
\hat\eta = \hat 1-\frac{\tilde \omega}{2}\hat p.
\label{vet6}
\end{equation}
From (\ref{me10}) we see that the multipole basis vectors $|l)$ and $\lvec l|$ are eigenvalues of $\hat \eta$, 
\begin{equation}
\lvec l|\hat\eta=\lvec l|\eta_l,\quad\hbox{and}\quad \hat\eta |l)=\eta_l|l),\quad\hbox{where}\quad \eta_l = 1-\tilde \omega p_l.
\label{vet6a}
\end{equation}
From  (\ref{mm14})  we note the special case of (\ref{vet6a}),
\begin{eqnarray}
 \lvec 0|\hat\eta 
= \lvec l|(1-\tilde\omega),\quad\hbox{and}\quad\hat\eta |0)= (1-\tilde\omega)|0).
\label{vet8}
\end{eqnarray}
The efficiency matrix (\ref{vet6}) is even under reflections
\begin{equation}
\hat r\hat\eta\hat r = \hat\eta
\label{vet7}
\end{equation}
Using (\ref{vet4}), (\ref{vet6}), (\ref{vet6a}) and (\ref{vet8}) we write the equation of transfer (\ref{vet2}) as
\begin{equation}
\left(\hat\mu\frac{\partial}{\partial \tau}+\frac{\partial}{\partial\vartheta} +\hat\eta\right) |I \}=\hat\eta |B\}.
\label{vet10}
\end{equation}
\subsubsection{Spatially uniform atmosphere}
An instructive special case of (\ref{vet10}) is a spatially uniform atmosphere, where neither $|I\}$ nor $|B\}$ depend on optical depth $\tau$. Then we can set $\partial/\partial\tau = 0$ in (\ref{vet10}) and write it as
\begin{equation}
\left(\frac{d}{d\vartheta} +\hat\eta\right) |I \}=\hat\eta |B\}.
\label{sua2}
\end{equation}
If the Planck intensity $B$ is independent of the relative time $\vartheta$, the solution of (\ref{sua2}) is 
\begin{equation}
|I (\vartheta)\}=e^{-\hat\eta\vartheta}|I(0)\}+\left(1-e^{-\eta_0\vartheta}\right)|B\}.
\label{sua4}
\end{equation}
Here $|I(0)\}$ is the intensity vector at relative time $\vartheta =0$. The exponential operator can be written as
\begin{equation}
e^{-\hat\eta\vartheta}=\sum_{l=0}^{\infty}e^{-\eta_l\vartheta}|l)\lvec l|.
\label{sua6}
\end{equation}
From (\ref{sua6}) we see that the eigenvalues $\eta_l=1-\tilde\omega p_l$ of (\ref{vet6a}) are the decay rates of the $l$th multipole $\lvec l|I(0)\}$ of the initial intensity distribution.
As the relative time approaches infinity, $\vartheta\to \infty$, we see from (\ref{sua4}) that $|I(\vartheta\}\to |B\}$, provided that $\eta_0=1-\tilde\omega >0$. For $\tilde\omega =1$, we have $\eta_0=0$ and 
$|I(\infty)\}=|0)\lvec 0|I(0)\}$. Without absorption, that is, with $\tilde\omega = 1$, there can be no coupling between the energy of the radiation and that stored in the scattering molecules or cloud particulates, which is characterized by the atmospheric temperature $T$, which sets the value of the Planck-brightness $B$ of (\ref{in18}).

\subsubsection{Time-independent atmosphere}
In practice, a
more important limit than that of spatial uniformity, discussed above, is negligibly slow changes in time. In this limit we can set $\partial/\partial\theta =0$ in (\ref{vet10}) and multiply both sides of the equation on the left by the direction-secant operator 
$\hat\varsigma =\hat\mu^{-1}$ to find the steady-state equation of transfer
\begin{equation}
\left(\frac{d}{d\tau}+\hat\kappa\right) |I \}=\hat\kappa|B\}.
\label{vet14}
\end{equation}
The {\it exponentiation-rate operator} $\hat\kappa$, is
\begin{equation}
\hat\kappa = \hat \varsigma\hat\eta,
\label{vet12}
\end{equation}
One can use (\ref{me28}) and (\ref{vet7}) to show that
$\hat \kappa$  is odd under reflections
\begin{equation}
\hat r \hat\kappa \hat r= \hat r \hat \varsigma\hat r \hat r \hat\eta\hat r = -\hat\varsigma\hat \eta = -\hat\kappa.
\label{vet16}
\end{equation}
The solution of (\ref{vet14}) has some formal similarities to (\ref{sua4}). But there are important differences, so we postpone further discussion until after we have outlined the properties of the $2n$-dimensional space that we will use to describe radiative transfer.
\section{2n-dimensional spaces\label{fdms}}
 For the special case of pure absorption with no scattering ($\tilde\omega = 0$), $\hat\eta\to \hat 1$ and $\hat\kappa\to \hat\varsigma$. Then we can multiply 
the  steady-state equation of transfer (\ref{vet14}) on the left with $\langle \mu|$ to find
\begin{equation}
\left(\frac{\partial}{\partial\tau}+\varsigma\right) I(\mu,\tau)=\varsigma B(\tau).
\label{lg2}
\end{equation}
Equation (\ref{lg2}) is often called the Schwarzschild equation. It describes emission and absorption of radiation in a non-scattering atmosphere, for example,  the transfer of thermal radiation through cloud-free air with greenhouse-gases\cite{WH}. It is relatively easy to solve (\ref{lg2}). It is much harder to solve the more general equation of transfer (\ref{vet14}) with arbitrary combinations of scattering and absorption.

A way to accurately calculate radiation transfer with any combination of absorption, emission and scattering  is to use the $2n$-stream method described in this paper.  Many of the basic ideas of this method  were outlined in Chapter II of Chandrasekhar\cite{Chandrasekhar}.

Instead of representing the angular dependence of the intensity exactly  with an infinite number of multipole moments, $I_l$, as in (\ref{mm4}),  one can generalize  Schuster's 2-stream method \cite{Schuster} and approximate  the intensity with the coefficients, $I_0,\,I_1, \ldots,\,I_{2n-1}$, of the expansion (\ref{mm4}).  Rather than specifying the value of $I(\mu)$ for all direction cosines, $|\mu|\le 1$, it is only necessary to specify $2n$ values of $I(\mu_i)$ at $2n$ {\it Gauss-Legendre cosines}, $\mu_i$ \cite{Gauss}. As mentioned in \S{\bf20} of Chandrasekhar\cite{Chandrasekhar}, the $\mu_i$ are the roots of the Legendre polynomial $P_{2n}$, that is,
\begin{equation}
P_{2n}(\mu_i)=0.
\label{lg4}
\end{equation}
We assume that the roots of (\ref{lg4}) are ordered such that
\begin{equation}
-1<\mu_1<\mu_2<\cdots<\mu_{2n}<1.
\label{lg6}
\end{equation}
 Because of the symmetry $P_{2n}(-\mu)=P_{2n}(\mu)$, we see that  the roots (\ref{lg4}) occur as equal and opposite pairs, $\mu_1=-\mu_{2n}$, $\mu_2=-\mu_{2n-1}$, or in general,
\begin{equation}
\mu_i=-\mu_{r(i)},
\label{lg8}
\end{equation}
where the reflection function for the indices, $i=1,2,\ldots, 2n$, is
\begin{equation}
r(i)=2n+1-i.
\label{lg10}
\end{equation}
We write the expansion of the abstract intensity vector $|I\}=|I(\tau)\}$ on the first $2n$ multipole bases as
\begin{equation}
| I\}=\sum_{l=0}^{2n-1}|l) \lvec l| I\}=\sum_{l=0}^{2n-1}|l)  I_l
\label{lg12}
\end{equation}
The basis vectors $|l)$ of (\ref{lg12}) were defined by (\ref{vc22}).  For $l=0,1,2,\ldots,2n-1$, the intensity multipoles $I_l$ of (\ref{mm2})  are given by the discrete, Gauss-Legendre quadratures\cite{Gauss} 
\begin{equation}
 I_l=\lvec l| I\}=\int_{-1}^1d\mu \,P_l(\mu)I(\mu) = \frac{1}{2}\sum_{i=1}^{2n}w_i P_l(\mu_i)I(\mu_i).
\label{lg14}
\end{equation}
There are many equivalent formulas for the weights $w_i$ of (\ref{lg14}). For our purposes, a useful one, that makes it clear that $w_i>0$, is
\begin{eqnarray}
\frac{1}{w_i}=\sum_{l=0}^{2n-1}\frac{2l+1}{2}P_l^2(\mu_i).
\label{lg16}
\end{eqnarray}
Chandrasekhar\cite{Chandrasekhar} uses the symbol $a_i$ to denote the weight $w_i$ and he evaluates them with his formula \S{\bf 20}(5), which gives the same numerical value as (\ref{lg16}).
 
Since $P_l(-\mu)=(-1)^lP_l(\mu)$, we see from (\ref{lg16}) and (\ref{lg8}) that 
\begin{equation}
w_i =w_{r(i)},
\label{lg18}
\end{equation}
where the index reflection function $r(i)$ was defined by (\ref{lg10}).
It is not hard to prove that the weights sum to 2,
\begin{equation}
\sum_{i=1}^{2n}w_i = 2.
\label{lg20}
\end{equation}
See, for example, \S{\bf 25}(8) of Chandraskhar\cite{Chandrasekhar}. For the limit $2n\to \infty$, the weights $w_i$ approach the infinitesimals $d\mu$ of the integral (\ref{mm2}), $w_i\to d\mu$.
The multipole moments evaluated with the Gauss-Legendre quadrature (\ref{lg14}) are identical to those of the continuous integral (\ref{mm2}) if $I_l=0$ for $l\ge 2n$ \cite{Gauss}.

Setting $\mu = \mu_i$ in (\ref{mm4}) we find the inverse of (\ref{lg14}),
\begin{equation}
I(\mu_i)
=\sum_{l=0}^{2n-1}(2l+1)P_l(\mu_i) I_l.
\label{lg22}
\end{equation}
Using (\ref{vc2}), (\ref{lg12})  and (\ref{lg14}) we find
\begin{eqnarray}
I(\mu)=\langle\mu|I\}=\sum_{i=1}^{2n}\left\{\sum_{l=0}^{2n-1}\frac{2l+1}{2}P_l(\mu)P_l(\mu_i)w_i\right\} I(\mu_i).
\label{lg24}
\end{eqnarray}
Setting $\mu = \mu_{i'}$ in (\ref{lg24}) implies the identity,
\begin{equation}
\sum_{l=0}^{2n-1}\frac{2l+1}{2}P_l(\mu_{i'})P_l(\mu_i)w_i =\delta_{i'\,i}.
\label{lg26}
\end{equation}
For $i'=i$,  (\ref{lg26}) gives the formula (\ref{lg16}) for the weight $w_i$. The inverse of (\ref{lg26}) is given later as (\ref{sdbv42}).
\subsection{Index  conventions \label{ic}}
In our subsequent discussions of the $2n$-stream model,
many  equations will involve sums over all $2n$ streams,  or separate sums over the $n$ upward streams or the $n$ downward streams. To simplify notation, we will label the complete set of $2n$ streams with the indices
\begin{equation}
i, i', i'',\ldots = 1,2,3,\ldots, 2n.
\label{ic2}
\end{equation}
The $n$ downward streams, with $\mu_j<0$, will be labeled with the indices
\begin{equation}
j, j', j'',\ldots = 1,2,3,\ldots,n.
\label{ic4}
\end{equation}
The $n$ upward streams, with $\mu_k>0$ will be labeled with the indices
\begin{equation}
k, k', k'',\ldots = n+1,n+2,n+3,\ldots,2n.
\label{ic6}
\end{equation}
We simplify the summation symbols to
\begin{eqnarray}
 \sum_{i=1}^{2n}\cdots  &\to& \sum_i\cdots\, ,\nonumber\\
\sum_{j=1}^{n}\cdots&\to&\sum_j\cdots\,  \nonumber\\
 \sum_{k=n+1}^{2n}\cdots&\to&\sum_k\cdots\, .
\label{ic8}
\end{eqnarray}
In like manner,  we will label multipole moments of the 
$2n$-stream model with the indices 
\begin{equation}
l, l', l'',\ldots =0,1,2,\ldots, 2n-1,
\label{ic10}
\end{equation}
and we simplify the summation symbols to
\begin{equation}
\sum_{l=0}^{2n-1}\cdots\to \sum_l\cdots\, .
\label{ic12}
\end{equation}
\begin{figure}[t]
\postscriptscale{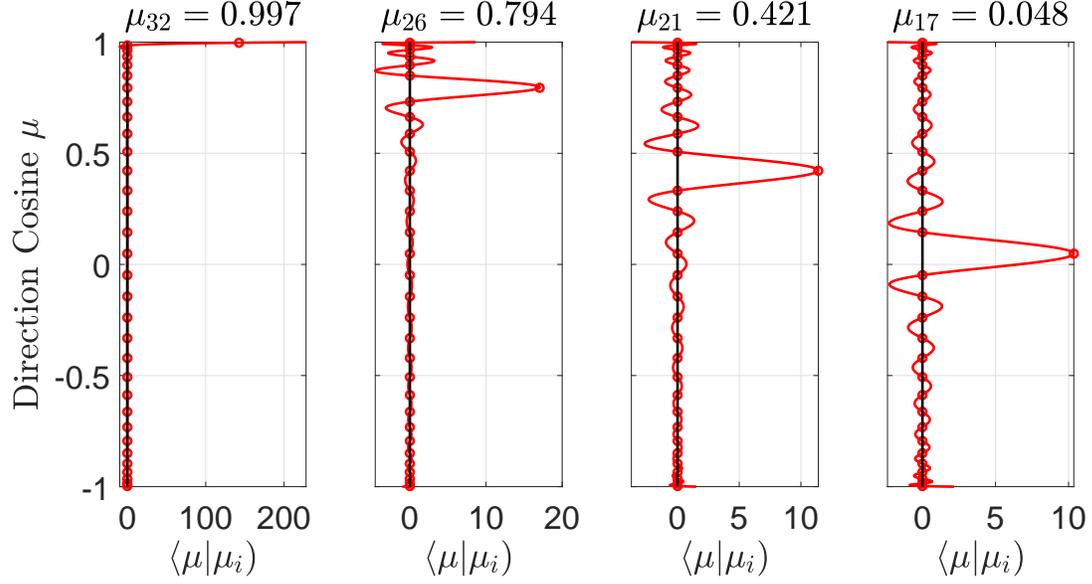}{1}
\caption{The red curves are the projections $\langle\mu|\mu_i)$ of the direction-cosine basis vectors $|\mu_i)$ into continuous  $\mu$-space, as given by (\ref{sdbv12}) or (\ref{sdbv20})  According to  (\ref{sdbv18}), at the sample points $\mu =\mu_i$ the projections have the values   $\langle\mu_i|\mu_{i'})=w_i^{-1}\delta_{ii'}$, which are shown as  the small
red circles.  There are $2n=32$ streams.    The sample points $\mu_i$ are given at the top of the panels. The corresponding weights are $[w_{32}, w_{26}, w_{21}, w_{17}] =$ $[0.0070, 0.0587, 0.0877, 0.0965]$.
\label{mumuj}}
\end{figure}
\begin{figure}[t]
\postscriptscale{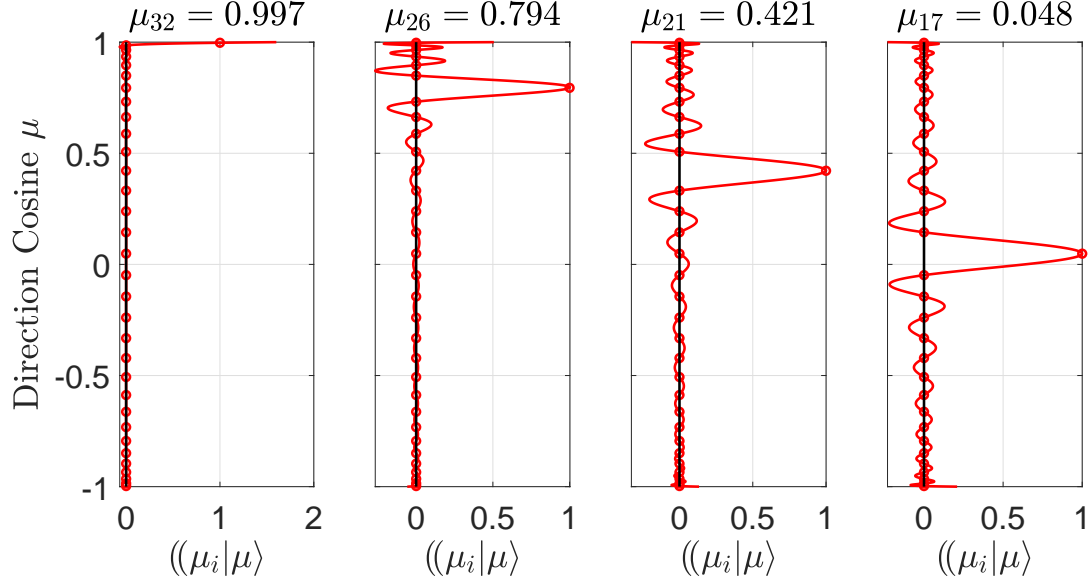}{1}
\caption{Like Fig. \ref{mumuj}, but here the red curves are the projections $\lvec\mu_i|\mu\rangle$ of the left  basis vectors $\lvec\mu_i|$ into $\mu$-space, as given by (\ref{sdbv14}) or (\ref{sdbv22}). According to  (\ref{sdbv18}), at the sample points $\mu =\mu_i$, the projections have the values   $\lvec \mu_{i'}|\mu_i\rangle = \delta_{ii'}$, which are shown as  the small
red circles.
\label{mujmu}}
\end{figure}
\subsection{Stream basis vectors $|\mu_i)$ and $\lvec\mu_i|$}
In analogy to (\ref{vc14}), we define  $\mu$-space column  vectors $|\mu_{i'})$ by
\begin{equation}
\lvec l|\mu_{i'}) =\frac{1}{2}P_l(\mu_{i'}).
\label{sdbv2}
\end{equation}
The multipole components of the corresponding row vector $\lvec\mu_i|$ are
\begin{equation}
\lvec \mu_i|l)=w_i(2l+1)P_l(\mu_i).
\label{sdbv4}
\end{equation}
From (\ref{sdbv2}) and (\ref{sdbv4}) we see that (\ref{lg26}) can be interpreted  as the orthonormality relation
\begin{equation}
\lvec \mu_i|\mu_{i'})=\sum_l\lvec \mu_i|l)\lvec l|\mu_{i'})=\delta_{i i'}.
\label{sdbv6}
\end{equation}
The {\it stream basis vectors} $|\mu_i)$ and $\lvec \mu_i|$ have the completeness property 
\begin{equation}
\hat 1=\sum_i|\mu_i)\lvec \mu_i|=\mathcal{M}_{\bf d}+ \mathcal{M}_{\bf u}.
\label{sdbv8}
\end{equation}
Here we have defined projection matrices $\mathcal{M}_{\bf d}$ and $\mathcal{M}_u$ for the downward and upward parts of $\mu$-space by
\begin{equation}
\mathcal{M}_{\bf d }=\sum_j|\mu_j)\lvec \mu_j|,\quad\hbox{and}\quad
\mathcal{M}_{\bf u} =\sum_k|\mu_k)\lvec \mu_k|.
\label{sdbv9}
\end{equation}
The projection operators have the multiplication property
\begin{equation}
\mathcal{M}_{\bf q}\mathcal{M}_{\bf q'}  =\mathcal{M}_{\bf q}\delta_{\bf qq'},
\label{sdbv9a}
\end{equation}
where the indices $q$ and $q'$ can have the values $u$ or $d$.
We can use (\ref{vc22}) and (\ref{sdbv2}) to write the projection of $|\mu_i)$ onto $\mu$-space as
\begin{equation}
\langle \mu|\mu_i) = \sum_l\langle \mu|l)\lvec l|\mu_{i}) =\sum_l\frac{2l+1}{2}P_l(\mu)P_l(\mu_{i}).
\label{sdbv12}
\end{equation}
In like manner
we can use (\ref{vc14}) and (\ref{sdbv4}) to write the projection of $\lvec \mu_i|$ onto $\mu$-space as
\begin{equation}
\lvec \mu_i|\mu\rangle =\sum _l\lvec  \mu_i|l)\lvec l|\mu\rangle =w_i \sum_l\frac{2l+1}{2} P_l(\mu_i)P_l(\mu).
\label{sdbv14}
\end{equation}
From (\ref{sdbv12}) 
we see that $\langle\mu|\mu_i)$ has unit area
\begin{eqnarray}
\int_{-1}^1d\mu\,\langle \mu|\mu_i) 
&=&\sum_l\int_{-1}^1d\mu\frac{2l+1}{2}P_l(\mu)P_l(\mu_i)\nonumber\\
&=&\sum_l\delta_{l0}P_l(\mu_i)\nonumber\\
&=&1.
\label{sdbv15}
\end{eqnarray}
From (\ref{sdbv12}) and (\ref{sdbv14}) we see that $\langle \mu|\mu_i)$ and $\lvec \mu_i|\mu\rangle$ are simple multiples of each other
\begin{equation}
\lvec \mu_i|\mu\rangle = w_i\langle\mu|\mu_i).
\label{sdbv16}
\end{equation}
From (\ref{lg26}), (\ref{sdbv12}) and (\ref{sdbv14}) we see that for $\mu = \mu_{i'}$,
\begin{equation}
\langle \mu_{i'}|\mu_i) =\frac{\delta_{ii'}}{w_i},\quad\hbox{and}\quad \lvec\mu_i|\mu_{i'}\rangle = \delta_{ii'}.
\label{sdbv18}
\end{equation}
Examples of the projections $\langle\mu|\mu_j)$ of (\ref{sdbv12}) are shown in Fig. \ref{mumuj} for an expansion on $2n=32$ Legendre polynomials. The projections, $\langle\mu|\mu_i)$, have their largest amplitudes for direction cosines $\mu\approx\mu_i$. The corresponding projections $\lvec \mu_i|\mu\rangle$ of (\ref {sdbv14}) are shown in Fig. \ref{mujmu}. 

The direction cosines $\mu_{i'}$ are defined by (\ref{lg4}) as the roots of the polynomial $P_{2n}(\mu)$.
The polynomial  $\langle\mu|\mu_i)$ of (\ref{sdbv12}) and $\lvec\mu_i|\mu\rangle$  of (\ref{sdbv14}) are of degree $2n-1$ in $\mu$. From (\ref{sdbv18}) we see that both $\langle\mu|\mu_i)$ and $\lvec\mu_i|\mu\rangle$  have roots at $\mu =\mu_{i'}$, with the exception of 
$\mu_{i'}=\mu_i$, for which $P_{2n}(\mu_{i}) = 0$, but for which (\ref{sdbv18}) show that $\langle\mu_i|\mu_i)=1/w_i$ and $\lvec\mu_i|\mu_i\rangle=1$.   The projections  (\ref{sdbv12})  and  (\ref{sdbv14}) of the stream basis states must therefore be given by the formulas
\begin{equation}
\langle \mu|\mu_i) =\frac{P_{2n}(\mu)}{w_i P_{2n}'(\mu_i)(\mu-\mu_i)},
\label{sdbv20}
\end{equation}
and
\begin{equation}
\lvec \mu_i|\mu\rangle =\frac{P_{2n}(\mu)}{ P_{2n}'(\mu_i)(\mu-\mu_i)}.
\label{sdbv22}
\end{equation}
The denominators of (\ref{sdbv20})  and (\ref{sdbv22}) ensure that $\langle\mu_i|\mu_i)=1/w_i$, and $\lvec\mu_i|\mu_i\rangle=1$, in agreement with (\ref{sdbv18}). 
Substituting (\ref{sdbv20}) into (\ref{sdbv15}) gives  Chandrasekhar's\cite{Chandrasekhar} formula \S{\bf 20}(5) for the weight $w_i $.

Multiplying the basis vector $|\mu_i)$  on the left by the reflection operator $\hat r$, and recalling from (\ref{me22}) that $\hat r|l)=(-1)^l|l)$, we find
\begin{eqnarray}
\hat r|\mu_i) &=&\sum_l\hat r |l)\lvec l|\mu_i)
=\sum_l(-1)^l |l)\frac{1}{2}P_l(\mu_i)
=\sum_l |l)\frac{1}{2}P_l(-\mu_i)\nonumber\\
&=&\sum_l |l)\frac{1}{2}P_l(\mu_{r(i)})
=\sum_l |l)\lvec l|\mu_{r(i)}),
\label{sdbv32}
\end{eqnarray}
or 
\begin{equation}
\hat r|\mu_i) =|\mu_{r(i)}).
\label{sdbv34}
\end{equation}
In like manner we find
\begin{equation}
\lvec \mu_i|\hat r =\lvec\mu_{r(i)}|.
\label{sdbv36}
\end{equation}
Using (\ref{sdbv2}) and (\ref{sdbv8}) we can write
\begin{equation}
\lvec l|=\sum_{i}\lvec l|\mu_i)\lvec\mu_i|=\frac{1}{2}\sum_{i}P_l(\mu_i)\lvec\mu_i|
\label{sdbv38}
\end{equation}
and using (\ref{sdbv4}) and (\ref{sdbv8}) we find
\begin{equation}
|l)=\sum_{i}|\mu_i)\lvec\mu_i|l)=\sum_{i}|\mu_i)w_i(2l+1)P_l(\mu_i)
\label{sdbv40}
\end{equation}
Multiplying (\ref{sdbv40}) on the left by (\ref{sdbv38}) and using (\ref{sdbv6}), we find the inverse of the othogonality relation (\ref{lg26}), and the discrete analog of (\ref{mm10}),
\begin{equation}
\lvec l|l')=\delta_{ll'} 
=\sum_{i} w_i\frac{2l+1}{2}P_l(\mu_i)P_{l'}(\mu_i).
\label{sdbv42}
\end{equation}
Setting $l = l' = 0$ in (\ref{sdbv42}) gives (\ref{lg20}): the weights $w_i$ sum to 2.

Useful special cases of (\ref{sdbv38}) and  (\ref{sdbv40}) for $l=0$ and $l=1$ are
\begin{eqnarray}
\lvec 0|=\frac{1}{2}\sum_{i}\lvec\mu_i|,\quad&\hbox{and}&\quad |0)=\sum_i|\mu_i)w_i.
\label{sdbv44}\\
\lvec 1|=\lvec 0|\hat\mu=\frac{1}{2}\sum_{i}\mu_i\lvec\mu_i|,\quad&\hbox{and}&\quad |1)
=3\hat\mu|0)=3\sum_i|\mu_i)\mu_i w_i.
\label{sdbv46}
\end{eqnarray}
\subsection{Eigenvectors and eigenvalues of $\hat\mu$}
The direction-cosine basis-vectors $|\mu_i)$ and $\lvec \mu_i|$ discussed above are right and left eigenvectors of the direction cosine operator $\hat\mu$, thought of as the $2n\times 2n$ upper left corner of the matrix (\ref{me3}),
\begin{equation}
\mu_i|\mu_i)-\hat\mu|\mu_i)=\hat 0,\quad\hbox{and}\quad \lvec\mu_i|\mu_i-\lvec \mu_i|\hat\mu=\hat 0.
\label{emu2}
\end{equation}
Here and subsequently we will use the symbol $\hat 0$ to represent an array of zeros. The dimensions of the array will be given by the context. 

We can verify that the first equation of (\ref{emu2})  is true by multiplying it on the left with each of the $2n$ left multipole basis vectors $\lvec l|$ for $l=0,1,2,\ldots,2n-1$. If the resulting $2n$ equations are valid, then so is (\ref{emu2}).
Multiplying (\ref{emu2}) on the left by $\lvec l|$, with $l=0,1,2,\ldots,2n-2$, and using (\ref{sdbv2}) and (\ref{mm18}) we find
\begin{eqnarray}
\mu_i\lvec l|\mu_i)-\lvec l|\hat\mu|\mu_i)&=& \frac{1}{2}\mu_iP_l(\mu_i)
-\frac{1}{2}\sum_{l'=0}^{2n-1}\lvec l|\hat\mu|l')P_l'(\mu_i)\nonumber\\
&=& \frac{1}{2}\bigg[\mu_iP_l(\mu_i)
-\sum_{l'=0}^{2n-1}\hat\mu_{ll'}P_l'(\mu_i)\bigg]\nonumber\\
&=& \frac{1}{2}\bigg[\mu_iP_l(\mu_i)
-\frac{lP_{l-1}(\mu_i)+(l+1)P_{l+1}(\mu_i)}{2l+1}\bigg]\nonumber\\
&=&0.
\label{emu4}
\end{eqnarray}
The final step follows from Bonnet's recursion formula (\ref{mm20}). Multiplying (\ref{emu2}) on the left by $\lvec 2n-1|$, the highest multipole of the $2n$-space, we find
\begin{eqnarray}
\mu_i\lvec 2n-1|\mu_i)-\lvec 2n-1|\hat\mu|\mu_i)
&=& \frac{1}{2}\bigg[\mu_iP_{2n-1}(\mu_i)
-\frac{(2n-1)P_{2n-2}(\mu_i)}{4n-1}\bigg]\nonumber\\
&=& \frac{1}{2}\bigg[\mu_iP_{2n-1}(\mu_i)
-\frac{(2n-1)P_{2n-2}(\mu_i)+2nP_{2n}(\mu_i)}{4n-1}\bigg]\nonumber\\
&=&0.
\label{emu6}
\end{eqnarray}
According to (\ref{lg4}), $P_{2n}(\mu_i)=0$ for any of the $2n$ possible values of $\mu_i$. So we have subtracted a term $2n P_{2n}(\mu_i)/2(4n-1)=0$ from the right side of the first line of (\ref{emu6}) to get
the second line, which is zero because of Bonnet's recursion formula (\ref{mm20}). One can use similar arguments to show that the second equation of (\ref{emu2}) is also valid. This completes the proof of (\ref{emu2}).

Multiplying (\ref{sdbv8}) on the right or left by $\hat \mu$ and using (\ref{emu2})  we find  that the representation of $\hat\mu$ in $\mu$-space is diagonal, 
\begin{equation}
\hat\mu=\sum_{i}\mu_i|\mu_i)\lvec \mu_i|=\hat\mu_{\bf d}+\hat\mu_{\bf u}.
\label{eobd4}
\end{equation}
In analogy to (\ref{sdbv8}) we have written the downward and upward parts of $\hat\mu$ by
\begin{eqnarray}
\hat\mu_{\bf d}&=&\hat\mu \mathcal{M}_{\bf d}= \mathcal{M}_{\bf d}\hat\mu=\sum_{j}\mu_j|\mu_j)\lvec \mu_j|\label{eobd5a}\\
\hat\mu_{\bf u}&=&\hat\mu\mathcal{M}_{\bf u} =\mathcal{M}_{\bf u}\hat\mu =\sum_{k}\mu_k|\mu_k)\lvec \mu_k|.
\label{eobd5}
\end{eqnarray}

Since the $2n\times 2n$  upper left corner of the direction-secant matrix $\hat\varsigma$  of (\ref{me17}) is the
inverse of the $2n\times 2n$  upper left corner direction cosine matrix $\hat\mu$ of (\ref{me3}), we can write
\begin{equation}
\hat\varsigma=\hat\mu^{-1}=\sum_{i}\frac{1}{\mu_i}|\mu_i)\lvec \mu_i|=
\sum_{i}\varsigma_i|\mu_i)\lvec \mu_i|=\hat\varsigma_{\bf d}+\hat\varsigma_{\bf u},
\label{eobd6}
\end{equation}
where
\begin{eqnarray}
\hat\varsigma_{\bf d}&=&\hat\varsigma \mathcal{M}_{\bf d}= \mathcal{M}_{\bf d}\hat\varsigma=\sum_{j}\varsigma_j|\mu_j)\lvec \mu_j|\label{eobd7a}\\ 
\hat\varsigma_{\bf u}&=&\hat\varsigma \mathcal{M}_{\bf u}=\hat\varsigma \mathcal{M}_{\bf u}=\sum_{k}\varsigma_k|\mu_k)\lvec \mu_k|.
\label{eobd7}
\end{eqnarray}
In analogy to (\ref{emu2}), the left and right eigenvectors of the direction-secant matrix are defined, aside from normalization, by 
\begin{equation}
\lvec \varsigma_i|\hat\varsigma=\lvec \varsigma_i|\varsigma_i,\quad\hbox{and}\quad \hat\varsigma|\varsigma_i)=\varsigma_i|\varsigma_i).
\label{eobd10}
\end{equation}
We see from (\ref{eobd6}) and (\ref{eobd10}) that the left and right eigenvectors of $\hat\mu$ and $\hat \varsigma$ can be chosen to be identical. The eigenvalues are inverses of each other
\begin{equation}
\lvec \varsigma_i|=\lvec \mu_i|,\quad |\varsigma_i) =|\mu_i),\quad\hbox{and}\quad \varsigma_i=\frac{1}{\mu_i}.
\label{eobd12}
\end{equation}

We recall that in $l$-space, the reflection matrix $\hat r$ has alternating 1's and -1's along the main diagonal and zeros elsewhere,
\begin{equation}
\hat r  =\sum_l(-1)^l|l)\lvec l|.
\label{eobd14}
\end{equation}
Multiplying (\ref{sdbv8}) on the right or left by the reflection operator $\hat r$ and using 
(\ref{sdbv34}) and (\ref{sdbv36}) we find
\begin{equation}
\hat r=\sum_{i}|\mu_i)\lvec \mu_{r(i)}|.
\label{eobd16}
\end{equation}
In the direction-cosine basis, the reflection matrix (\ref{eobd16}) is a $2n\times 2n$  anti-diagonal matrix, for which the non-zero elements, extending from the lower left to the upper right, are all 1's. 

Multiplying (\ref{eobd4}) on the left and right by $\hat r$, and using (\ref{sdbv34}), (\ref{sdbv36})  and (\ref{lg8}),  we find
\begin{eqnarray}
\hat r \hat\mu \hat r &=&\sum_{i}\mu_i\hat r |\mu_i)\lvec \mu_i|\hat r\nonumber\\
&=&\sum_{i}\mu_i|\mu_{r(i)})\lvec \mu_{r(i)}|\nonumber\\
&=&-\sum_{i}\mu_{r(i)}|\mu_{r(i)})\lvec \mu_{r(i)}|.
\label{eobd18}
\end{eqnarray}
or since the set of indices $r(i)$ is the same as the set $i$, 
\begin{eqnarray}
\hat r \hat\mu \hat r =-\hat \mu.
\label{eobd20}
\end{eqnarray}
In like manner we find
\begin{eqnarray}
\hat r \hat\varsigma \hat r =-\hat \varsigma.
\label{eobd22}
\end{eqnarray}
The direction-cosine operator $\hat \mu$, and its inverse, the direction-secant operator $\hat\varsigma = \hat\mu^{-1}$ are odd under reflection.
\subsection{Weighted and unweighted variables}
We can expand variables that depend on the direction cosine $\mu$  on the $\mu$-space basis vectors $|\mu_i)$. For example, multiplying the abstract vector $|I\}$ that represents the intensity, $I=I(\mu)$, on the left by (\ref{sdbv8}),
we find
\begin{eqnarray}
|I\}=\sum_i|\mu_i)\lvec \mu_i|I\}.
\label{sdbv10a}
\end{eqnarray}
We can use
 (\ref{lg12}), (\ref{sdbv4})  (\ref{lg14}) and  (\ref{lg26})  to write the amplitudes of (\ref{sdbv10a}) as
\begin{eqnarray}
\lvec \mu_i|I\}&=&\sum_l\lvec \mu_i|l)I_l\nonumber\\
&=&\sum_l w_i(2l+1)P_l(\mu_i)\frac{1}{2}\sum_{i'}w_{i'}P_l(\mu_{i'})I(\mu_{i'})\nonumber\\
&=&w_i\sum_{i'}\delta_{ii'}I(\mu_{i'})
\nonumber\\
&=&w_iI(\mu_i).
\label{sdbv10}
\end{eqnarray}
For any function, $f=f(\mu)$ of the direction cosine $\mu$, represented by the abstract vector $|f\}$ we will call
\begin{eqnarray}
\lvec \mu_i|f\}&=&w_i f(\mu_i) = \hbox{the weighted value of $|f\}$ at $\mu = \mu_i$},\nonumber\\
w_i^{-1}\lvec \mu_i|f\}&=& f(\mu_i) =\hbox{the unweighted value of $|f\}$ at $\mu = \mu_i$}.
\label{sdbv11}
\end{eqnarray}
\section{Phase Functions  }
We assume that the diagonal elements, $2p_l$ of the scattering phase matrix $\hat p$ of  (\ref{me12}) are zero if $l\ge 2n$. Then we can write $\hat p$ as
\begin{equation}
\hat p =2\sum_{l}p_l |l)\lvec l|.
\label{eobd36}
\end{equation}
Multiplying (\ref{eobd36}) on the left and right by the reflection operator $\hat r$, and using (\ref{eobd14}) we see that the scattering-phase matrix is even under reflections
\begin{equation}
\hat r \hat p\hat r =\hat p.
\label{eobd38}
\end{equation}
Using (\ref{sdbv2}) and (\ref{sdbv4}) with (\ref{eobd36}) in (\ref{mm6}) we see that
the matrix elements of $\hat p$ in $\mu$-space are 
\begin{eqnarray}
\lvec \mu_i|\hat p|\mu_{i'}) &=&2\sum_{l}p_l\lvec\mu_i |l)\lvec l|\mu_{i'})
\nonumber\\
&=&\sum_{l}p_lw_i(2l+1)P_l(\mu_i) P_l(\mu_{i'})\nonumber\\
&=&w_i p(\mu_i,\mu_{i'}).
\label{eobd40}
\end{eqnarray}
Using (\ref{eobd40}), we write the scattering-phase matrix as
\begin{equation}
\hat p =\sum_{i i'}w_i p(\mu_i,\mu_{i'})|\mu_i)\lvec \mu_{i'}|.
\label{eobd42}
\end{equation}
Summing (\ref{eobd40}) over all stream indices $i$ and using the Gauss-Legendre quadrature  property of (\ref{lg14}) with (\ref{in16}) we find
\begin{eqnarray}
\sum_i\lvec \mu_i|\hat p|\mu_{i'}) 
&=&\sum_i w_i p(\mu_i,\mu_{i'})\nonumber\\
&=&\int_{-1}^1d\mu\, p(\mu,\mu_{i'})\nonumber\\
&=&2.
\label{eobd44}
\end{eqnarray}
Since the weights $w_i$ of (\ref{lg16}) are positive, and the matrix elements $p(\mu_i,\mu_{i'})$ are non-negative according to (\ref{in14}), we see from (\ref{eobd40}) that $\lvec \mu_i|\hat p|\mu_{i'})\ge 0$. This limit, together with (\ref{eobd44}), implies that the elements of the scattering-phase matrix are bounded by
\begin{equation}
0\le \frac{1}{2}\lvec \mu_i|\hat p|\mu_{i'})\le 1. 
\label{eobd46}
\end{equation}
From (\ref{eobd46}) we see that the probability that a  photon is scattered from stream $i'$ to stream $i$ is 
$\lvec \mu_i|\hat p|\mu_{i'})/2$.
For axially symmetric radiation transfer, the scattering phase functions $p(\mu,\mu')$ of (\ref{mm6}) can be fully determined from the  simpler phase function $p(\mu)=p(\mu, 1)$ for scattering vertically propagating light with $\mu'=1$,
\begin{equation}
p(\mu)=p(\mu,1)=\sum_lP_l(\mu)(2l+1)p_l.
\label{rp2}
\end{equation}
The function $p(\mu,\mu')$ of two direction cosines and the function $p(\mu)$ of a single direction cosine are so closely related that we will use the same symbol $p$ for  both.  The context will make clear which is meant.
Here we show that if $p(\mu)$ is non-negative for any value of $|\mu|\le 1$, then $p(\mu,\mu')$ is also nonnegative for any values of $|\mu|\le 1$ and $|\mu'|\le 1$,
\begin{equation}
p(\mu,\mu') \ge 0\quad \hbox{if}\quad  p(\mu)\ge 0.
\label{rp4}
\end{equation}
To be sure a phase function $p(\mu,\mu')$ satisfies the nonnegativity constraint (\ref{in14}) for all allowed values of $\mu$ and $\mu'$, it is sufficient to verify the second inequality of (\ref{rp4}) for the simpler, one-variable function $p(\mu)$ of (\ref{rp2}). More discussion of (\ref{rp2}) can be found in \S{\bf 48} of Chandrasekhar\cite{Chandrasekhar}.
\begin{figure}[t]
\postscriptscale{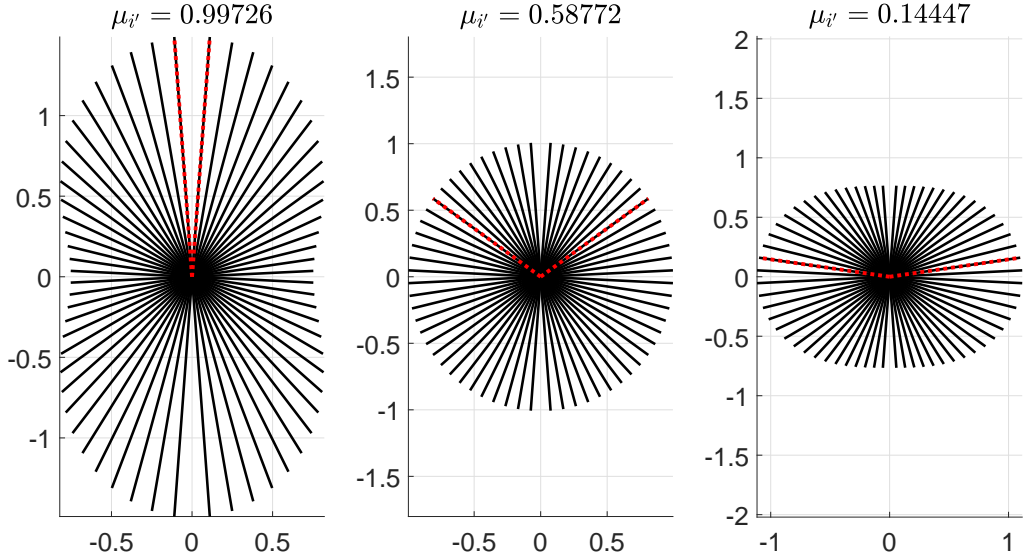}{1}
\caption{The lengths of the black rays are the numerical values of the phase functions $p(\mu_{i},\mu_{i'})$ of (\ref{mm6}) for Rayleigh scattering, discussed in Section \ref{rs}. $\mu_{i}$ and $\mu_{i'}$ are Gauss-Legendre sample values of the direction cosines, defined by (\ref{lg4}).   For this example, there are $2n=32$ streams.
{\bf Left.} A nearly vertical incident stream, indicated by red dotted lines, with zenith angle $\theta_{32} = \cos^{-1}(.9973) = 4.24^{\circ}$. {\bf Middle.} An incident stream, indicated by red dotted lines, with zenith angle $\theta_{23} = \cos^{-1}(.5877) = 54.0^{\circ}$, close to the ``magic angle" $\theta_m =\cos^{-1}(1/\sqrt{3}) = 54.7^{\circ}$, where the Rayleigh scattering  changes from prolate to oblate\cite{Magic}. Here the phase is slightly prolate.  {\bf Right.} A nearly horizontal incident stream, indicated by red dotted lines, with zenith angle $\theta_{17} = \cos^{-1}(.0483) = 87.2^{\circ}$.
\label{ph2}}
\end{figure}
\begin{figure}[t]
\postscriptscale{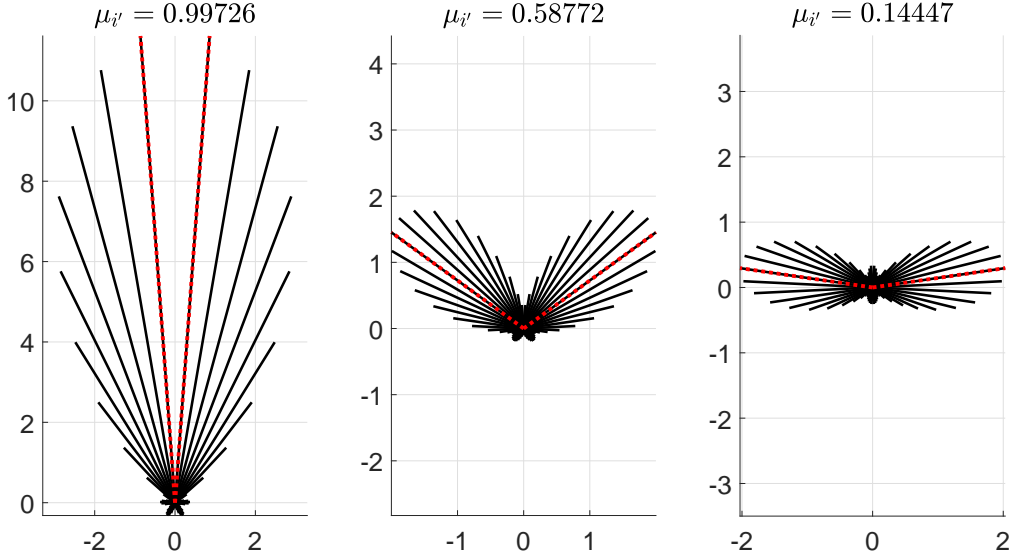}{1}
\caption{The lengths of the black rays are the values of the forward-scattering  phase functions $p(\mu_{i},\mu_{i'})=
\varpi^{\{3\}}(\mu_{i},\mu_{i'})$ of (\ref{pfb16}).  The multipole amplitudes of (\ref{pfb14}) are $[\varpi^{\{3\}}_{0},\varpi^{\{3\}}_{1},
\varpi^{\{3\}}_{2},  \varpi^{\{3\}}_{3}, \varpi^{\{3\}}_{4}, \varpi^{\{3\}}_{5}]$
=$[1, 5/7, 4/7, 8/21, 5/21, 25/231]$. These are listed as decimal fractions in Table \ref{pln}. There are $2n=32$ sampling streams. The direction cosines $\mu_i$ of scattered streams and $\mu_{i'}$ of the incident stream  have the same values as for Fig. \ref{ph2}.
\label{ph3}}
\end{figure}
\begin{figure}[t]
\postscriptscale{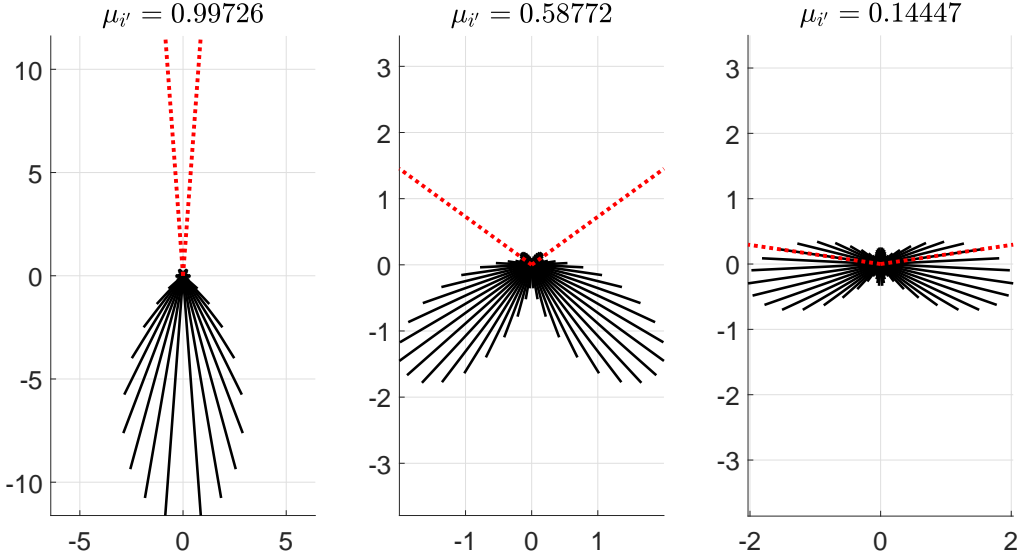}{1}
\caption{The lengths of the black rays are the values of the backward-scattering  phase functions $p(\mu_{i},\mu_{i'})=
 r\varpi^{\{3\}}(\mu_{i},\mu_{i'})$ of (\ref{pfb18}). The multipole amplitudes  are $[\varpi^{\{3\}}_{0},-\varpi^{\{3\}}_{1},
\varpi^{\{3\}}_{2},  -\varpi^{\{3\}}_{3}, \varpi^{\{3\}}_{4}, -\varpi^{\{3\}}_{5}]$=$[1, -5/7, 4/7, -8/21, 5/21, -25/231]$.  There are $2n=32$ sampling streams. The direction cosines $\mu_i$ of scattered streams and $\mu_{i'}$ of the incident stream  have the same values as for Fig. \ref{ph2}.
\label{ph4}}
\end{figure}
\begin{figure}[t]
\postscriptscale{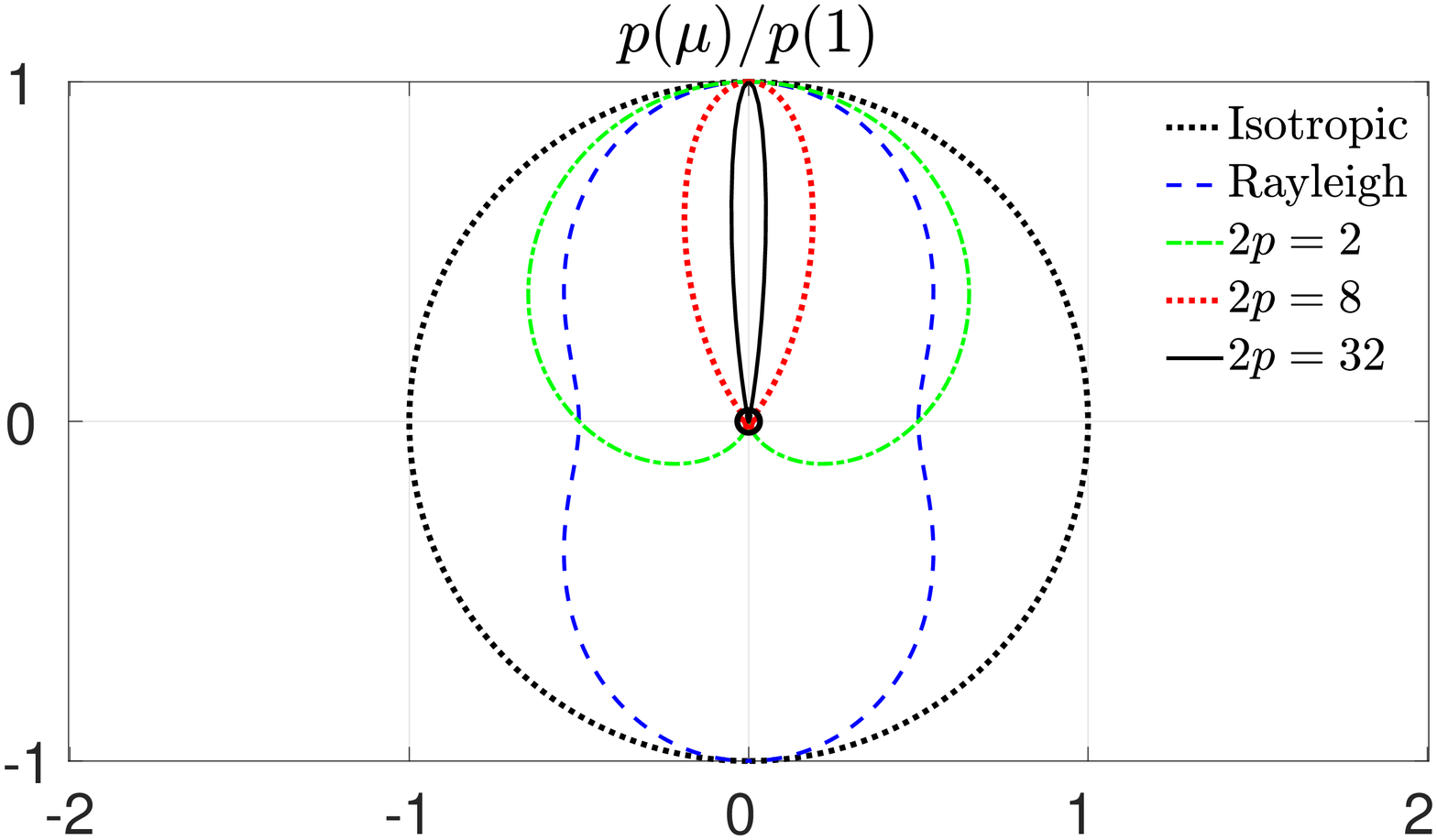}{1} \caption{Phase functions $p(\mu)$ of (\ref{rp2}) for isotropic scattering, and Rayleigh scattering (\ref{rs2}), compared to representative, forward-scattering basis phase functions $\varpi^{\{p\}}(\mu)$ of (\ref{pfb2}). The distance from the center, marked by the  small black circle, to the curves is $p(\mu)/p(1)$. The forward-scattering phases are: 
$p(1) =1$ for isotropic scatter; $p(1)=1.5$ for Rayleigh scatter; $p(1)=\varpi^{\{p\}}(1) = p(p+1) = 2, 20$ and $272$ for forward-scattering  basis phase functions (\ref{pfb12}) constructed from the first $2p=2, 8$ and $32$ Legendre polynomials.
\label{maxphase}}
\end{figure}
\subsection{Rayleigh scattering\label{rs}}
The phase function (\ref{rp2}) for Rayleigh scattering is
\begin{equation}
p(\mu)= \frac{3}{4}(\mu^2+1)= P_0(\mu)+\frac{1}{2}P_2(\mu).
\label{rs2}
\end{equation}
The non-zero coefficients of the multipole expansion are
\begin{equation}
p_0=1,\quad\hbox{and}\quad p_2=\frac{1}{10}.
\label{rs4}
\end{equation}
Representative examples of the Rayleigh-scattering phase functions, $p(\mu_i,\mu_{i'})$ are shown in Fig. \ref{ph2}.  Rayleigh scattering differs little from completely isotropic scattering.

Rayleigh scattering  describes the polarization-averaged  resonance scattering of radiation by an atom with electronic angular momentum $J=0$ in the ground state and $J =1$ in the excited state. It is an excellent approximation for the scattering of light by polarizable molecules at frequencies far below their electronic resonance frequencies. Rayleigh scattering (mostly single scattering) of short-wave sunlight by nitrogen and oxygen molecules makes Earth's sky blue. Eq. (\ref{rs2}) is also the phase function  for  Thomson scattering of radiation by free electrons, and it therefore plays an important role for radiation transfer inside the Sun and stars.   Cloud particulates that are much smaller than the wavelengths of thermal radiation produce Rayleigh scattering, for example, scattering of radar beams.  But cloud particle sizes can be comparable or larger than wavelengths of thermal radiation. This is even more true for shorter-wave sunlight, where the scattering phase functions for cloud particulates are  often strongly peaked in the forward direction. In the next section we discuss how to model phase functions for strong forward or backward scattering in clouds.
\begin{table}[t]
\begin{center}
\begin{tabular}{|l|r|r|r|c|c|c|c|c|c|c|c|}
 \hline
$ p=$ & 1 & 2 & 3 & 4 & 5 & 6 & 7 & 8 & 9 &10& 11\\ [0.5ex]
 \hline\hline
$\varpi^{\{p\}}_{0}$ & 1 & 1 & 1 & 1 & 1 & 1 & 1 & 1 & 1 & 1 &1\\
 \hline
$\varpi^{\{p\}}_{1}$&.3333&.6000&.7143&.7778&.8182&.8462&.8667&.8824&.8947&.9048&.9130\\
 \hline
$\varpi^{\{p\}}_{2}$&       &.4000&.5714&.6667&.7273&.7692&.8000&.8235&.8421&.8571&.8696\\
 \hline
$\varpi^{\{p\}}_{3}$&       &.1714&.3810&.5108&.5967&.6573&.7023&.7368&.7643&.7865&.8050\\
 \hline
$\varpi^{\{p\}}_{4}$&       &      &.2381&.3939&.4988&.5734&.6290&.6718&.7059&.7336&.7565\\
 \hline
 $\varpi^{\{p\}}_{5}$&       &      &.1082&.2681&.3869&.4747&.5413&.5932&.6348&.6688&.6970\\
 \hline
 $\varpi^{\{p\}}_{6}$&       &      &       &.1632&.2937&.3924&.4682&.5277&.5756&.6148&.6474\\
 \hline
 $\varpi^{\{p\}}_{7}$&       &      &      &.0761&.2016&.3058&.3888&.4553&.5092&.5538&.5911\\
 \hline
 $\varpi^{\{p\}}_{8}$&       &      &      &       &.1209&.2301&.3194&.3918&.4512&.5005&.5418\\
 \hline
 $\varpi^{\{p\}}_{9}$&       &      &      &       &.0573&.1586&.2495&.3259&.3896&.4431&.4884\\
 \hline
$\varpi^{\{p\}}_{10}$&       &      &      &       &      &.0943&.1866&.2665&.3342&.3916&.4404\\
 \hline
$\varpi^{\{p\}}_{11}$&       &      &      &       &      &.0451&.1290&.2085&.2782&.3383&.3901\\
 \hline
$\varpi^{\{p\}}_{12}$&       &      &      &       &      &     &.0762&.1553&.2268&.2895&.3440\\
 \hline
$\varpi^{\{p\}}_{13}$&       &      &      &       &     &      &.0367&.1076&.1776&.2411&.2973\\
 \hline
$\varpi^{\{p\}}_{14}$&       &      &      &      &     &       &      &.0633&.1319&.1961&.2539\\
 \hline
$\varpi^{\{p\}}_{15}$&       &      &      &      &     &       &      &.0306&.0915&.1536&.2115\\
 \hline
$\varpi^{\{p\}}_{16}$&       &      &      &      &     &       &      &       &.0536&.1138&.1717\\
 \hline
$\varpi^{\{p\}}_{17}$&       &      &      &      &     &      &      &       &.0260&.0790&.1346\\
 \hline
$\varpi^{\{p\}}_{18}$&       &      &      &      &     &      &      &       &       &.0462&.0995\\
 \hline
 $\varpi^{\{p\}}_{19}$&       &      &      &      &     &       &      &       &       &.0225&.0692\\
 \hline
 $\varpi^{\{p\}}_{20}$&       &      &      &      &     &       &      &       &       &       &.0404\\
 \hline
 $\varpi^{\{p\}}_{21}$&       &      &      &      &     &       &      &       &       &       &.0197\\
 \hline
\end{tabular}
\end{center}
\caption{ Numerical values of the coefficients  $\varpi^{\{p\}}_l$ of (\ref{pfb14})  for the multipole expansion (\ref{pfb12}) of the maximum possible forward-scattering phase functions $\varpi^{\{p\}}(\mu)$ (\ref{pfb2}).
\label{pln}}
\end{table}
\subsection{Maximum forward-scattering\label{pfb}}
As we prove in the Appendix, the phase function $p(\mu)$ of (\ref{rp2}) that is constructed from the first $2p$ Legendre polynomials, that satisfies the constraint (\ref{in14}) of nonnegativity and the area constraint (\ref{in16}), and that has the maximum possible forward-scattering value, $p(1)$, is 
\begin{equation}
\varpi^{\{p\}}(\mu) =\frac{2(1+\mu)}{p(p+1)}\left[\frac{dP_p}{d\mu}(\mu)\right]^2.
\label{pfb2}
\end{equation}
The nonnegativity of (\ref{pfb2})  is obvious, since it is the product of a nonnegative factor $(1+\mu)\ge 0$, and the nonnegative square of a polynomial, $\left[dP_p(\mu)/d\mu\right]^2$. 
The polynomial $\varpi^{\{p\}}(\mu)$ is of degree $2p-1$ and has $2p-1$ roots. There is a single root at $\mu=-1$. The remaining $2p-2$  roots are double and are defined by $dP_p/d\mu =0$.  The double roots lie on the interval $-1<\mu<1$. So the ``antenna pattern" of (\ref{pfb2}) has $p$ nulls  for direction cosines starting at $\mu=-1$ and ending just shy of $\mu = 1$, the direction cosine for maximum  scattering. The forward-scattering  lobe is much larger than the sidelobes.

The factor $\left[dP_p(\mu)/d\mu\right]^2$ of (\ref{pfb2}) is even in $\mu$, so we can write the area as
\begin{equation}
\int_{-1}^1 d\mu\, \varpi^{\{p\}}(\mu) =\frac{2}{p(p+1)}\int_{-1}^1 d\mu\left[\frac{dP_p}{d\mu}(\mu)\right]^2=2,
\label{pfb4}
\end{equation}
as required by (\ref{in16}).  In writing (\ref{pfb4}) we used the identity
\begin{equation}
\int_{-1}^1 d\mu\left[\frac{dP_p}{d\mu}(\mu)\right]^2=p(p+1),
\label{pfb6}
\end{equation}
which can be proved using elementary properties of Legendre polynomials. It is also not hard to prove the identity
\begin{equation}
\frac{dP_p}{d\mu}(1)=\frac{p(p+1)}{2},
\label{pfb8}
\end{equation}
which we can use with (\ref{pfb2}) to find that the forward-scattering and backward-scattering values of (\ref{pfb2}) are
\begin{equation}
\varpi^{\{p\}}(1) = p(p+1),\quad\hbox{and}\quad \varpi^{\{p\}}(-1) =0.
\label{pfb10}
\end{equation}
We can write the maximum foward-scattering phase function (\ref{pfb2}) as the multipole expansion
\begin{equation}
\varpi^{\{p\}}(\mu)=\sum_{l}P_l(\mu)(2l+1)\varpi^{\{p\}}_{l}.
\label{pfb12}
\end{equation}
From  (\ref{mm2}) and (\ref{pfb2}) we see that the multipole coefficients of (\ref{pfb12}) are
\begin{eqnarray}
\varpi^{\{p\}}_{l}&=&\frac{1}{2}\int_{-1}^1 d \mu\, P_l(\mu)\varpi_p(\mu)\nonumber\\
 &=&\int_{-1}^1 d\mu \frac{P_l(\mu)[1+\mu]}{p(p+1)}\left[\frac{dP_p}{d \mu}(\mu)\right]^2.
\label{pfb14}
\end{eqnarray}
The values of  the coefficients $\varpi^{\{p\}}_{l}$  from (\ref{pfb14}) are rational numbers, as mentioned in the caption of Fig. \ref{ph3}.
But for convenience, we have 
listed them as four-place decimal fractions in Table \ref{pln} for $p=1,2,3,\ldots, 11$, and for $l=0, 1, 2,\ldots,2p-1$.

If the phase function (\ref{pfb12}) is rotated by the colatitude angle $\theta'=\cos^{-1}\mu'$, and the result is averaged over all azimuthal angles, the resulting  phase function becomes
\begin{equation}
\varpi^{\{p\}}(\mu,\mu')=\sum_{l}P_l(\mu)(2l+1)\varpi^{\{p\}}_{l}P_l(\mu').
\label{pfb16}
\end{equation}
The phase function with the maximum possible backscattering is the reflected version of (\ref{pfb16})
\begin{equation}
 r \varpi^{\{p\}}(\mu,\mu')=\sum_{l}P_l(\mu)(2l+1)\varpi^{\{p\}}_{l}P_l(\mu')(-1)^l.
\label{pfb18}
\end{equation}

Fig. \ref{ph3} shows the forward-scattering basis phase function $\varpi^{\{3\}}(\mu,\mu')$ of (\ref{pfb16}) constructed from  $2p= 6$ Legendre polynomials. In accordance with (\ref{pfb10}), the maximum forward scattering in Fig. \ref{ph3} from exactly vertical light is $\varpi^{\{3\}}(1)=3(4) = 12$. Fig. \ref{ph4} shows  the reflected version $r\varpi^{\{3\}}(\mu,\mu')$ of (\ref{pfb18}).
In Fig. \ref{maxphase} we have compared isotropic and Rayleigh  scattering phase functions with forward-scattering basis phase functions $\varpi^{\{p\}}(\mu)$ (\ref{pfb2})   constructed from $2p = 2, 8,$ and $32$ Legendre polynomials.

One can construct superpositions of phase functions like those of Fig. \ref{ph3}, and their ``mirror images,"   to form more complicated phase functions, for example, phase functions strongly peaked in both the forward and backward directions. If the superposition coefficients are positive and sum to 1, the compound phase functions will satisfy the two contraints (\ref{in14}) and (\ref{in16}). For a $2n$-stream model, any phase function can contain no more than the first $2n$ Legendre polynomials, and this limits how sharp the angular features can be.

\begin{figure}[t]
\postscriptscale{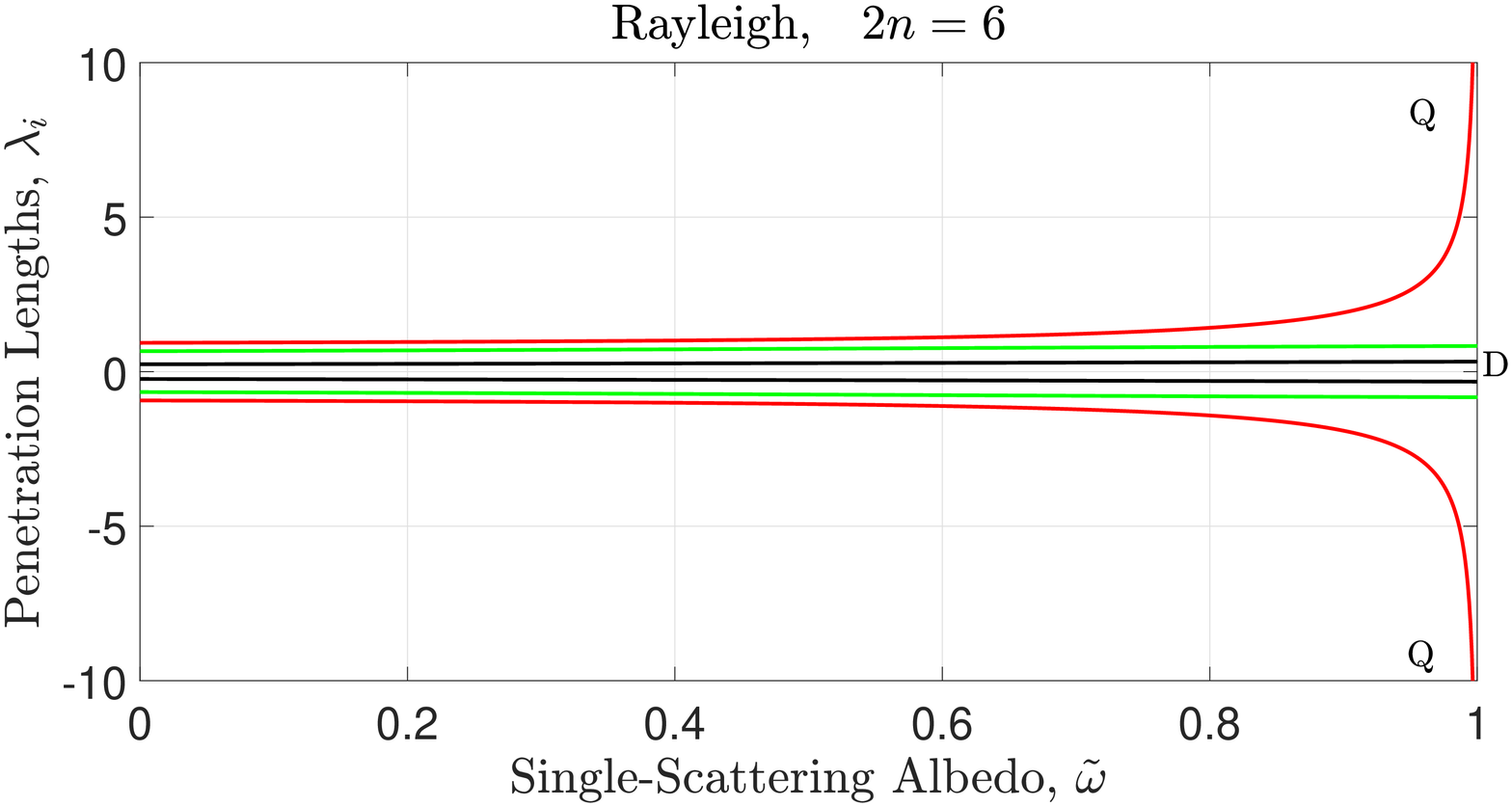}{1}
\caption{The penetration lengths $\lambda_i$ of (\ref{plm4})
for $2n = 6$ penetration modes,  $|\lambda_i)$, and for Rayleigh scattering. The results are qualitiatively similar for any other scattering phase function. 
For a vanishing single-scattering albedo, $\tilde\omega \to 0$, the modes become identical to the direction-cosine basis functions, $|\lambda_i)\to |\mu_i)$, and the penetration lengths become identical to the direction cosines, $\lambda_i\to\mu_i$.  As the single-scattering albedo approaches unity,
$\tilde \omega\to 1$, the absorption becomes very small and the penetration lengths are ``stretched," $|\lambda_i|>|\mu_i|$. There is unlimited stretching of the quasi-isotropic modes, with $i=1$ and $i=2n$, where $\lambda_1\to -\infty$ and $\lambda_{2n}\to +\infty$.
These modes, marked by the letter Q in the figure, consist of nearly isotropic  light, like diffuse sunlight deep inside a cloud.  An example is the quasi-isotropic upward mode shown in the left panel of Fig. \ref{mulamj}.
As $\tilde\omega \to 1$ the values of the two diverging, quasi-isotropic penetration lengths, $\lambda_{2n}=-\lambda_1$, are given by the formula (\ref{lim8}).
The properties of {\it directional} modes, $|\lambda_i)$, with $2\le i\le 2n-1$, and marked with the letter D in the figure, are much less dependent on $\tilde\omega$.
\label{miom}}
\end{figure}
\begin{figure}[t]
\postscriptscale{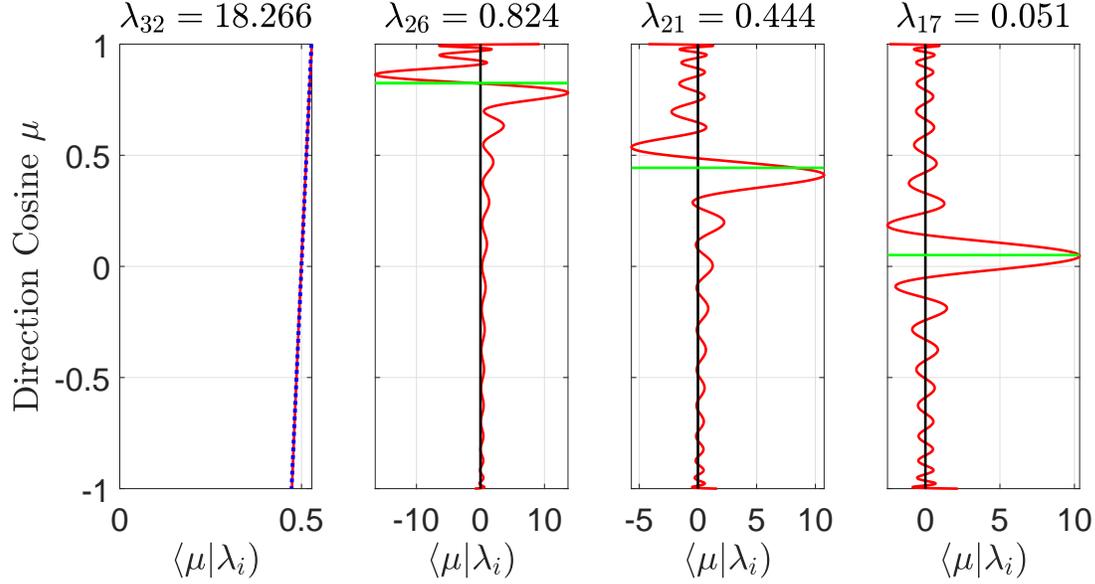}{1}
\caption{ Mode profiles $\langle\mu|\lambda_i)$ with the same indices   $i=32, 26, 21, 17$ as the stream profiles $\langle\mu|\mu_i)$ of the purely absorbing atmosphere of
Fig. \ref{mumuj}. There are $2n=32$ streams.  But in this case there is Rayleigh scattering and only weak absorption with $ \tilde\omega = 0.999$.
The penetration lengths $\lambda_i$ (the eigenvalues  of $\hat\lambda$)  are shown (where possible) as horizontal green lines.  For the directional modes, with $i =$ 26, 21 and 17 the penetration lengths  $\lambda_i$ are only slightly larger than the corresponding direction cosines $\mu_i$, the penetration lengths for a purely absorbing atmosphere. The angular distributions $\langle\mu|\lambda_i)$ and $\langle \mu|\mu_i)$ are similar for  $i = $ 26, 21 and 17.  The quasi-isotropic mode, $\langle\mu|\lambda_{32})$, differs drastically from the highly forward-directed stream $\langle\mu|\mu_{32})$, that is shown on the left panel of Fig. \ref{mumuj}. The blue dots are the analytic approximation  (\ref{lim12}) of $\langle\mu|\lambda_{2n})$.
\label{mulamj}}
\end{figure}
\begin{figure}[t]
\postscriptscale{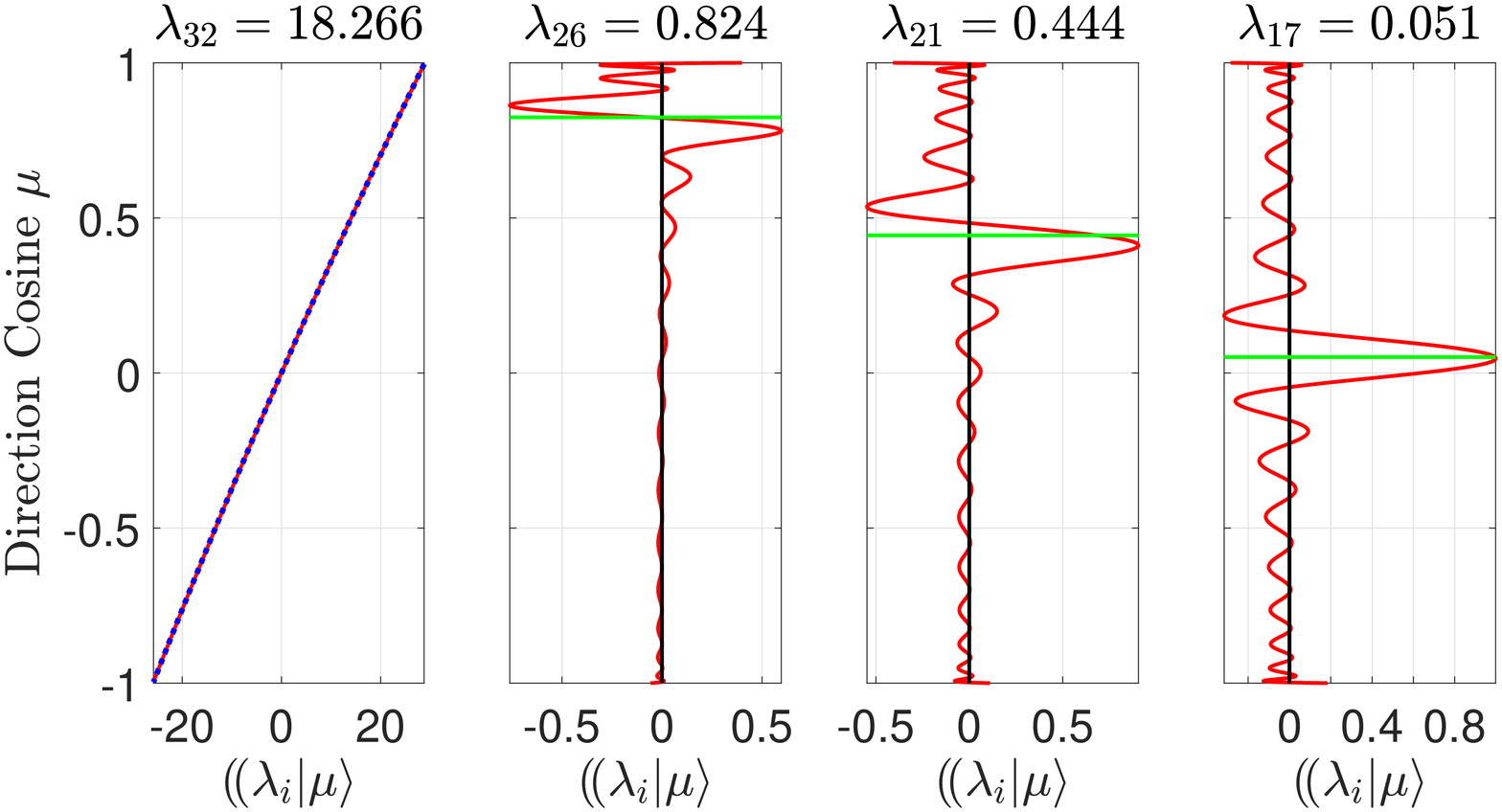}{1}
\caption{ Mode profiles $\lvec \lambda_i|\mu\rangle$ with $ \tilde\omega = 0.999$ and Rayleigh scattering, and with the same indices   $i=32, 26, 21,17$ as the stream profiles $\lvec\mu_i|\mu\rangle$ of the purely absorbing atmosphere of
Fig. \ref{mujmu} with $\tilde\omega =0$. These are the left eigenvectors of $\hat\lambda$, that correspond to the right  eigenvectors of Fig. \ref{mulamj}. The profiles $\lvec \lambda_i|\mu\rangle$ of the upward, directional modes, with $i =$ 26, 21 and 17 are qualitatively similar to the stream profiles  $\langle\mu|\lambda_i)$ with the same indices $i$. The profile $\langle \lambda_{32}|\mu\rangle$ of  the  left quasi-isotropic mode is very nearly proportional to $\mu$. It differs completely from the profile $\lvec \lambda_{32}|\mu\rangle$ of the right quasi-isotropic mode of Fig \ref{mulamj}, which is  very nearly independent of $\mu$. The blue dots are the analytic aproximation  (\ref{lim20}) of $\lvec\lambda_{2n}|\mu\rangle$.
\label{lamjmu}}
\end{figure}
\section{Penetration Modes \label{plm}}
When scattering cannot be neglected, neither the $l$-space basis vectors $\lvec l|$ and $|l)$, nor the $\mu$-space basis vectors $\lvec\mu_i|$ and $|\mu_i)$ are convenient for calculations. The natural choice of basis vectors are the left and right eigenvectors $\lvec\lambda_i|$ and $|\lambda_i)$ of the penetration-length matrix 
\begin{equation}
\hat\lambda =\hat\kappa^{-1}=\hat \eta^{-1}\hat \mu.
\label{plm2}
\end{equation}
the inverse of the exponentiation-rate matrix $\hat\kappa$ of (\ref{vet12}). Projections of vectors on the basis vectors $\lvec\lambda_i|$ and $|\lambda_i)$ will be called  mode amplitudes.
The penetration lengths $\lambda_i$, and the penetration modes,  $|\lambda_i)$ and 
$\lvec \lambda_i|$, are determined, aside from normalization, by the eigenvalue equations
\begin{equation}
\lvec \lambda_i|\hat\lambda=\lvec \lambda_i|\lambda_i,\quad\hbox{and}\quad \hat\lambda|\lambda_i)=\lambda_i|\lambda_i).
\label{plm4}
\end{equation}
We will call the eigenvector $|\lambda_i)$ a  {\it penetration mode}.  Here we assume that the penetration lengths $\lambda_i$ are real numbers, and we will prove this assumption in the next section. For $\lambda_k>0$ a penetration mode will be attenuated by a factor of $e$ for an optical-depth increase $\Delta \tau =\lambda_k$.  For $\lambda_j<0$, a penetration mode will be amplified by a factor of $e$ for an increase of optical depth $\Delta \tau =|\lambda_j|$.

Fig. \ref{miom} shows how the penetration lengths $\lambda_i$ depend on the single-scatter albedo $ \tilde\omega$. For this simple example, there are $2n=6$ streams and Rayleigh scattering. For $\tilde\omega\to 1$, the magnitudes of the first and last penetration lengths become very large, $|\lambda_1|=|\lambda_{2n}|\gg 1$.  We will call the stream with $i=1$ the {\it downward quasi-isotropic mode} and the stream with $i=2n$ the {\it upward quasi-isotropic mode}. In the example of Fig. \ref{mulamj}, with $2n=32$, the penetration lengths of the quasi-isotropic modes are $|\lambda_1|=|\lambda_{2n}| = 18.266\gg 1$ for $\tilde \omega = 0.999$ and Rayleigh scattering. The  streams, with $2\le i \le 2n-1$, will be called {\it directional modes}, since $|\lambda_i|<1$ for most of them, and they can threfore be described, approximately, as streams with directions making angles $\theta_i=\cos^{-1}\lambda_i$ with the vertical. 

The eigenvalue equations (\ref{plm4}) do not define the normalization of $|\lambda_i)$ and $\lvec\lambda_i|$. We will let the normalization of $|\lambda_i)$ be defined by the conditions 
\begin{equation}
\lvec 0|\lambda_i)=\frac{1}{2},\quad\hbox{and}\quad \lvec \lambda_i|\lambda_{i'})=\delta_{i i'}.
\label{plm6}
\end{equation}
In analogy to (\ref{sdbv8}), the modes are complete, that is,
\begin{equation}
\hat 1=\sum_i|\lambda_i)\lvec \lambda_i|=\mathcal{L}_{\bf d}+\mathcal{L}_{\bf u}.
\label{plm22}
\end{equation}
The projection operators for the lower and upper halves of $\lambda_i$-space are
\begin{equation}
\mathcal{L}_{\bf d}=\sum_j|\lambda_j)\lvec \lambda_j|,\quad\hbox{and}\quad\mathcal{L}_{\bf u}=\sum_k|\lambda_k)\lvec \lambda_k|.
\label{plm24}
\end{equation}
Here the index $j=1,2,3,\ldots,n$ labels the $n$ penetration modes with negative penetration lengths, $\lambda_j<0$, and the index $k=n+1,n+2,n+3,\ldots, 2n$ labels the penetration modes with positive lengths, $\lambda_k>0$.
The projection operators have the multiplication property
\begin{equation}
\mathcal{L}_{\bf q}\mathcal{L}_{{\bf q}'}  =\mathcal{L}_{\bf q}\delta_{{\bf q}{\bf q}'},
\label{plm29}
\end{equation}
where the indices ${\bf q}$ and ${\bf q}'$ can have the values ${\bf u}$ or ${\bf d}$.

The left and right eigenvectors $\lvec\lambda_i|$ and $|\lambda_i)$ of the penetration-length matrix $\hat \lambda$ can be chosen to be identical to the corresponding eigenvectors $\lvec \kappa_i|$ and $|\kappa_i)$ of  the exponentiation-rate matrix  $\hat \kappa=\hat\lambda^{-1}$. The eigenvalues $\lambda_i$ and $\kappa_i$ are inverses of each other, that is,
\begin{equation}
\lvec \lambda_i|=\lvec \kappa_i|,\quad |\lambda_i)=|\kappa_i),\quad\hbox{and}\quad\lambda_i=\frac{1}{\kappa_i}.
\label{plm26}
\end{equation}
In analogy  to (\ref{plm22}) we write the exponentiation-rate operator, $\hat\kappa$, as
\begin{equation}
\hat\kappa=\sum_i\kappa_i|\lambda_i)\lvec \lambda_i|=\hat\kappa_{\bf d}+\hat\kappa_{\bf u}.
\label{plm28}
\end{equation}
As indicated in (\ref{plm28}), it is useful to write $\hat \kappa$ as the sum of  a downward part $\hat \kappa_d$ and an upward part $\hat\kappa_u$,
\begin{equation}
\hat\kappa_{\bf d} =\sum_j\kappa_j|\lambda_j)\lvec \lambda_j|,\quad\hbox{and}\quad \hat\kappa_{\bf u} =\sum_k\kappa_k|\lambda_k)\lvec \lambda_k|.
\label{plm30}
\end{equation}
The penetration length matrix (\ref{plm2}) is odd under reflection
\begin{equation}
\hat r \hat\lambda=-\hat\lambda\hat r.
\label{plm32}
\end{equation}
Therefore
\begin{equation}
\hat\lambda\hat r|\lambda_i)=-\hat r\hat\lambda|\lambda_i)=-\lambda_i\hat r|\lambda_i)
\label{plm34}
\end{equation}
According to (\ref{plm34}), if $|\lambda_i)$ is a right eigenvector of $\hat\lambda$, with the eigenvalue $\lambda_i$, then $\hat r|\lambda_i)$ is also a right eigenvector with eigenvalue $-\lambda_i$.  The eigenvalues $\lambda_i$ occcur as equal and opposite pairs, corresponding to pairs of reflected eigenvectors $|\lambda_i)$ and $\hat r|\lambda_i)=|\lambda_{r(i)})$. The index-reflection function $r(i)$ was defined by (\ref{lg10}). We will order the eigenvalues such that
\begin{equation}
\lambda_1\le \lambda_2\le \lambda_3\le \cdots\le \lambda_{2n}.
\label{plm36}
\end{equation}
Half of the penetration lengths $\lambda_i$ and their inverses $\kappa_i=1/\lambda_i$ are negative and half are positive,
\begin{eqnarray}
\lambda_j&<&0,\quad\hbox{and}\quad \kappa_j<0,\quad\hbox{for}\quad j=1,2,3,\ldots, n,\nonumber\\
\lambda_k&>&0,\quad\hbox{and}\quad \kappa_j>0,\quad\hbox{for}\quad k=n+1,n+2,n+3,\ldots, 2n.
\label{plm38}
\end{eqnarray}
\subsection{Real penetration lengths}
The penetration lengths $\lambda_i$ defined by the eigenvalue equations (\ref{plm4}) are real numbers because we can write the penetration-length matrix $\hat\lambda$ as the similarity transformation
\begin{equation}
\hat \lambda= \hat s^{-1}\hat h \hat s.
\label{rpl2}
\end{equation}
The eigenvalues of the Hermitian matrix $\hat h=\hat h(\tilde\omega)$ are real. Eigenvalues are not changed by a similarity transformation. So the eigenvalues of $\hat \lambda$ are the same as those of $\hat h$.

To construct the Hermitian matrix $\hat h$ of  (\ref{rpl2}), we introduce a $2n\times 2n$ statistical-weight matrix, $\hat g$, that is diagonal  in $l$-space and has the elements
\begin{equation}
\lvec l|\hat g|l')=\delta_{l l'}g_l, \quad\hbox{where}\quad g_l = 2l+1.
\label{rpl4}
\end{equation}
We use $\hat g$ and the efficiency matrix, $\hat\eta =\hat\eta(\tilde\omega)$ of (\ref{vet6}), which is also diagonal in $l$-space, to write 
\begin{equation}
\hat h =\hat\eta^{-1/2}\hat g^{1/2}\hat\mu \hat g^{-1/2}\hat\eta^{-1/2}.
\label{rpl6}
\end{equation}
In $l$-space, only the elements of $\hat h$ above and below the main diagonal are non-zero,  and from (\ref{rpl6}) and (\ref{mm18}) we find that the values of these elements are
\begin{equation}
\lvec l|\hat h|l+1) =\lvec l+1|\hat h | l) = \frac{l+1}{\sqrt{\eta_l\eta_{l+1}g_l g_{l+1}}}.
\label{rpl8}
\end{equation}
The similarity transformation of (\ref{rpl2}) is effected by the diagonal matrix $\hat s=\hat s(\omega)$ with the elements
\begin{equation}
\hat s =(\hat g \hat \eta)^{1/2}, \quad\hbox{with}\quad \lvec l|\hat s|l')=\delta_{l l'}(g_l \eta_l)^{1/2}.
\label{rpl10}
\end{equation}
Using (\ref{rpl8}) and (\ref{rpl10}), or using (\ref{plm2}) directly, we find that the non-zero matrix elements in $\hat \lambda$ in $l$-space are
\begin{eqnarray}
\lvec l|\hat \lambda|l+1) &=& \frac{(l+1)\eta_l^{-1}}{2l+1},\nonumber\\
\lvec l+1|\hat \lambda|l) &=& \frac{(l+1)\eta_{l+1}^{-1}}{2l+3}.
\label{rpl12}
\end{eqnarray}
In $l$-space, the  matrix representation of $\hat\lambda$ is not Hermitian.
\subsection{Overlap operators}
For discussions of scattering and thermal emission, it will be convenient to use overlap operators 
$\mathcal{C}_{\bf q q'}$  which we define by the product of a projection operator $\mathcal{M}_{\bf q}$ of $\mu$-space and a projection operator $\mathcal{L}_{\bf q'}$ of $\lambda$-space,

\begin{equation}
\mathcal{C}_{\bf q q'} = \mathcal{M}_{\bf q}\mathcal{L}_{\bf q'}.
\label{olp2}
\end{equation}
From (\ref{olp2}), (\ref{plm22}) and (\ref{sdbv8})  we see that the overlap operators  satisfy the sum rule
\begin{eqnarray}
\hat 1&=&\hat 1 \hat 1\nonumber\\
&=&(\mathcal{M}_{\bf d}+\mathcal{M}_{\bf u})(\mathcal{L}_{\bf d}+\mathcal{L}_{\bf u})\nonumber\\
&=&\mathcal{C}_{\bf dd}+\mathcal{C}_{\bf ud}+\mathcal{C}_{\bf du}+\mathcal{C}_{\bf uu}.
\label{olp4}
\end{eqnarray}
\subsection{The limit $\tilde \omega\to 0$  \label{lim}} 
The efficiency matrix $\hat \eta$ of  (\ref{vet6})  becomes a $2n\times 2n$ unit matrix $\hat 1$ as the single-scattering albedo vanishes, $\tilde \omega \to 0$, and the cloud becomes purely absorptive.  Then the penetration-length matrix $\hat\lambda$ of (\ref{plm2}) becomes the $2n\times 2n$ upper left corner of the direction-cosine matrix $\hat \mu$ of (\ref{me3}), and the exponentiation-rate matrix $\hat \kappa$ of (\ref{vet12}) becomes the $2n\times 2n$ upper left corner of the direction-secant matrix $\hat\varsigma$ of (\ref{me17}). The normalization condition (\ref{plm6}) ensures that  the penetration-length basis vectors $|\lambda_i)$ and $\lvec\lambda_i|$  of (\ref{plm4}) approach the direction-cosine basis vectors $|\mu_i)$ of (\ref{sdbv2}) and $\lvec\mu_i|$ of (\ref{sdbv4}) as $\tilde \omega \to 0$.  For future reference, we summarize the pure absorption limits, for $\tilde\omega \to 0$ as
\begin{eqnarray}
\tilde\omega&\to& 0.\nonumber\\
\hat \eta &\to& \hat 1,\nonumber\\
\hat \lambda &\to& \hat \mu,\nonumber\\
\hat \kappa &\to& \hat\varsigma,\nonumber\\
|\lambda_i) &\to& |\mu_i),\nonumber\\
\lvec\lambda_i| &\to& \lvec \mu_i|,\nonumber\\
\mathcal{L}_{\bf q}&\to& \mathcal {M}_{\bf q}\nonumber\\
\mathcal{C}_{\bf q q'}&\to& \mathcal{M}_{\bf q}\delta_{\bf q q'}.
\label{lim2}
\end{eqnarray}
In the last lines of (\ref{lim2}) the projection operator $\mathcal{L}_{\bf q}$ of $\lambda$-space was given by (\ref{plm24}) and the projection operator $\mathcal{M}_{\bf q}$ of $\mu$-space was given by (\ref{sdbv9}).
\subsection{The limit $\tilde \omega\to 1$  \label{lim2n}} 
In \S{\bf 8}(57) Chandrasekhar\cite{Chandrasekhar} calls pure scattering, when $\tilde\omega = 1$,  {\it perfect} or {\it conservative} scattering.
For a unit single-scattering albedo $\tilde \omega =1$, the penetration-length matrix $\hat \lambda$ of (\ref{plm2}) becomes singular and some of the numerical procedures we have discussed earlier fail. Most of the numerical methods work well for single-scattering albedos as close to 1 as $\tilde\omega = 1-10^{-6}$. With minor modifications it is possible to adapt the $2n$-stream model to $\tilde\omega = 1$, as we will show in a subsequent paper.
 
 Here we will discuss a few important aspects of scattering in the limit $\tilde\omega \to 1$. We can use (\ref{me3}) and (\ref{vet6}) to write the penetration-length matrix (\ref{plm2}) as

\begin{equation}
\lvec l|\hat\lambda|l')
	=\left[\begin{array}{llll} 0&(1-\tilde\omega)^{-1}&0&\cdots \\ (1-\tilde\omega p_1)^{-1}3^{-1}&0&2(1-\tilde\omega p_1)^{-1}3^{-1}&\cdots \\ 
0&2 (1-\tilde\omega p_2)^{-1}5^{-1}&0&\cdots \\ 
 \vdots&\vdots &\vdots&\ddots\\ \end{array}\right]
\label{lim4}
\end{equation}
From inspection of the left panel of Fig. \ref{mulamj} we see that the quasi-isotropic eigenvector $\langle\mu|\lambda_{2n})$ for weak absorption can be written very accurately as the sum of the first two Legendre polynomials, $P_0(\mu)=1$ and $P_1(\mu)=\mu$. So to good approximation, the second eigenvalue equation of (\ref{plm4}) must reduce to 
\begin{equation}
\hat 0
	=\left[\begin{array}{ll} -\lambda_{2n}&(1-\tilde\omega)^{-1} \\ (1-\tilde\omega p_1)^{-1}3^{-1}
&-\lambda_{2n}  \end{array}\right]
\left[\begin{array}{ll}\lvec 0 |\lambda_{2n}) \\ \lvec 1|\lambda_{2n})  \end{array}\right]
\label{lim6}
\end{equation}
Setting the determinant of the $2\times 2 $ matrix of (\ref{lim6}) equal to zero, we find that  for the weak-absorption limit, $\tilde\omega \to 1$, the eigenvalue $\lambda_{2n}$ must approach the limit, 
\begin{equation}
\lambda_{2n}\to[3(1-\tilde\omega p_1)(1-\tilde\omega)]^{-1/2}.
\label{lim8}
\end{equation}
For the Rayleigh-scattering example of Fig. \ref{mulamj}, where $\tilde \omega = 0.999$ and $p_1=0$, exact diagonalization of the eigenvalue equations (\ref{plm4}) give $\lambda_{2n} = 18.266$, and (\ref{lim8}) gives $\lambda_{2n} = 18.257$.

Using the normalization convention $\lvec 0|\lambda_{2n}) = 1/2$ of (\ref{plm6}) in (\ref{lim6}) we see that for the weak absorption limit, the right eigenvector $|\lambda_{2n})$ approaches the limit
\begin{equation}
|\lambda_{2n})\to\left[\begin{array}{l}\lvec 0 |\lambda_{2n}) \\ \lvec 1|\lambda_{2n})  \end{array}\right]=
\left[\begin{array}{l} 1/2 \\  (1-\tilde\omega)^{1/2}(12[1-\tilde \omega p_1])^{-1/2}  \end{array}\right].
\label{lim10}
\end{equation}
In continuous $\mu$-space, (\ref{lim10}) becomes
\begin{equation}
\langle \mu|\lambda_{2n})\to\frac{1}{2}
+\mu\sqrt{\frac {3(1-\tilde\omega)}{4(1-\tilde \omega p_1)}}.
\label{lim12}
\end{equation}
The left panel of Fig. \ref{mulamj} shows  values of $\langle\mu|\lambda_{2n})$ from (\ref{lim12}),  the blue dots, compared to values of the exact solution of the second eigenvalue equation (\ref{plm4}), the continuous red line.

From inspection of the left panel of Fig. \ref{lamjmu} we see that the quasi-isotropic eigenvector $\lvec\lambda_{2n}|\mu\rangle$ for weak absorption can be written very accurately as the sum of the first three Legendre polynomials, $P_0(\mu)=1$, $P_1(\mu)=\mu$ and $P_2(\mu)=(3\mu^2-1)/2$. 
Then the first and third columns of the first eigenvalue equation (\ref{plm4}) become
\begin{equation}
\left[\begin{array}{lll}\lvec\lambda_{2n}|0)&\lvec\lambda_{2n}|1)&\lvec\lambda_{2n}|2)\end{array}\right]\left[\begin{array}{ll} -\lambda_{2n}&0 \\ (1-\tilde\omega p_1)^{-1}3^{-1}&2(1-\tilde\omega p_1)^{-1}3^{-1} \\ 
0&-\lambda_{2n}\end{array}\right]=[0\quad 0].
\label{lim14}
\end{equation}
Solving (\ref{lim14}) for $\lvec\lambda_{2n}|0)$ and $\lvec\lambda_{2n}|2)$ in terms of $\lvec\lambda_{2n}|1)$, and requiring that $\lvec\lambda_{2n}|\lambda_{2n}) = 1$  we find 
\begin{equation}
\lvec\lambda_{2n}|1)\to\sqrt{\frac{3(1-\tilde\omega p_1)}{1-\tilde\omega}}
\label{lim16}
\end{equation}
and 
\begin{equation}
\lvec\lambda_{2n}|2)=2\lvec\lambda_{2n}|0)\to 2.
\label{lim18}
\end{equation}
In continuous $\mu$-space, (\ref{lim16}) and (\ref{lim18}) give 
\begin{equation}
\lvec\lambda_{2n}|\mu\rangle \to\mu\sqrt{\frac{3(1-\tilde\omega p_1)}{4(1-\tilde\omega)}}+\frac{3}{2}\mu^2.
\label{lim20}
\end{equation}
The left panel of Fig. \ref{lamjmu} shows  values of $(\lambda_{2n} |\mu\rangle $ from (\ref{lim20}),  the blue dots, compared to the values of the solution of the first eigenvalue equation (\ref{plm4}), the continuous red line.
Reflecting $|\lambda_{2n})$ gives the other quasi-isotropic eigenfunction $|\lambda_1)=\hat r|\lambda_{2n})$ or $\langle \mu|\lambda_{1})=\langle -\mu|\lambda_{2n})$. The corresponding eigenvalue is $\lambda_1=-\lambda_{2n}$.

\section{Clouds}
For optical depths $\tau$ inside a cloud of optical thickness $\tau_c$, that is,  for $0\le\tau\le \tau_c$, the intensity $|I(\tau)\}$ varies because of absorption, thermal emission and scattering, as described by (\ref{vet14}).
We denote the radiation coming into the cloud with the $2n\times 1$ {\it incoming intensity vector}, 
\begin{equation}
|I^{\{\rm in\}} \} =\mathcal{M}_{\bf u}|I(0)\}+\mathcal{M}_{\bf d}|I(\tau_c)\} .
\label{c10}
\end{equation}
The projection operators $\mathcal{M}_{\bf u}$ and $\mathcal{M}_{\bf d}$ for upward and downward streams were defined by
(\ref{sdbv9}).
The intensity  going out of the cloud is denoted  by the {\it outgoing intensity vector},
\begin{equation}
|I^{\{\rm out\}} \} =
\mathcal{M}_{\bf d}|I(0)\} +\mathcal{M}_{\bf u} |I(\tau_c)\}.
\label{c12}
\end{equation}
\subsection{Optically thin clouds \label{tnc}}
For a cloud of infinitesimal thickness $\tau_c=d\tau\ll 1$, we can take  $\kappa$ and  $|B\}$ as constant, and write the solution of the equation of transfer (\ref{vet14}), to order $d\tau>0$, as
\begin{equation}
|I (d\tau)\}=(\hat 1-d\tau\hat\kappa)|I (0)\}+d\tau\hat\kappa|B\},
\label{c20}
\end{equation}
Noting from (\ref{sdbv8}) that $\mathcal{M}_{\bf u}+\mathcal{M}_{\bf d}=\hat 1$, we use (\ref{c20}) to write
the incoming intensity (\ref{c10}) as
\begin{eqnarray}
|I^{\{\rm in\}} \} &=&\mathcal{M}_{\bf u}|I(0)\}+\mathcal{M}_{\bf d}|I(d\tau)\} \nonumber\\
&=&\mathcal{M}_{\bf u}|I(0)\}+\mathcal{M}_{\bf d}\left[(\hat 1-d\tau\hat\kappa)|I (0)\}+d\tau\hat\kappa|B\}\right]\nonumber\\
&=&(\hat 1-d\tau\mathcal{M}_{\bf d}\hat\kappa)|I (0)\}+d\tau\mathcal{M}_{\bf d}\hat\kappa|B\}.
\label{c22}
\end{eqnarray}
In like manner we write
the outgoing intensity (\ref{c12}) as
\begin{eqnarray}
|I^{\{\rm out\}} \} &=&\mathcal{M}_{\bf d}|I(0)\}+\mathcal{M}_{\bf u}|I(d\tau)\} \nonumber\\
&=&\mathcal{M}_{\bf d}|I(0)\}+\mathcal{M}_{\bf u}\left[(\hat 1-d\tau\hat\kappa)|I (0)\}+d\tau\hat\kappa|B\}\right]\nonumber\\
&=&(\hat 1-d\tau\mathcal{M}_{\bf u}\hat\kappa)|I (0)\}+d\tau\mathcal{M}_{\bf u}\hat\kappa|B\}.
\label{c24}
\end{eqnarray}
Solving (\ref{c22}), to order $d\tau$, for $|I(0)\}$, we find
\begin{equation}
 |I (0)\}
=(\hat 1+d\tau\mathcal{M}_{\bf d}\hat\kappa)|I^{\{\rm in\}} \}-d\tau\mathcal{M}_{\bf d}\hat\kappa|B\}.
\label{c26}
\end{equation}
Substituting (\ref{c26}) into (\ref{c24}) we find, to order $d\tau$, that the output intensity from the thin cloud is
\begin{equation}
|I^{\{\rm out\}} \}=\mathcal{S}|I^{\{\rm in\}} \}+\mathcal{E}|B\}.
\label{c28}
\end{equation}
To order $d\tau$, the scattering matrix for the thin cloud is
\begin{equation}
\mathcal{S}=\hat 1-d\tau(\mathcal{M}_{\bf u}-\mathcal{M}_{\bf d})\hat\kappa.
\label{c30}
\end{equation}
The emissivity matrix is
\begin{equation}
\mathcal{E}=d\tau(\mathcal{M}_{\bf u}-\mathcal{M}_{\bf d})\hat\kappa.
\label{c32}
\end{equation}
From inspection of (\ref{c30}) and (\ref{c32}) we see that the scattering and emissivity matrices for a thin cloud obey Kirchhoff's law,
\begin{equation}
\mathcal{S}+\mathcal{E}=\hat 1.
\label{c34}
\end{equation}
As we shall show below, Kirchhoff's law (\ref{c34})  remains valid for an isothermal cloud of arbitrary thickness.

Substituting the efficiency operator $\hat\eta$ of (\ref{vet6}) into the growth rate operator $\hat\kappa = \hat \varsigma\hat\eta$ of (\ref{vet12}) we write the scattering matrix (\ref{c30}) of a thin cloud as

\begin{eqnarray}
\mathcal{S}&=&\hat 1-d\tau(\mathcal{M}_{\bf u}-\mathcal{M}_{\bf d})\hat\varsigma(\hat 1-\frac{\tilde\omega}{2}\hat p)\nonumber\\
&=&\hat 1-d\tau(\hat\varsigma_{\bf u}-\hat\varsigma_{\bf d})(\hat 1-\frac{\tilde\omega}{2}\hat p)\nonumber\\
&=&\mathcal{S}^{\rm C}+\mathcal{S}^{\rm D}.
\label{c36}
\end{eqnarray}
Here the coherent scattering operator for the thin cloud is
\begin{equation}
\mathcal{S}^{\rm C}=\hat 1-d\tau(\hat\varsigma_{\bf u}-\hat\varsigma_{\bf d}).
\label{c38}
\end{equation}
The diffuse scattering operator is
\begin{equation}
\mathcal{S}^{\rm D}=\frac{d\tau\,\tilde\omega}{2}(\hat\varsigma_{\bf u}-\hat\varsigma_{\bf d})\hat p.
\label{c40}
\end{equation}
The  matrix elements of the scattering matrix in $\mu$-space can be written as
\begin{eqnarray}
\lvec\mu_i|\mathcal{S}|\mu_{i'})
&=&\lvec\mu_i|\mathcal{S}^{\rm C}|\mu_{i'})+\lvec\mu_i|\mathcal{S}^{\rm D}|\mu_{i'})\nonumber\\
&=&\delta_{ii'}(1-d\tau|\varsigma_i|)+\frac{d\tau\tilde\omega}{2}|\varsigma_i|\lvec\mu_i|\hat p|\mu_{i'}).
\label{c42}
\end{eqnarray}
From (\ref{c42}) we see that the matrix elements of $\mathcal{S}$ are bounded by
\begin{equation}
0\le\lvec\mu_i|\mathcal{S}|\mu_{i'})<1.
\label{c44}
\end{equation}
For off diagonal elements, with $i'\ne i$, the second line of (\ref{c42}) and (\ref{eobd40}) imply that for small enough $d\tau>0$, 
\begin{eqnarray}
\lvec\mu_i|\mathcal{S}|\mu_{i'})=\frac{d\tau\tilde\omega}{2}|\varsigma_i|\lvec\mu_i|\hat p|\mu_{i'})\ge 0,\quad\hbox{and}\quad \lvec\mu_i|\mathcal{S}|\mu_{i'})<1.
\label{c46}
\end{eqnarray}
For diagonal elements, with $i'= i$, the second line of (\ref{c42}) and (\ref{eobd46}) imply that for small enough $d\tau$,
\begin{eqnarray}
\lvec\mu_i|\mathcal{S}|\mu_{i})=1-d\tau|\varsigma_i|\left[1-\frac{\tilde\omega}{2}\lvec\mu_i|\hat p|\mu_{i})\right]<1,\quad\hbox{and}\quad \lvec\mu_i|\mathcal{S}|\mu_{i})>0.
\label{c48}
\end{eqnarray}
\subsection{Scattering in clouds of finite thickness\label{sc}}
We now consider non-emissive clouds of finite thickness, which can scatter or absorb, but which are too cold to emit  radiation at the frequency of interest.  An example is visible sunlight in Earth's clouds.  The clouds, at a temperature of around $300$ K,  are much cooler than the temperature, around 5700 K, of the Sun's photosphere, where the sunlight was generated. 

For a non-emissive cloud we can set $|B\} =\hat 0$ in (\ref{vet14}), which then becomes  the homogeneous equation  of transfer 
\begin{equation}
\left(\frac{d}{d\tau}+\hat\kappa\right) |I(\tau) \}=\hat 0.
\label{sct2}
\end{equation}

A general solution of (\ref{sct2}) that is useful for our purposes is 
\begin{equation}
|I(\tau) \}=\mathcal{U}(\tau)|A\}.
\label{sct4}
\end{equation}
Here we have introduced the operator
\begin{eqnarray}
\mathcal{U}(\tau)&=&\sum_i e^{-\kappa_i(\tau-\tau_i)}|\lambda_i)\lvec\lambda_i|\nonumber\\
&=&e^{-\hat\kappa_{\bf d} (\tau-\tau_c)}+e^{-\hat\kappa_{\bf u} \tau}.
\label{sct6}
\end{eqnarray}
The reference optical depths are
\begin{equation}
\tau_i=\left \{\begin{array}{rl}\tau_c, &\mbox{if $i\le n$, }\\
0,&\mbox{if $i>n$.} \end{array}\right . 
\label{sct6a}
\end{equation}
The upward and downward parts of $\mathcal{U}(\tau)$ are
\begin{equation}
e^{-\hat\kappa_{\bf d} (\tau-\tau_c)}=\sum_j e^{-\kappa_i(\tau-\tau_c)}|\lambda_j)\lvec\lambda_j|
\quad\hbox{and}\quad
e^{-\hat\kappa_{\bf u} \tau}=\sum_ke^{-\kappa_k\tau}|\lambda_k)\lvec\lambda_k|.
\label{sct6b}
\end{equation}
For a cloud of finite thickness, $\tau_c>0$, for optical depths $\tau$ within the cloud,  $0\le \tau\le \tau_c$, and for at least some single-scattering absorption, $1-\tilde \omega >0$, the diagonal matrix elements of $\mathcal{U}(\tau)$ are positive and no larger than 1,
\begin{equation}
0<\lvec\lambda_i|\mathcal{U}(\tau)|\lambda_i)\le 1.
\label{sct7}
\end{equation}
The amplitude $|A\}$ of (\ref{sct4}) is a $2n\times 1$ column vector that is independent of the optical depth $\tau$.  As we shall show presently, the value of $|A\}$ is determined by boundary conditions.

From (\ref{sct6}) we see that the rate of change of $\mathcal{U}(\tau)$ with optical depth $\tau$ is
\begin{equation}
\frac{d}{d\tau}\mathcal{U}(\tau)=-\sum_i e^{-\kappa_i(\tau-\tau_i)}\kappa_i|\lambda_i)\lvec\lambda_i|
=-\hat\kappa\mathcal{U}(\tau).
\label{sct8}
\end{equation}
According to (\ref{sct8}) the evolution operator $\mathcal{U}(\tau)$ is a solution of the homogeneous equation of transfer (\ref{sct2}) and therefore the intensity (\ref{sct4}) is also a solution.

From (\ref{sct4}) we see that the incoming intensity $|I^{\{\rm in\}} \}$ of (\ref{c10}) must be related to the amplitude $|A\}$ by
\begin{equation}
|I^{\{\rm in\}} \} = \mathcal{I}|A\},
\label{sct10}
\end{equation}
where the
{\it  incoming matrix} is
\begin{eqnarray}
 \mathcal{I} &=&
\mathcal{M}_{\bf u}\mathcal{U}(0)+\mathcal{M}_{\bf d}\mathcal{U}(\tau_c)\nonumber\\
&=&\mathcal{C}_{\bf u  d}e^{\hat\kappa_{\bf d}\tau_c}+\mathcal{C}_{\bf u u} + \mathcal{C}_{\bf d d}+\mathcal{C}_{\bf d u}e^{-\hat\kappa_{\bf u} \tau_c}.
\label{sct12}
\end{eqnarray}
The overlap matrices $\mathcal{C}_{\bf q q'}$ were defined by (\ref{olp2}).

Similarly,
from  (\ref{sct4}) and  (\ref{c12}) we see that the outgoing intensity $|I^{\{\rm out\}} \}$ is the product of the {\it outgoing matrix} $\mathcal{O}$ and the amplitude $|A\}$
\begin{equation}
|I^{\{\rm out\}} \} = \mathcal{O}|A\},
\label{sct24a}
\end{equation}
where
\begin{eqnarray}
 \mathcal{O}&=&\mathcal{M}_{\bf d}\mathcal{U}(0) +\mathcal{M}_{\bf u}\mathcal{U}(\tau_c) \nonumber\\
&=&\mathcal{C}_{\bf d  d}e^{\hat\kappa_{\bf d}\tau_c}+\mathcal{C}_{\bf d u} + \mathcal{C}_{\bf u d}+\mathcal{C}_{\bf u u}e^{-\hat\kappa_{\bf u} \tau_c}.
\label{sct22}
\end{eqnarray}
Solving (\ref{sct10}) for $|A\}$ we find
\begin{equation}
\quad|A \}=\mathcal{I}^{-1}|I ^{\{\rm in\}}\}.
\label{sct14}
\end{equation}
Substituting (\ref{sct14}) into (\ref{sct4}) we find that the intensity within the cloud is
\begin{equation}
|I (\tau)\} =\mathcal{U}(\tau)\mathcal{I}^{-1}|I ^{\{\rm in\}}\}.
\label{sct16}
\end{equation}
We can use (\ref{sct6}) to write the projection of the intensity  (\ref{sct16}) on the stream $i$ at an optical depth $\tau$ inside the cloud as
\begin{equation}
\lvec \mu_i|I (\tau)\} 
=\sum_{i'}e^{-\kappa_{i'}(\tau-\tau_{i'})}\lvec \mu_{i}|\lambda_{i'})\lvec \lambda_{i'}|\mathcal{I}^{-1}|I ^{\{\rm in\}}\}.
\label{sus6}
\end{equation}

Using (\ref{sct14}) in (\ref{sct24a}) we find that the outgoing intensity can be written as
\begin{equation}
|I ^{\{\rm out\}}\} =\mathcal{S}|I ^{\{\rm in\}}\},
\label{sct18}
\end{equation}
where the scattering-matrix is
\begin{equation}
\mathcal{S}=\mathcal{O}\mathcal{I}^{-1}.
\label{sct20}
\end{equation}
For an infinitesimally thin cloud, $\tau_c=d\tau \to 0$, one can show that (\ref{sct20}) approaches the limiting expression (\ref{c30}).  

For a cloud of finite thickness $\tau_c$, the coherent scattering matrix, denoted by the symbol $\mathcal{S}^{\rm C}$, is
\begin{equation}
\mathcal{S}^{\rm C}=e^{-\tau_c(\hat \varsigma_{\bf u}-\hat\varsigma_{\bf d})}.
\label{sct22a}
\end{equation}
$\mathcal{S}^{\rm C}$ describes the fractions of outgoing streams that have not been absorbed or scattered, and simply passed through the cloud.
The diffuse scattering matrix, denoted by the symbol $\mathcal{S}^{\rm D}$, is
\begin{equation}
\mathcal{S}^{\rm D}=\mathcal{S}-\mathcal{S}^{\rm C} = \mathcal{O}\mathcal{I}^{-1}-e^{-\tau_c(\hat \varsigma_{\bf u}-\hat\varsigma_{\bf d})},
\label{sct24}
\end{equation}
The diffuse scattering matrix, $\mathcal{S}^{\rm D}$, describes outgoing radiation  that has been scattered one or more times.
\begin{figure}[t]
\postscriptscale{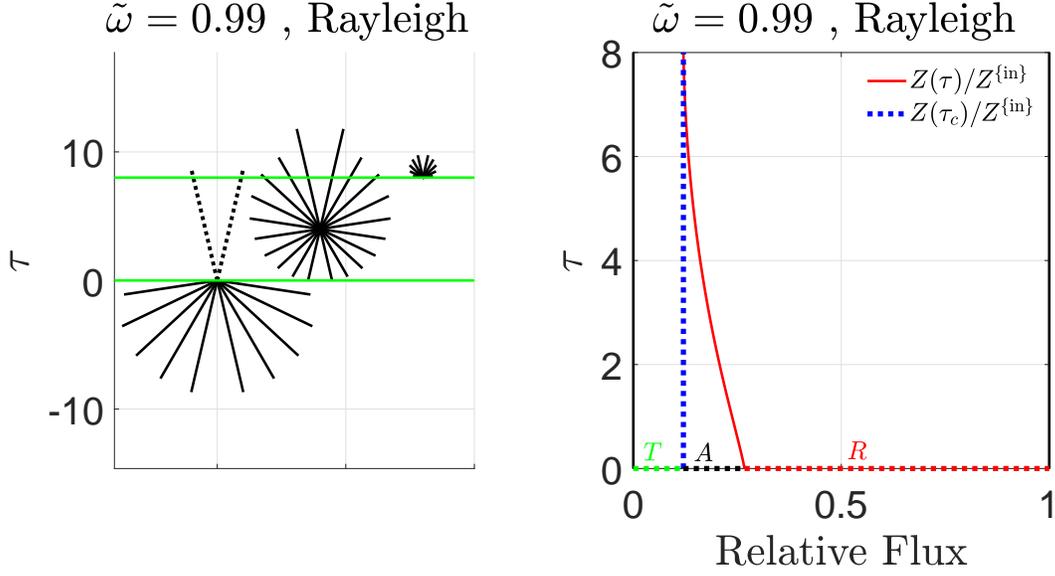}{1}
\caption{Transmission, absorption and reflection of an initial upward stream, $|I^{\{\rm in\}}\}=|\mu_{2n})$ by a cloud with weak single-scatter absorption, $1-\tilde\omega = 0.01$, with a Rayleigh-scattering phase function like that of Fig. \ref{ph2}, and with an optical depth $\tau_c = 8$. There are $2n = 10$ streams. 
The lengths of the black rays on the left panel are the sample intensities $I(\mu_i,\tau)=w_i^{-1}\lvec \mu_i| I(\tau)\}$ at the optical depth $\tau$, as calculated with (\ref{sus6}). The dotted rays at the bottom of the cloud are $1/10$ the value of $I(\mu_{2n},\tau=0)$, the intensity of the incident stream which is too long to fit on the graph.  Indicated on the bottom of the figure are fractions of the incident flux that are transmitted: $T=0.1204$, absorbed:  $A= 0.1472$, and reflected: $R = 0.7324$. 
\label{fin2}}
\end{figure}
\begin{figure}[t]
\postscriptscale{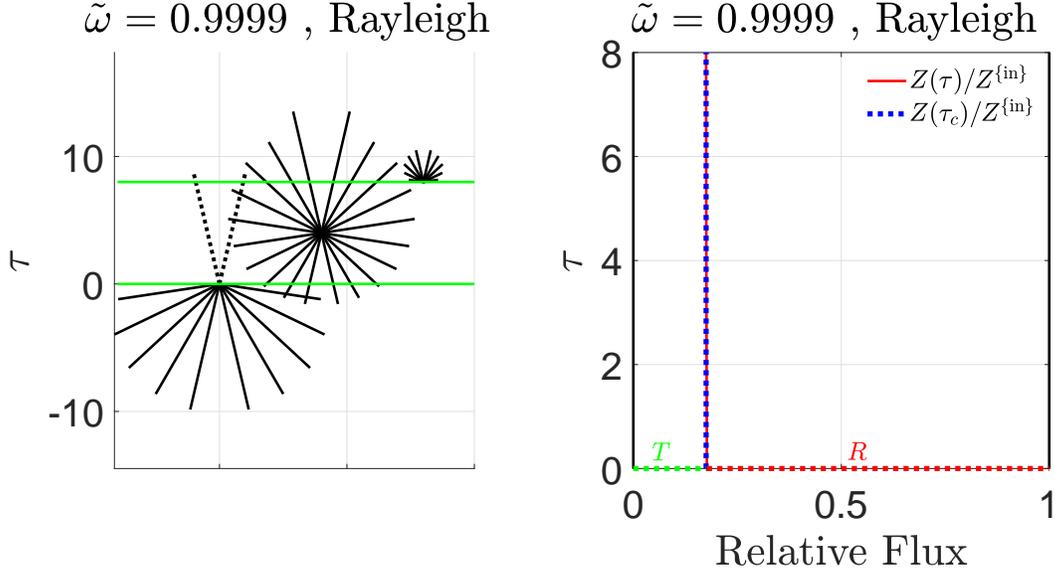}{1}
\caption{Like Fig. \ref{fin2} but with 100 times smaller single-scattering absorption $1-\tilde\omega = 0.0001$. Indicated on the bottom of the figure are fractions of incident flux transmitted: $T= 0.1751$, absorbed: $A=0.0018$, and reflected: $R = 0.8231$. For such small single-scattering absorption, the cloud absorption cannot be clearly displayed in the graph.
\label{fin2a}}
\end{figure}
\begin{figure}[t]
\postscriptscale{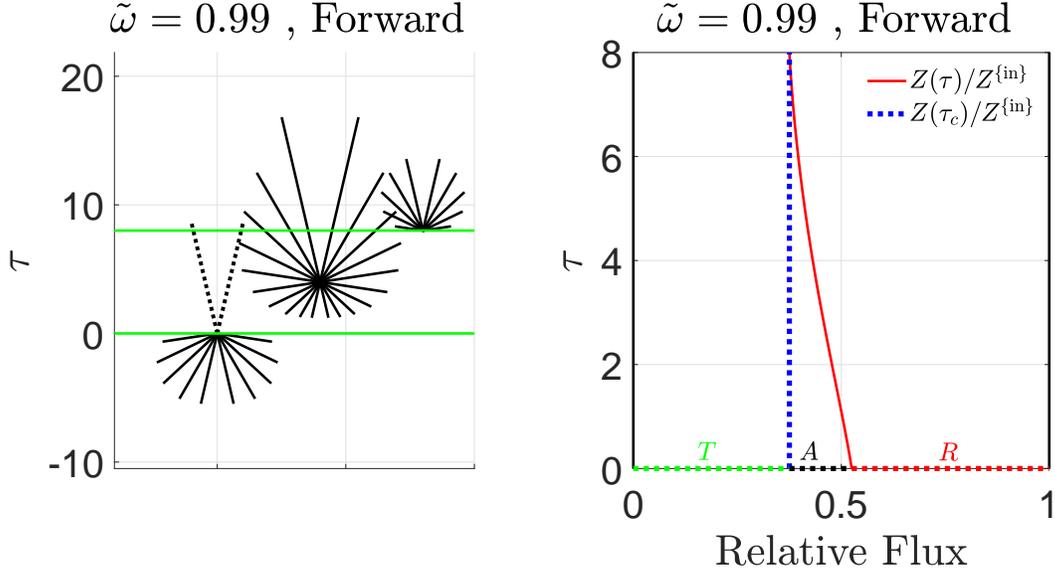}{1}
\caption{ Transmission and reflection  as in Fig. \ref{fin2} but for a forward-scattering phase function like that of Fig. \ref{ph3}. Because of the strong forward scattering, more of the light is transmitted and less is reflected than for the Rayleigh scattering of  Fig. \ref{fin2}. The  fractions of incident flux are:  transmitted, $T= 0.3752$; absorbed, $A=0.1499$; and reflected $R = 0.4749$.
\label{fin3}}
\end{figure}
\begin{figure}[t]
\postscriptscale{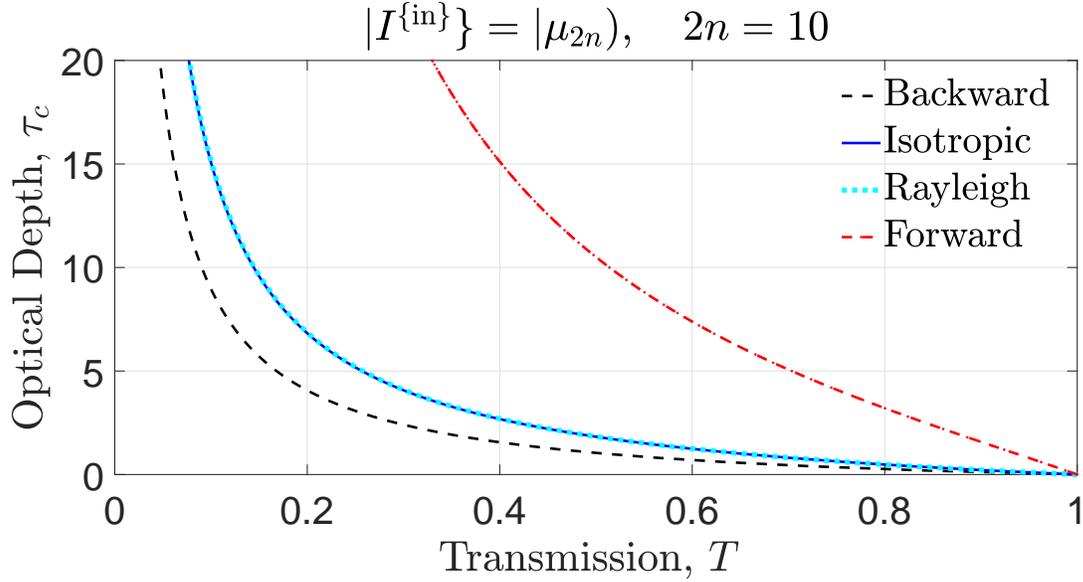}{1}
\caption{ Transmission fraction $T$ versus optical depth $\tau_c$ for a pure-scattering cloud ($\tilde\omega \to 1$) and modeled with $2n = 10$ streams. The scattering phases are isotropic, Rayleigh, maximum forward with  $p(\mu)=\varpi^{\{n\}}(\mu)$ from (\ref{pfb2}) and maximum backward with   $p(\mu)=\varpi^{\{n\}}(-\mu)$. The input intensity, $|I^{\{\rm in\}}=|\mu_{2n})$, is nearly vertical,  with a direction cosine $\mu_{2n} = 0.9739$.  The forward-scattering is peaked at $\varpi^{\{n\}}(\mu=1)=n(n+1)=30$.  For pure scattering, the cloud reflection is $R=1-T$ and there is no absorption, $A=0$. For an optical depth of $\tau_c = 20$, the transmitted fractions for backward, isotropic, Rayleigh and forward scattering are: 0.0469, 0.0770, 0.0773, and 0.3299. 
 See the text for more discussion.
\label{trans1}}
\end{figure}
\begin{figure}[t]
\postscriptscale{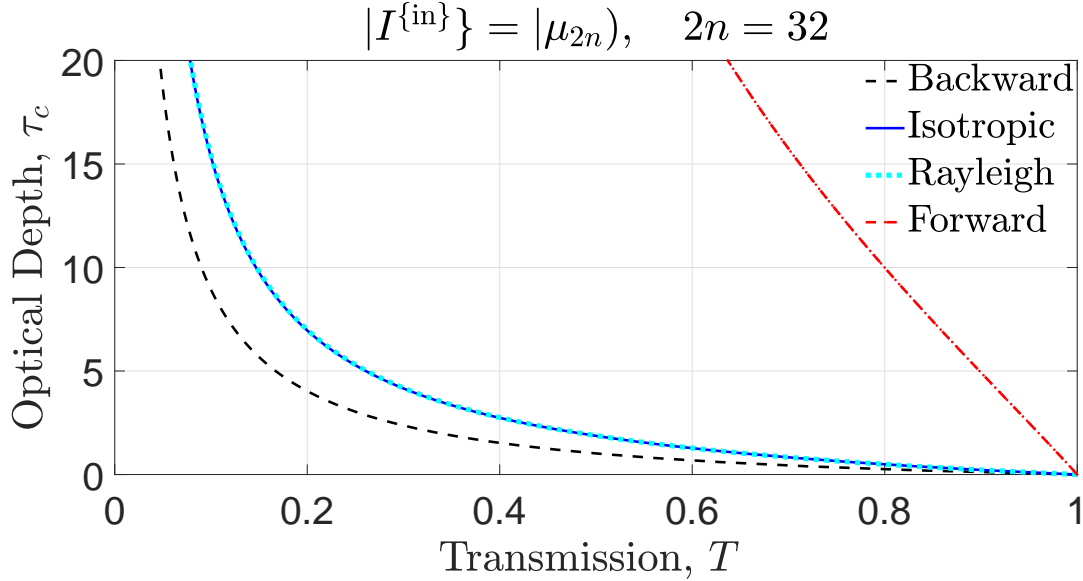}{1}
\caption{ Transmission fraction $T$ versus optical depth $\tau_c$ for a pure-scattering cloud like Fig. \ref{trans1},  but with $2n = 32$ streams. The input intensity, $|I^{\{\rm in\}}=|\mu_{2n})$, is more  vertical,  with a direction cosine $\mu_{2n} = 0.9973$, and the forward-scattering is more strongly peaked at $\varpi^{\{n\}}(\mu=1)=n(n+1)=272$. The transmission for the maximum forward scattering phase function increases from 33\% to 64\%. The transmission  for isotropic, Rayleigh or maximum backward scattering is not changed much. For an optical depth of $\tau_c = 20$, the transmitted fractions for backward, isotropic, Rayleigh and forward scattering are: 0.0467, 0.0782, 0.0786, and 0.6369. 
\label{trans2}}
\end{figure}
\begin{figure}[t]
\postscriptscale{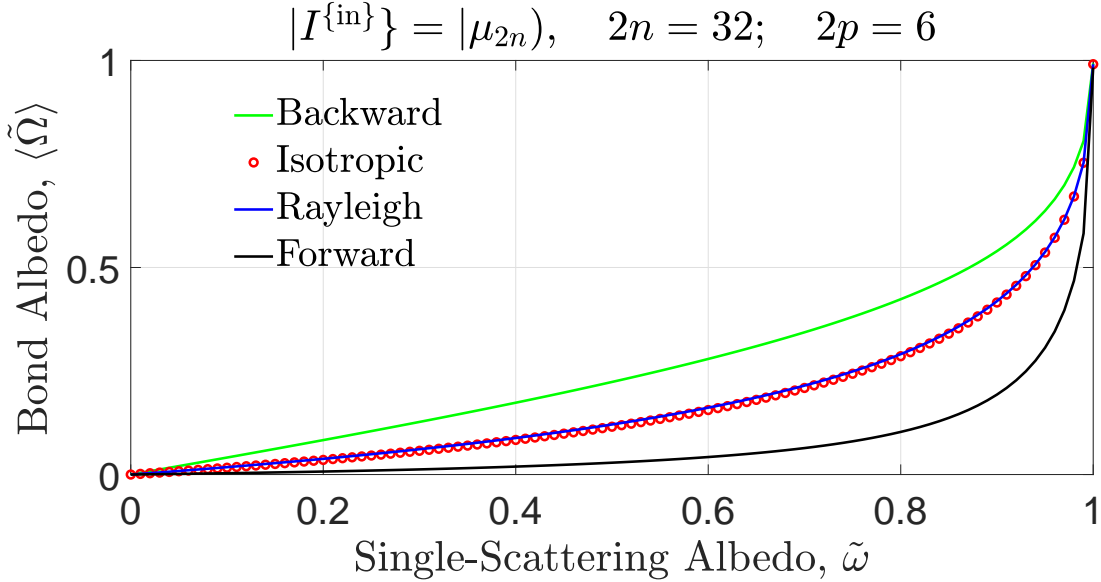}{1}
\caption{Cloud albedos $\langle\Omega\rangle$ of (\ref{alb14})  for optically thick clouds with the albedo matrices $\Omega$ of  (\ref{otc6}) and with the most vertical possible incident intensity $|I^{\{\rm in\}}\}=|\mu_{2n})$ , versus single-scattering albedo $\tilde \omega$. Albedos are shown for four different scattering phase functions which are discussed further in the text.   The results depend little on the number $n$ of stream pairs. In this plot $2n=32$. For $\tilde \omega = 0.8$ the cloud albedos $\langle\Omega\rangle$ for forward, isotropic, Rayleigh, and backward scattering are:  0.1029, 0.2857, 0.2907 and 0.4234. For all the same conditions except $2n = 6$, the values of $\langle\Omega\rangle$ become: 0.1111, 0.2976, 0.3011, and 0.4277. In accordance with (\ref{alb18}), all of the cloud albedos approach unity, $\langle \Omega\rangle\to 1$ as the single-scattering albedo approaches unity, $\tilde\omega \to 1$. See the text for more discussion.
\label{alb0}}
\end{figure}

\subsection{Cloud albedo}
We can use  (\ref{sdbv46}) and (\ref{vc10}) to write the vertical flux (\ref{mn32})  as
\begin{eqnarray}
Z(\tau)&=&\lvec 0|Z(\tau)\}
\label{alb0a}
\end{eqnarray}
Here the upward flux vector $|Z(\tau)\}$ is related to the intensity vector $|I(\tau)\}$ by
\begin{eqnarray}
|Z(\tau)\}&=&4\pi\hat\mu|I(\tau)\}.
\label{alb0b}
\end{eqnarray}
The projections of intensity, $\lvec \mu_i|I\}\ge 0$ are positive (or zero) for all streams $i$, but  the projections of vertical flux, $\lvec\mu_k|Z\}=4\pi\mu_k \lvec \mu_k|I\}\ge 0$ are positive for upward streams, with $\mu_k>0$, but negative. $ \lvec\mu_j|Z\}=4\pi\mu_j \lvec \mu_k|I\}\le 0$ for downward streams with $\mu_j<0$.
 
We write the net flux coming into the cloud as the sum of the flux 
into the bottom of the cloud carried by upward streams, and the flux into the top of the cloud carried by downward streams,
\begin{eqnarray}
Z^{\{\rm in\}} 
&=&\lvec 0|Z^{\{\rm in\}} \},
\label{alb0c}
\end{eqnarray}
where the input flux vector is 
\begin{eqnarray}
|Z^{\{\rm in\}} \}&=&\mathcal{M}_{\bf u}|Z(0)\}-\mathcal{M}_{\bf d}|Z(\tau_c)\}\nonumber\\
&=&4\pi\left(\mathcal{M}_{\bf u}\hat\mu|I(0)\}-\mathcal{M}_{\bf d}\hat\mu|I(\tau_c)\}\right)\nonumber\\
&=&4\pi(\hat\mu_{\bf u}-\hat\mu_{\bf d})\left(\mathcal{M}_{\bf u}|I(0)\}+\mathcal{M}_{\bf d}|I(\tau_c)\}\right)\nonumber\\
&=&4\pi(\hat\mu_{\bf u}-\hat\mu_{\bf d})|I^{\{\rm in\}}\}.
\label{alb0d}
\end{eqnarray}
In like manner we write the outgoing flux as the sum of the upward flux out of the top of the cloud and the downward flux out of the bottom
\begin{eqnarray}
Z^{\{\rm out\}}
&=&\lvec 0|Z^{\{\rm out\}}\},
\label{alb0e}
\end{eqnarray}
where the output flux vector is
\begin{eqnarray}
|Z^{\{\rm out\}} \}=4\pi(\hat\mu_{\bf u}-\hat\mu_{\bf d})|I^{\{\rm out\}}\}.
\label{alb0f}
\end{eqnarray}
For a non-emissive cloud we can use (\ref{alb0d}) and (\ref{alb0f}) with (\ref{sct18}) to write as
\begin{equation}
|Z^{\{\rm out\}} \}=\Omega |Z^{\{\rm in\}}\}.
\label{alb10}
\end{equation}
Here the cloud albedo matrix is
\begin{eqnarray}
\Omega &=& (\hat\mu_{\bf u}-\hat\mu_{\bf d}) \mathcal{S} (\hat\mu_{\bf u}-\hat\mu_{\bf d})^{-1}\nonumber\\ 
&=& (\hat\mu_{\bf u}-\hat\mu_{\bf d}) \mathcal{S} (\hat\varsigma_{\bf u}-\hat\varsigma_{\bf d}).
\label{alb12}
\end{eqnarray}
From the first line of (\ref{alb12}) we see that the cloud albedo matrix, $\Omega$, is a similarity transformation of the scattering matrix, $\mathcal{S}$.
We define the cloud albedo $\langle \Omega\rangle$ as the ratio of the outgoing flux (\ref{alb0e}) to the  incoming flux (\ref{alb0c}),
\begin{eqnarray}
\langle \Omega\rangle&=&\frac{Z^{\{\rm out\}}}{Z^{\{\rm in\}}}\nonumber\\
&=&\frac{\lvec 0|\Omega |Z^{\{\rm in\}}\}}{\lvec 0|Z^{\{\rm in\}}\}}.
\label{alb14}
\end{eqnarray}
Since the cloud albedo $\langle  \Omega\rangle$ of (\ref{alb14}) is a ratio of an outgoing flux to an incoming flux, it is equivalent to an astronomical Bond albedo  -- as is the single-scattering albedo $\tilde \omega$.

In a subsequent paper we will show that 
\begin{equation}
\lvec 0|\Omega \to \lvec 0|,\quad\hbox{as}\quad \tilde\omega \to 1.
\label{alb16}
\end{equation}
The limiting behavior (\ref{alb16}) implies that the cloud albedo (\ref{alb14}) approaches unity as $\omega \to 1$, 
\begin{equation}
\langle\Omega\rangle \to 1,\quad\hbox{as}\quad \tilde\omega \to 1,
\label{alb18}
\end{equation}
irrespective of the cloud thickness $\tau_c$, the scattering phase function $p(\mu,\mu')$, or  the nature of the incident flux, $|Z^{\{\rm in\}}\}=4\pi (\hat \mu_{\bf u}-\hat\mu_{\bf d}) |I^{\{\rm in\}}\}$.  An example of (\ref{alb18}) can be seen on Fig. \ref{alb0}.
\subsection{Examples of radiative transfer in clouds \label{esc}}
In Fig. \ref{fin2} we show $Z(\tau)/Z^{\{\rm in\}}$, calculated with (\ref{sct16}), (\ref{alb0a}) and (\ref{alb0b}), and with the most vertical possible input intensity $|I^{\{\rm in\}}\}=|\mu_{2n})$   for  a non-emissive (cold) cloud of optical depth $\tau_c = 8$. The cloud has a Rayleigh-scattering phase function like that of Fig. \ref{ph2}, and it has very weak single-scattering absorption, $1-\tilde\omega = 0.01$, comparable to that of Earth's clouds for sunlight. To facilitate plotting Fig. \ref{fin2}, only $2n=10$ streams were used. The fluxes change very little with the number of streams. For example, the transmissivities are $T=Z(\tau_c)/Z^{\{\rm in\}} = 0,1167,\,0.1204,$ and $0.1225$ for $2n=6,\,10,$ and $32$ streams. The increased transmission for more streams is mainly due to the  shorter relative slant path of the input stream through the cloud, $\varsigma_{2n}=1/\mu_{2n} = 1.0724,\, 1.0268,$ and $1.0027$, for $2n = 6,\,10$ and $32$.

Shown on the left panel of the figure are the unweighted  intensities  $I(\mu_i,\tau)= w_i^{-1} \lvec \mu_i|I (\tau)\}$ at various optical depths $\tau$. At the bottom of the cloud  the only non-zero amplitude of upward intensity is  $I(\mu_{2n},0)=w_{2n}^{-1}\lvec \mu_{2n}|I^{\{\rm in\}}\}$, that of the single, nearly vertical input stream.  The downward intensity at the bottom of the cloud $I(\mu_j,0)$, comes from diffuse reflection.     In the middle of the cloud, at optical depth $\tau=4$, the intensity $I(\mu_i,4)$ is roughly isotropic, but has an upward bias. At the top of the cloud, there is no downward intensity, $I(\mu_j, 8)=0$, but there is  diffusely transmitted upward intensity,  $I(\mu_k, 8)>0$. Because of the isotropization by multiple scattering, the outgoing light from the top of the cloud retains no trace of the directionality  of the nearly vertical incoming stream  at the bottom of the cloud.

The right panel of Fig. \ref{fin2} shows the relative total flux $Z(\tau)/Z^{\{\rm in\}}$.
For the large optical depth of the cloud, $\tau_c = 8$, the coherent transmission  $\approx e^{-8} = 3.35 \times 10^{-4}$, is much too small to  display  on the graph. But the weak absorption and multiple scattering allows about 12\% of the radiation to reach the top of the cloud as diffuse transmission. 

Fig. \ref{fin2a} has all the same parameters as Fig. \ref{fin2}, except for 100 times weaker single-scattering absorption, $1-\tilde\omega = 0.0001$. This increases the transmission of the cloud to about 18\% and it reduces the absorption to only 0.18\%, too small to be recognizable on the figure.

The plots of Fig. \ref{fin3} are like those of Fig. \ref{fin2}, but the Rayleigh-scattering phase function has been replaced by the forward-scattering phase function, $p(\mu)=\varpi^{\{3\}}(\mu)$, of Fig. \ref{ph3}, constructed from the first $2p=6$ Legendre polyomials. This modestly peaked forward-scattering, ($p(1)=12$), more than triples  the transmitted fraction of flux, from 12\% to 38\%. The reflected flux decreases from  73\% to 47\%.
 
Fig. \ref{trans1} shows the fraction of a nearly vertical stream that is transmitted through a cloud with negligible single-scattering absorption, $\tilde\omega \to 1$, as a function of the cloud's optical depth $\tau_c$. The radiative transfer is modeled with $2n = 10$ streams. Four different scattering phase functions were  assumed, isotropic, Rayleigh,  maximum forward scattering with $p(\mu)=\varpi^{\{n\}}(\mu)$ from (\ref{pfb2})  and the corresponding maximum backward scattering phase $p(\mu) = \varpi^{\{n\}}(-\mu)$.

Fig. \ref{trans2} is the same as Fig. \ref{trans1}, except that the transmission is modeled with $2n = 32$ streams.  This makes very little difference to the transmission for clouds with isotropic, Rayleigh or backscattering phase functions, but it greatly increases the transmission of the cloud with maximum forward-scattering phase function, $p(\mu)=\varpi^{\{n\}}(\mu)$ of (\ref{pfb2}).

Fig. \ref{alb0} shows cloud albedos $\langle\Omega\rangle$ of  (\ref{alb14}) for optically thick clouds, with $\tau_c = \infty$, versus the single-scattering albedo $\tilde \omega$ for four different scattering phase functions:   isotropic scattering, with the simplest scattering phase function $p(\mu) = 1$;  Rayleigh scattering  like that of Fig. \ref{ph2} with the phase function $p(\mu)= (3/4)(1+\mu^2)$;  forward scattering like that of Fig. \ref{ph3} with the phase function $p(\mu)=\varpi^{\{3\}}(\mu)$, the maximum forward scattering that can be modelled with the first six Legendre polynomials;  and  backward   scattering like that of Fig. \ref{ph4} with the phase function  $p(\mu)=\varpi^{\{3\}}(-\mu)$. The phase functions  $ \varpi^{\{p\}}(\mu)$ are defined by (\ref{pfb2}).  Not surprisingly, clouds with backward-scattering phase functions have significantly larger cloud albedos than those with forward-scattering phase functions, where photons penetrate deeply into the cloud and require many more collisions to reverse direction and ``random walk" out of the cloud before being absorbed.

\begin{figure}[t]
\postscriptscale{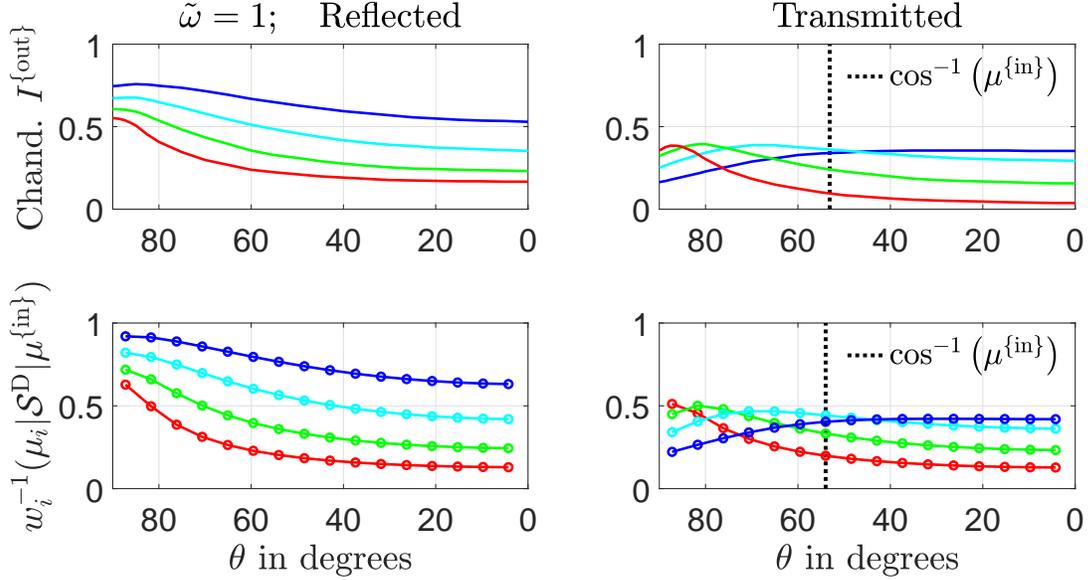}{1}
\caption{The top panels are reproductions of Fig. 21 from Chandrasekhar's book {\it Radiative Transfer}\cite{Chandrasekhar}. Diffuse reflections and transmissions were calculated for conservative isotropic scattering in clouds of optical depths $\tau_c =$ 0.25 (red), 0.5 (green), 1 (light blue) and 2 (dark blue). The diffuse reflections increase with increasing optical depth. Assuming radiation incident on the bottom of the cloud, the diffuse reflections on the left are plotted versus ``south-latitude" angles $\theta$. Diffuse  transmissions on the right are plotted versus ``north-latitude" angles $\theta$. The bottom two panels are the same quantities calculated with the $2n=32$-stream model,  at the latitude angles corresponding to the Gauss-Legendre cosines $\mu_i$ with near unity single-scattering albedo, $\tilde \omega = 1-10^{-6}$, and for isotropic scattering.  The diffuse part, $\mathcal{S}^{\rm D}$, of the scattering matrix is given by (\ref{sct24}). The incident stream was chosen to have a Gauss-Legendre cosine, $\mu^{\{\rm in\}}=\mu_{23}= 0.5877$, as close as possible to the value $\mu^{\{\rm in\}} = 0.6$ used for Chandrasekhar's calculations.
\label{Ch21}}
\end{figure}
\begin{figure}[t]
\postscriptscale{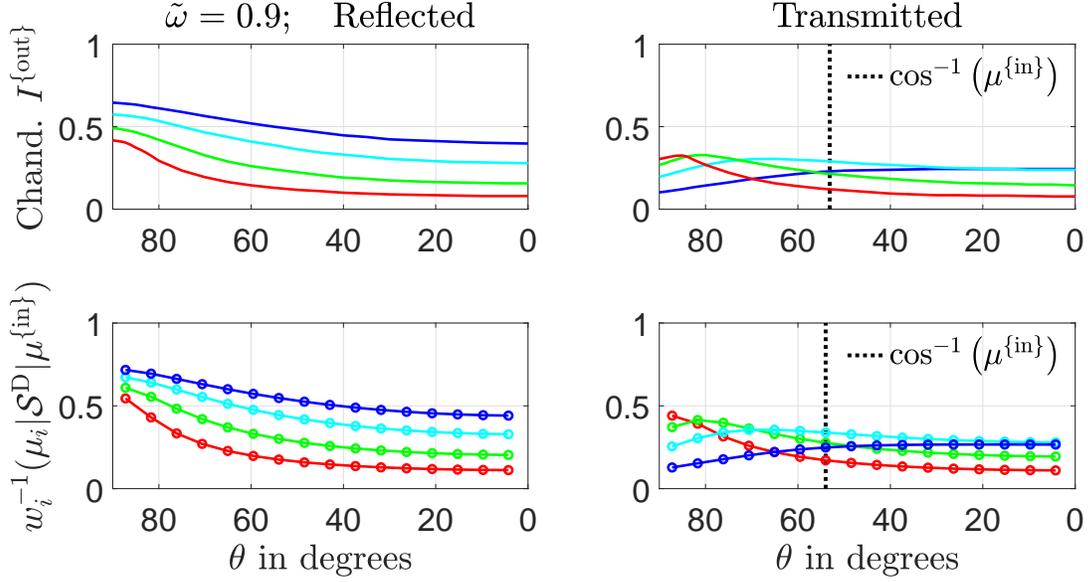}{1}
\caption{Like Fig. \ref{Ch21}, but the top two panels are reproductions of Fig. 22 from Chandrasekhar's book, and the single-scattering albedo was $\tilde \omega = 0.9$. The finite single-scattering absorption probability, $1-\tilde\omega=0.1$, has the largest effect on clouds with the largest optical depth.
\label{Ch22}}
\end{figure}
\subsection{Comparison with Chandrasekhar}
In Fig. 21 and Fig. 22 of  {\it Radiative Transfer}\cite{Chandrasekhar}, Chandrasekhar displays diffuse transmission and reflection for isotropic scattering, calculated for  relatively thin clouds with optical depths $\tau_c = 0.25, 0.5, 1$ and $2$.

In Fig. \ref{Ch21} we compare diffuse reflection and transmission, calculated with (\ref{sct24}) with the results shown in Chandrasekhar's Fig. 21, conservative isotropic scattering with single-scattering albedo $\tilde \omega = 1$.  Chandrasekhar's Fig. 22, calculated with $\tilde\omega = 0.9$, should be compared with our Fig \ref{Ch22}. The results from the $2n$-stream model, with $2n=32$ can hardly be distinguished from those of Chandrasekhar.  

With the $2n$-stream model it is easy to make calculations analogous to those of Fig. \ref{Ch21} and Fig. \ref{Ch22}  for diffuse reflection and transmission with any nonnegative  phase function that can be expanded on the first $2n$ Legendre polynomials. It appears to be much harder to extend the methods outlined by  Chandrasekhar beyond isotropic scattering.
\subsection{Block Matrices}
To facilitate subsequent discussions, we will represent  some of the operators defined above as blocks of matrix elements between upward or downward,  left or right basis vectors. As illustrated below, the basis vectors may be  in either $\mu$-space or $\lambda$-space or both, depending on the task at hand.

It is natural to write  the scattering matrix  as the sum of four blocks of elements between left and right basis vectors in $\mu$-space,
\begin{equation}
\mathcal{S}
=\mathcal{S}_{\bf dd}+\mathcal{S}_{\bf ud}+\mathcal{S}_{\bf du}+\mathcal{S}_{\bf uu},
\label{blm2}
\end{equation}
where
\begin{eqnarray}
\mathcal{S}_{\bf dd}&=&\sum_{jj'}|\mu_j)\lvec \mu_j|S|\mu_{j'})\lvec\mu_{j'}|=\left[\begin{array}{ll}\mathcal{S}_{\bf d d}&\hat 0\\ \hat 0&\hat 0\end{array}\right],\label{blm4}\\
\mathcal{S}_{\bf ud}&=&\sum_{kj'}|\mu_k)\lvec \mu_k|S|\mu_{j'})\lvec\mu_{j'}|=\left[\begin{array}{ll}\hat 0&\hat 0\\ 
\mathcal{S}_{\bf u d}&\hat 0\end{array}\right],\label{blm6}\\
\mathcal{S}_{\bf du}&=&\sum_{j k'}|\mu_j)\lvec \mu_j|S|\mu_{k'})\lvec\mu_{k'}|=\left[\begin{array}{ll}\hat 0&\mathcal{S}_{\bf  d u}\\ 
\hat 0&\hat 0\end{array}\right],\label{blm8}\\
\mathcal{S}_{\bf uu}&=&\sum_{kk'}|\mu_k)\lvec \mu_k|S|\mu_{k'})\lvec\mu_{k'}|
=\left[\begin{array}{ll}\hat 0&\hat 0\\ 
\hat 0&\mathcal{S}_{\bf  u u}\end{array}\right].
\label{blm10}
\end{eqnarray}
As shown in (\ref{blm4}) - (\ref{blm10}), we will use the same symbol $\mathcal{S}_{\bf  q q'}$ to denote either: (1) an $n\times n$ block of matrix elements, or (2) the full $2n\times 2n$ matrix, padded with three $n\times n$ arrays of zeros, denoted by the symbol $\hat 0$.  The context will normally make clear which quantity is meant.

In an analogous way, we write
the incoming matrix $\mathcal{I}$ of (\ref{sct12})  as 
\begin{equation}
\mathcal{I}
=\mathcal{I}_{\bf dd}+\mathcal{I}_{\bf ud}+\mathcal{I}_{\bf du}+\mathcal{I}_{\bf uu}.
\label{blm12}
\end{equation}
The blocks are  arrays of matrix elements between the left basis vectors $\lvec\mu_i|$ and the right basis vectors $|\lambda_{i'})$,
\begin{equation}
\mathcal{I}_{\bf dd}=\sum_{jj'}|\mu_j)\lvec \mu_j|\mathcal{I}|\lambda_{j'})\lvec\lambda_{j'}|
=\left[\begin{array}{ll}\mathcal{I}_{\bf d d}&\hat 0\\ \hat 0&\hat 0\end{array}\right].
\label{blm14}
\end{equation}
The block matrices $\mathcal{I}_{\bf ud}$, $\mathcal{I}_{\bf du}$ and $\mathcal{I}_{\bf uu}$ are defined in analogy to (\ref{blm4}) - (\ref{blm10}) and (\ref{blm14}).

In like manner we write the outgoing matrix $\mathcal{O}$ of (\ref{sct22})  as 
\begin{equation}
\mathcal{O}
=\mathcal{O}_{\bf dd}+\mathcal{O}_{\bf ud}+\mathcal{O}_{\bf du}+\mathcal{O}_{\bf uu}.
\label{blm16}
\end{equation}
where
\begin{equation}
\mathcal{O}_{\bf dd}=\sum_{jj'}|\mu_j)\lvec \mu_j|\mathcal{O}|\lambda_{j'})\lvec\lambda_{j'}|
=\left[\begin{array}{ll}\mathcal{O}_{\bf d d}&\hat 0\\ \hat 0&\hat 0\end{array}\right].
\label{blm18}
\end{equation}
The block matrices $\mathcal{O}_{\bf ud}$, $\mathcal{O}_{\bf du}$ and $\mathcal{O}_{\bf uu}$ are defined in analogy to (\ref{blm4}) - (\ref{blm10}) and (\ref{blm18}).

We will write $\mathcal{I}^{-1}$, the inverse of the incoming matrix (\ref{sct12}), as arrays of matrix elements between the left basis vectors $\lvec\lambda_i|$ and the right basis vectors $|\mu_{i'})$.
\begin{equation}
\mathcal{I}^{-1}
=(\mathcal{I}^{-1})_{\bf dd}+(\mathcal{I}^{-1})_{\bf ud}+(\mathcal{I}^{-1})_{\bf du}+(\mathcal{I}^{-1})_{\bf uu},
\label{blm20}
\end{equation}
where
\begin{eqnarray}
(\mathcal{I}^{-1})_{\bf dd}&=&\sum_{jj'}|\lambda_j)\lvec \lambda_j|\mathcal{I}^{-1}|\mu_{j'})\lvec\mu_{j'}|=\left[\begin{array}{cc}(\mathcal{I}^{-1})_{\bf d d}&\hat 0\\ \hat 0&\hat 0\end{array}\right],
\label{blm22}
\end{eqnarray}
The block matrices $(\mathcal{I}^{-1})_{\bf ud}$, $(\mathcal{I}^{-1})_{\bf du}$ and $(\mathcal{I}^{-1})_{\bf uu}$ are defined in analogy to (\ref{blm4}) - (\ref{blm10}) and (\ref{blm22}).
\subsection{Optically thick clouds}
A simple limiting case  of radiative transfer is an optically thick cloud, where $\tau_c\to \infty$ and  
$e^{-\hat \kappa_{\bf u}\tau_c}\to \hat 0 $ and $e^{\hat\kappa_{\bf d}\tau_c}\to \hat 0$. Then the incoming matrix $\mathcal{I}$ of (\ref{sct12}) and the outgoing matrix $\mathcal{O}$ of (\ref{sct22}) approach the limits
\begin{equation}
\mathcal{I}\to
\left[\begin{array}{ll}\mathcal{C}_{\bf d  d}&\hat 0 \\ 
\hat 0&\mathcal{C}_{\bf u u}\end{array}\right],\quad\hbox{and}\quad
\mathcal{O}\to
\left[\begin{array}{ll}\hat 0&\mathcal{C}_{\bf d  u}\\ 
\mathcal{C}_{\bf u d}&\hat 0 \end{array}\right],
\label{otc2}
\end{equation}
Then the scattering matrix (\ref{sct20}) approaches the limit
\begin{equation}
\mathcal{S}\to
\left[\begin{array}{ll}\hat 0&\mathcal{C}_{\bf d  u}(\mathcal{C}_{\bf u u})^{-1} \\ 
\mathcal{C}_{\bf u d}(\mathcal{C}_{\bf d d})^{-1}&\hat 0\end{array}\right],
\label{otc4}
\end{equation}
For optically thick clouds there is no transmission from the bottom to the top of the cloud  or vice-versa, so $\mathcal{S}_{\bf u u}=\hat 0$ and $\mathcal{S}_{\bf d d}=\hat 0$.  Diffuse reflection from the bottom and top of the cloud are represented by the nonzero, $n\times n$ matrices $\mathcal{S}_{\bf d u}=\mathcal{C}_{\bf d  u}(\mathcal{C}_{\bf u u})^{-1}$  and  $\mathcal{S}_{\bf u d}=\mathcal{C}_{\bf u d}(\mathcal{C}_{\bf d d})^{-1}$. Chandrasekhar\cite{Chandrasekhar} discusses optically thick clouds (which he calls {\it semi-infinite})  for isotropic and Rayleigh scattering phase functions in his Chapter VI.  

The cloud albedo matrix (\ref{alb12}) for an optically thick cloud approaches
\begin{equation}
\Omega\to
-\left[\begin{array}{ll}\hat 0&\hat\mu_{\bf d}\mathcal{C}_{\bf d  u}(\mathcal{C}_{\bf u u})^{-1} \hat\varsigma_{\bf u}\\ \hat\mu_{\bf u}
\mathcal{C}_{\bf u d}(\mathcal{C}_{\bf d d})^{-1}\hat\varsigma_{\bf d}&\hat 0\end{array}\right],
\label{otc6}
\end{equation}
\begin{figure}[t]
\postscriptscale{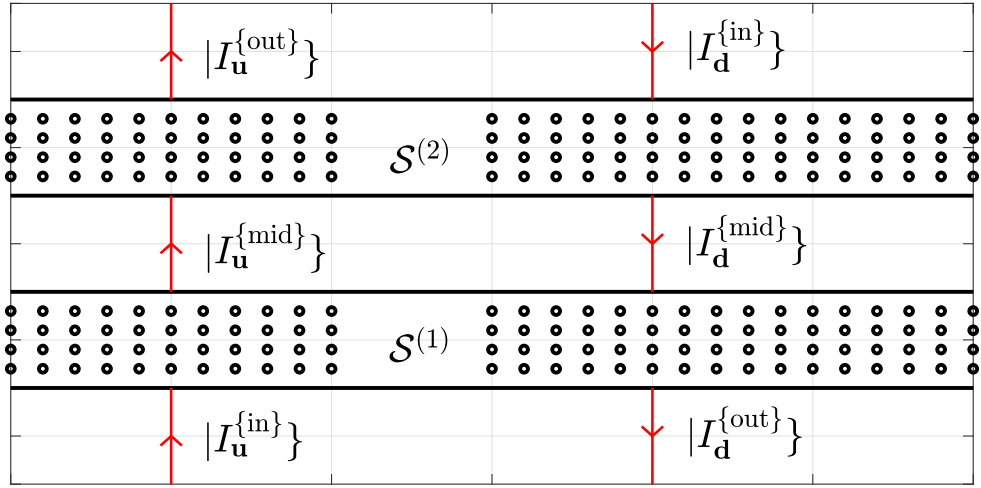}{1}
\caption{ Intensities for two clouds in series. The atmosphere is completely transparent in the mid-layer between the lower cloud with scattering matrix $\mathcal{S}^{(1)}$  and  the upper cloud with scattering matrix $\mathcal{S}^{(2)}$.
\label{seriesclouds}}
\end{figure}
\subsection{Clouds in series}
The scattering matrices, $\mathcal{S}^{\{1\}}$ and $\mathcal{S}^{\{2\}}$,  of two clouds ``in series," like those of Fig. \ref{seriesclouds} can be combined to form a single scattering matrix $\mathcal{S}$. If we write the scattering matrices of the lower and upper clouds as
\begin{equation}
\mathcal{S}^{(1)}=
\left[\begin{array}{ll}\mathcal{S}_{\bf d  d}^{(1)}&\mathcal{S}_{\bf d  u}^{(1)}\\ 
\mathcal{S}_{\bf u d}^{(1)}&\mathcal{S}_{\bf u u}^{(1)} \end{array}\right],\quad\hbox{and}\quad
\mathcal{S}^{(2)}=
\left[\begin{array}{ll}\mathcal{S}_{\bf d  d}^{(2)}&\mathcal{S}_{\bf d  u}^{(2)}\\ 
\mathcal{S}_{\bf u d}^{(2)}&\mathcal{S}_{\bf u u}^{(2)} \end{array}\right],
\label{trc0}
\end{equation}
then the addition formula is
\begin{equation}
\mathcal{S}= \left[\begin{array}{lr} \mathcal{S}^{(1)}_{\bf  d d}\left (\hat 1-\mathcal{S}^{(2)}_{\bf d u} \mathcal{S}^{(1)}_{\bf  u d}\right) ^{-1}\!\!\mathcal{S}^{(2)}_{\bf  d d}
&\mathcal{S}^{(1)}_{\bf  d d}\mathcal{S}_{\bf d u}^{(2)}\left (\hat 1-\mathcal{S}^{(1)}_{\bf u d} \mathcal{S}^{(2)}_{\bf d u}\right) ^{-1}\!\!\mathcal{S}^{(1)}_{\bf  u u}+\mathcal{S}^{(1)}_{\bf  d u}\\ 
\mathcal{S}^{(2)}_{\bf  u u}\mathcal{S}_{\bf u  d}^{(1)}\left (\hat 1-\mathcal{S}^{(2)}_{\bf d u} \mathcal{S}^{(1)}_{\bf  u d}\right) ^{-1}\!\!\mathcal{S}^{(2)}_{\bf  d d}+\mathcal{S}^{(2)}_{\bf  u d}\,\,\,
&\mathcal{S}^{(2)}_{\bf  u u}\left (\hat 1-\mathcal{S}^{(1)}_{\bf u d} \mathcal{S}^{(2)}_{\bf d u}\right) ^{-1}\!\!\mathcal{S}^{(1)}_{\bf  u u}
\end{array}\right].\label{sad2}
\end{equation}
To prove (\ref{sad2}) we see from inspection of Fig. \ref{seriesclouds} that we can write 4 equations that relate intensities and scattering matrices at the four cloud boundaries:

\begin{eqnarray}
|I ^{\{\rm out\}}_{\bf u}\} &=&\mathcal{S}_{\bf ud}^{(2)}
|I ^{\{\rm in\}}_{\bf d}\}+\mathcal{S}_{\bf uu}^{(2)}
|I _{\bf u}^{\{\rm mid\}}\},\label{sad4}\\
|I_{\bf d}^{\{\rm mid\}} \} &=&\mathcal{S}_{\bf dd}^{(2)}
|I ^{\{\rm in\}}_{\bf d}\}+\mathcal{S}_{\bf du}^{(2)}
|I_{\bf u}^{\{\rm mid\}} \},
\label{sad6}\\
|I_{\bf u}^{\{\rm mid\}} \} &=&\mathcal{S}_{\bf uu}^{(1)}
|I ^{\{\rm in\}}_{\bf u}\}+\mathcal{S}_{\bf ud}^{(1)}
|I_{\bf d}^{\{\rm mid\}} \},
\label{sad8}\\
|I ^{\{\rm out\}}_{\bf d}\} &=&\mathcal{S}_{\bf du}^{(1)}
|I ^{\{\rm in\}}_{\bf u}\}+\mathcal{S}_{\bf dd}^{(1)}
|I _{\bf d}^{\{\rm mid\}}\}.
\label{sad10}
\end{eqnarray}
Here (\ref{sad4}) relates the intensities at the top of the upper cloud; (\ref{sad6}) relates intensities at the bottom of the upper cloud; (\ref{sad8}) relates intensities at the top of the lower cloud, and (\ref{sad10}) relates intensities at the bottom of the lower cloud.  

In analogy to (\ref{c10}), we  write the  downward and upward input intensities as the column  vector
\begin{equation}
|I^{\{\rm in\}}\}=
\left |\begin{array}{r}|I _{\bf d}^{\{\rm in\}}\} \\ |I_{\bf u}^{\{\rm in\}}\} \end{array}\right\}
=
\left |\begin{array}{l}|I _{\bf d}(\tau_1+\tau_2)\} \\ |I_{\bf u}(0)\} \end{array}\right\}.
\label{sad12}
\end{equation}
We have assumed that the mid interval beween the clouds is transparent, with negligibly small optical depth, so the total optical depth from the bottom of the lower cloud to the top of the upper  cloud is $\tau_1+\tau_2$, the sum of the optical depth $\tau_1$ of the lower cloud and the optical depth $\tau_2$ of the upper cloud.

The downward and upward parts of the mid intensity  can be written  as 
\begin{equation}
|I^{\{\rm mid\}}\}=
\left |\begin{array}{r}|I _{\bf d}^{\{\rm mid\}}\} \\ |I_{\bf u}^{\{\rm mid\}}\} \end{array}\right\}
=
\left |\begin{array}{l}|I _{\bf d}(\tau_1)\} \\ |I_{\bf u}(\tau_1)\} \end{array}\right\}.
\label{sad14}
\end{equation}

In analogy to (\ref{c12}) we write
the downward and upward parts of the outward intensity as
\begin{equation}
|I^{\{\rm out\}}\}=
\left |\begin{array}{r}|I _{\bf d}^{\{\rm out\}}\} \\ |I_{\bf u}^{\{\rm out\}}\} \end{array}\right\}=
\left |\begin{array}{l}|I _{\bf d}(0)\} \\ |I_{\bf u}(\tau_1+\tau_2)\} \end{array}\right\}.
\label{sad16}
\end{equation}

From (\ref{sad4}) and (\ref{sad10}) we can write the outward intensity in terms of the mid intensity and input  intensity as
\begin{equation}
|I^{\{\rm out\}}\}=
\mathcal{M}|I^{\{\rm mid\}}\}+\mathcal{Q}|I^{\{\rm in\}}\}.\label{sad18}
\end{equation}
The coupling matrices are
\begin{equation}
\mathcal{M}
=\left[\begin{array}{ll}\mathcal{S}_{\bf d  d}^{(1)}&\hat 0\\ 
\hat 0&\mathcal{S}_{\bf u u}^{(2)} \end{array}\right].
\label{sad20}
\end{equation}
and
\begin{equation}
\mathcal{Q}= \left[\begin{array}{ll}\hat 0&\mathcal{S}^{(1)}_{\bf d u} \\ 
\mathcal{S}_{\bf u  d}^{(2)}&\hat{0} \end{array}\right].
\label{sad22}
\end{equation}
From (\ref{sad6}) and (\ref{sad8}) we see that the mid intensity and input  intensity are related by
\begin{equation}
\mathcal{P}|I^{\{\rm mid\}}\}=\mathcal{N}|I^{\{\rm in\}}\}.\label{sad24}
\end{equation}
where the matrices $\mathcal{P}$ and $\mathcal{N}$ are
\begin{equation}
\mathcal{P}= \left[\begin{array}{ll}\hat 1&-\mathcal{S}^{(2)}_{\bf d u} \\ 
-\mathcal{S}_{\bf u  d}^{(1)}&\hat{1} \end{array}\right].
\label{sad26}
\end{equation}
and
\begin{equation}
\mathcal{N}
=\left[\begin{array}{ll}\mathcal{S}_{\bf d  d}^{(2)}&\hat 0\\ 
\hat 0&\mathcal{S}_{\bf u u}^{(1)} \end{array}\right].
\label{sad28}
\end{equation}
Combining (\ref{sad18}) with (\ref{sad24}) we find
\begin{equation}
|I^{\{\rm out\}}\}=\mathcal{S}|I^{\{\rm in\}}\}.\label{sad30}
\end{equation}
where the overall scattering matrix for the two clouds is
\begin{equation}
\mathcal{S}=\mathcal{M}\mathcal{P}^{-1}\mathcal{N}+\mathcal{Q}.\label{sad32}
\end{equation}
One can verify that the inverse of the matrix $\mathcal{P}$ of (\ref{sad26}), which is needed to evaluate (\ref{sad32}) can be written as
\begin{equation}
\mathcal{P}^{-1}= \left[\begin{array}{rr}\left (\hat 1-\mathcal{S}^{(2)}_{\bf d u} \mathcal{S}^{(1)}_{\bf  u d}\right) ^{-1}
&\mathcal{S}_{\bf d u}^{(2)}\left (\hat 1-\mathcal{S}^{(1)}_{\bf u d} \mathcal{S}^{(2)}_{\bf d u}\right) ^{-1}\\ 
\mathcal{S}_{\bf u  d}^{(1)}\left (\hat 1-\mathcal{S}^{(2)}_{\bf d u} \mathcal{S}^{(1)}_{\bf  u d}\right) ^{-1}
&\left (\hat 1-\mathcal{S}^{(1)}_{\bf u d} \mathcal{S}^{(2)}_{\bf d u}\right) ^{-1}
\end{array}\right].
\label{sad34}
\end{equation}
Using  (\ref{sad20}), (\ref{sad22}), (\ref{sad28}) and (\ref{sad34}) in (\ref{sad32}) we find (\ref{sad2}).

One can divide the atmosphere into vertical layers of clear air or cloudy air. Each layer will have a scattering matrix. Starting  from the bottom, one can use (\ref{sad2}) to combine the scattering matrices of the first two layers. This net scattering matrix for the first two layers can then be  combined with the scattering matrix of the third layer, using  (\ref{sad2}) once again, to get a single scattering matix for the first three layers. This procedure can be continued to get a net scattering matrix for all atmospheric layers, from the surface to the top of the atmosphere. 
\subsection{Nonnegative $\mathcal{S}$}
The elements $\mathcal{S}_{ii'}$ of the scattering matrix are the  probabilities of scattering from an initial stream $i'$ to a final stream $i$. We therefore expect the elements of the scattering matrix to be nonnegative.
Suppose that all elements of the  scattering matrices, $\mathcal{S}^{\{1\}}$ and $\mathcal{S}^{\{2\}}$ of the two clouds of (\ref{trc0}) are nonnegative
\begin{equation}
\mathcal{S}^{(1)}_{ii'} \ge 0,\quad\hbox{and}\quad\mathcal{S}^{(2)}_{ii'} \ge 0.
\label{nns2}
\end{equation}
Then (\ref{nns2}) implies that the compound scattering matrix of the two clouds, with elements given by (\ref{sad2}), is also nonnegative
\begin{equation}
\mathcal{S}_{ii'} \ge 0.
\label{nns4}
\end{equation}
To prove (\ref{nns4}) we note that we can write the  $n\times n$ inverse matrix in the upper left corner of (\ref{sad2}) as the geometric series
\begin{equation}
\left (\hat 1-\mathcal{S}^{(2)}_{\bf d u} \mathcal{S}^{(1)}_{\bf  u d}\right) ^{-1}=\hat 1 +\left (\mathcal{S}^{(2)}_{\bf d u} \mathcal{S}^{(1)}_{\bf  u d}\right) 
+\left (\mathcal{S}^{(2)}_{\bf d u} \mathcal{S}^{(1)}_{\bf  u d}\right) ^2+\cdots
\label{nns6}
\end{equation}
Since products and sums of nonnegative matrices are non-negative, the inverse matrix (\ref{nns6}) is non-negative. To evaluate the $n\times n$ matrix  $\mathcal{S}_{\bf  d d}$ in the upper left corner of (\ref{sad2}), the non-negative matrix (\ref{nns6})  is multiplied on the left by the nonnegative matrix $\mathcal{S}^{(1)}_{\bf  d d}$ and on the right by the nonnegative matrix  $\mathcal{S}^{(2)}_{\bf  d d}$.  Therefore,    $\mathcal{S}_{\bf  d d}$ is nonnegative.  Analogous arguments show that the matrices   $\mathcal{S}_{\bf  d u}$,   $\mathcal{S}_{\bf  u d}$ and   $\mathcal{S}_{\bf  uu }$ of the other three corners of (\ref{sad2}) are nonnegative. This completes the proof of (\ref{nns4}).

Since the scattering matrix for a cloud of finite optical thickness can be compounded from optically thin layers, and since we showed in Section \ref{tnc} that the elements of the scattering matrix of an optically thin cloud are non-negative,  we conclude that (\ref{nns4}) is true in general. 
\section{Thermal Emission} 
Greenhouse molecules continuously make transitions between vibration-rotation energy levels  as they  absorb or emit radiation.   Small water droplets or ice crystallites in clouds also emit and absorb thermal radiation with corresponding changes in internal energy.  However, under atmospheric conditions, radiative heat transfer to and from greenhouse molecules is orders of magnitude slower than energy transfer by collisions with other atmospheric molecules.  Radiative heating or cooling of cloud particulates is also orders of magnitude slower than heat conduction from the surrounding air, or  the latent heat fluxes of condensing  or evaporating water molecules.

For thermal radiation with strongly absorbed frequencies, greenhouse-gas molecules, cloud particulates and the air that surrounds them are very nearly in local thermal equilibrium at the same temperature.  But for frequencies where the optical depth between the cloud and outer space is of order 1 or less the radiation is not in equilibrium with the air. More thermal radiation goes upward toward space than comes back downward. In thermal equilibrium, radiation must be isotropic, and its dependence on frequency $\nu$ must be given by (\ref{in18}). Only in the center of optically thick clouds can this be true.

Over most of the troposphere, more thermal radiation is emitted than absorbed from air parcels. This cools the parcel and decreases its entropy. In contrast, adiabatic cooling due to expansion of air parcels maintains constant entropy but it decreases the internal energy due to the $p\, dV$ work done. For this reason, radiative cooling is sometimes called diabatic (entropy-changing) cooling, to highlight the difference from adiabatic cooling.

The intensity $|T\}$ of thermal radiation in a cloud must satisfy the equation of radiation transfer (\ref{vet14}), which becomes
\begin{equation}
\left(\frac{d}{d\tau}+\hat\kappa\right) |T \}=\hat\kappa|B \},
\label{te0}
\end{equation}
The boundary conditions needed for a complete solution of (\ref{te0}) are that  thermal radiation emitted inside the cloud must flow out of the top or bottom. No internally generated radiation can flow into the top or bottom. 
\section{Green's function}
A convenient way to solve (\ref{te0}) is with a point-response function or Green's function, $|G(\tau,\tau')\}$ ,
\begin{equation}
|T(\tau)\}= \int_0^{\tau_c}d\tau'|G(\tau,\tau')\}B(\tau').
\label{gf4}
\end{equation}
Substituting (\ref{gf4}) into (\ref{te0}) we find that the Green's function must satisfy equation
\begin{equation}
\int_0^{\tau_c}d\tau'\left[\left(\frac{\partial}{\partial\tau}+\hat\kappa\right)|G(\tau,\tau')\}-\delta(\tau-\tau')\hat\kappa|0)\right]B(\tau')=0.
\label{gf6}
\end{equation}
For (\ref{gf6}) to be true for arbitrary Planck intensities $B(\tau')$, the Green's function must satisfy the differential equation
\begin{equation}
\left(\frac{\partial}{\partial\tau}+\hat\kappa\right)|G(\tau,\tau')\}=\delta(\tau-\tau')\hat\kappa|0).
\label{gf8}
\end{equation}
It will be convenient subsequently to write the Green's function as 
\begin{equation}
|G(\tau,\tau')\}=|G^{\{\infty\}}(\tau-\tau')\}+|\Delta G(\tau,\tau')\}.
\label{gf10}
\end{equation}
One can verify that a particular solution to (\ref{gf8}) is
\begin{eqnarray}
&&|G^{\{\infty\}}(\tau-\tau')\} =H(\tau-\tau')\hat \kappa_{\bf u}e^{-\hat \kappa_{\bf u}(\tau-\tau')}|0)
-H(\tau'-\tau)\hat \kappa_{\bf d}e^{-\hat \kappa_{\bf d}(\tau-\tau')}|0)\nonumber\\
&=&\delta(\tau-\tau')|0)+\frac{\partial}{\partial \tau'}\bigg[H(\tau-\tau')e^{-\hat\kappa_{\bf u}(\tau-\tau')}-
H(\tau'-\tau)e^{-\hat \kappa_{\bf d}(\tau-\tau')}\bigg]|0).
\label{gf12}
\end{eqnarray}
Here $H(\tau)$ is the Heaviside unit step function.
\begin{equation}
H(\tau)=\left \{\begin{array}{rl}0, &\mbox{if $\tau<0$, }\\
1/2, &\mbox{if $\tau=0$,}\\
1,&\mbox{if $\tau>0$.} \end{array}\right . 
\label{gf13}
\end{equation}
In writing the second line of (\ref{gf12}) we noted that the derivative of the Heaviside function is the Dirac delta function,
\begin{equation}
\frac{d}{d\tau}H(\tau)=\delta(\tau).
\label{gf13a}
\end{equation}
The Green's function $|G^{\{\infty\}}(\tau-\tau')\}$  for a hypothetical infinite cloud is the intensity that would be generated at the optical depth  $\tau$ by a thin layer of emitters at the optical depth $\tau'$. The cloud is assumed to have the same single-scattering albedo $\tilde\omega$ and scattering phase function, $p(\mu,\mu')$,  from optical depths $\tau=-\infty$ to $\tau=+\infty$.  Therefore,
$|G^{\{\infty\}}(\tau-\tau')\}$  is invariant to equal displacements of the optical depths $\tau$ and $\tau'$.

The Green's function (\ref{gf10}) includes a part,  $|\Delta G(\tau,\tau') \}$, 
that  is a solution of the homogeneous version of (\ref{gf8}),  
\begin{equation}
\left(\frac{\partial}{\partial\tau}+\hat\kappa\right)|\Delta G(\tau,\tau')\}=\hat 0.
\label{gf16}
\end{equation}
Using (\ref{sct4}) we write
\begin{equation}
|\Delta G(\tau,\tau') \}=\mathcal{U}(\tau)|A(\tau')\}.
\label{gf14}
\end{equation}
The amplitude $|A(\tau')\}$, which depends on the optical depth $\tau'$ of the point source,  can be determined from the boundary condition that no thermal radiation generated by the cloud comes into  the top or bottom.

The value of the Green's function (\ref{gf10}) for observation points  at the top of the cloud, where   $\tau = \tau_c$, is
\begin{eqnarray}
|G(\tau_c,\tau') \}&=&|G^{\{\infty\}}(\tau_c-\tau')\}+|\Delta G(\tau_c,\tau') \}\nonumber\\
&=&\hat\kappa_{\bf u}e^{-\hat\kappa_{\bf u}(\tau_c-\tau')}|0)+\left[\mathcal{L}_{\bf d}+e^{-\hat\kappa_{\bf u}\tau_c}\right]|A(\tau')\}.
\label{gf18}
\end{eqnarray}
For observation points at the bottom of the cloud, where $\tau =0$, we find
\begin{eqnarray}
|G(0,\tau') \}&=&|G^{\{\infty\}}(-\tau')\}+|\Delta G(0,\tau') \}\nonumber\\
&=&-\hat\kappa_{\bf d}e^{\hat\kappa_{\bf d}\tau'}|0)+\left[e^{\hat\kappa_{\bf d}\tau_c}+\mathcal{L}_{\bf u}\right]|A(\tau')\}.
\label{gf20}
\end{eqnarray}
Using (\ref{gf18}) with the boundary condition, $\mathcal{M}_{\bf d} |G(\tau_c,\tau')\}=\hat 0$, of no downward internally generated thermal radiation at the top of the cloud, and using (\ref{gf20}) with the boundary condition, $\mathcal{M}_{\bf u} |G(0,\tau')\}=\hat 0$ of no internally generated upward thermal radiation at the bottom of the cloud, we find
\begin{eqnarray}
\hat 0&=&\mathcal{C}_{\bf d u}\hat\kappa_{\bf u}e^{-\hat\kappa_{\bf u}(\tau_c-\tau')}|0)+\left[\mathcal{C}_{\bf d d}+\mathcal{C}_{\bf d u}e^{-\hat\kappa_{\bf u}\tau_c}\right]|A(\tau')\}
,\label{gf21}\\
\hat 0&=&-\mathcal{C}_{\bf u d}\hat\kappa_{\bf d}e^{\hat\kappa_{\bf d}\tau'}|0)+\left[\mathcal{C}_{\bf u d}e^{\hat\kappa_{\bf d}\tau_c}+\mathcal{C}_{\bf u u}\right]|A(\tau')\}.
\label{gf22}
\end{eqnarray}
Summing the left and right sides of  equations (\ref{gf21}) and (\ref{gf22}) we find
\begin{eqnarray}
\hat 0&=&\left[\mathcal{C}_{\bf d u}\hat\kappa_{\bf u}e^{-\hat\kappa_{\bf u}(\tau_c-\tau')}-\mathcal{C}_{\bf u d}\hat\kappa_{\bf d}e^{\hat\kappa_{\bf d}\tau'}\right]|0)\nonumber\\
&+&\left[\mathcal{C}_{\bf d d}+\mathcal{C}_{\bf d u}e^{-\hat\kappa_{\bf u}\tau_c}
+\mathcal{C}_{\bf u d}e^{\hat\kappa_{\bf d}\tau_c}+\mathcal{C}_{\bf u u}\right]|A(\tau')\}.
\label{gf23}
\end{eqnarray}
Introducing the matrix
\begin{equation}
\mathcal{R}(\tau')
=-\mathcal{C}_{\bf d u}e^{-\hat\kappa_{\bf u}(\tau_c-\tau')}+\mathcal{C}_{\bf u d}e^{\hat\kappa_{\bf d}\tau'},
\label{gf24}
\end{equation}
and noting the definition (\ref{sct12}) of $\mathcal{I}$, 
we rewrite (\ref{gf23}) as
\begin{equation}
\frac{\partial}{\partial \tau'}\mathcal{R}(\tau')|0)
=\mathcal{I}|A(\tau')\}.
\label{gf26}
\end{equation}
Since $\mathcal{I}$ is independent of $\tau'$ we can solve (\ref{gf26})  for $|A(\tau')\}$ to find
\begin{equation}
|A(\tau')\}=\frac{\partial}{\partial \tau'}\mathcal{I}^{-1}\mathcal{R}(\tau')|0)
\label{gf30}
\end{equation}
Substituting (\ref{gf30}) into (\ref{gf14}) we find
\begin{equation}
|\Delta G(\tau,\tau') \}=\frac{\partial}{\partial \tau'}\mathcal{U}(\tau)\mathcal{I}^{-1} \mathcal{R}(\tau')|0).
\label{gf32}
\end{equation}
Adding (\ref{gf12}) to (\ref{gf32}) gives an explicit expression for $|G(\tau,\tau')\}$, the total Green's function (\ref{gf10}) of the cloud.  

\begin{eqnarray}
&&|G(\tau,\tau')\} =\delta(\tau-\tau')|0)\nonumber\\
&&+\frac{\partial}{\partial \tau'}\bigg[H(\tau-\tau')e^{-\hat\kappa_{\bf u}(\tau-\tau')}-
H(\tau'-\tau)e^{-\hat \kappa_{\bf d}(\tau-\tau')}+\mathcal{U}(\tau)\mathcal{I}^{-1} \mathcal{R}(\tau')\bigg]|0).
\label{gf34}
\end{eqnarray}
\begin{figure}[t]
\postscriptscale{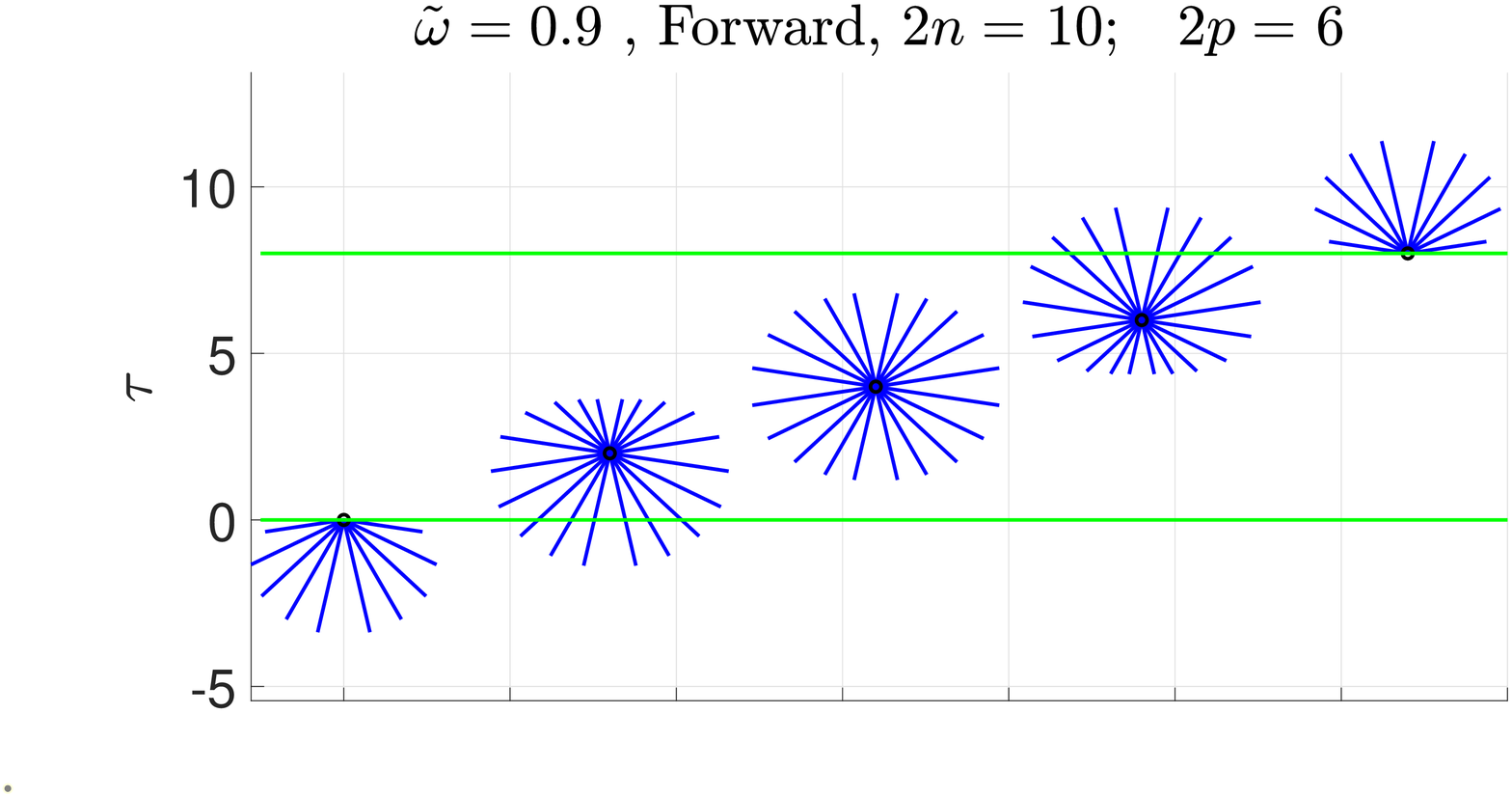}{1}
\caption{Thermal emission, computed with (\ref{iec6}) for a forward-scattering cloud of optical depth $\tau_c=8$. The cloud is isothermal with a constant value of the Planck intensity $B=1$. The radiation is modeled with $2n = 10$ streams. The lengths of the blue rays are proportional to the intensities of the thermal radiation, $T(\mu_i,\tau)=w_i^{-1}\lvec\mu_i|T(\tau)\}$, generated in the cloud. The forward-scattering phase function is constructed from the first $2p = 6$ Legendre polynomials, like that of Fig. \ref{ph3}. The single-scattering albedo is $\tilde\omega = 0.9$. At the center of the cloud, at an optical depth $\Delta \tau =4$ from both the top and bottom, the thermal radiation is nearly isotropic, but slightly more intense in horizontal than vertical directions. At the top and bottom  of the cloud, marked by the horizontal black lines, there is only outgoing, ``limb-darkened'' thermal radiation.
\label{therm1}}
\end{figure}
\begin{figure}[t]
\postscriptscale{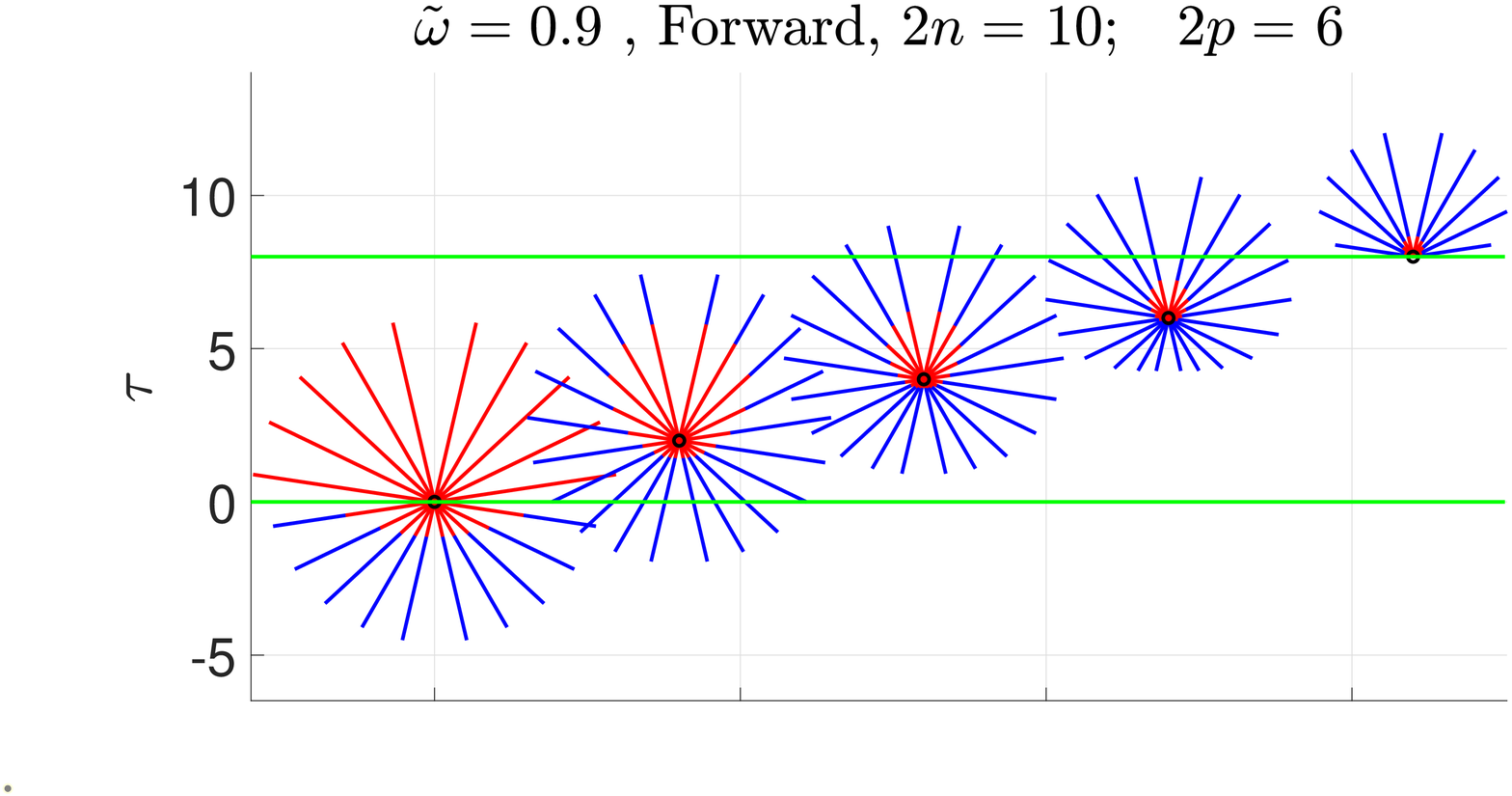}{1}
\caption{Like Fig. \ref{therm1}, but for an isothermal cloud with optical depth $\tau_c = 8$, located above a warmer black surface with Planck intensity $1.2B$ that is 20\% larger than the Planck intensity $B$ of the cloud.  The intensity in the cloud is a combination of ``blue'' thermal emission by cloud particulates, computed with (\ref{iec6}), and attenuated ``red'' thermal radiation from the warmer surface, calculated with (\ref{sct16}). The total intensity is the sum of the lengths of the two colored segments. The upward radiation from the surface has the  incident intensity $|I^{\{\rm in\}}\} = \mathcal{M}_{\bf u}|0)1.2B$ and is Lambertian.   At increasing optical depths above the surface, the slant rays of the red surface radiation are more attenuated than the more vertical radiation, which leads to increasing limb darkening with altitude. Very little ``red" surface radiation emerges from the cloud top, where the outgoing radiation is almost all ``blue'' thermal radiation generated in the cloud.  
A relatively small amount of ``red" surface radiation is diffusely reflected from the bottom of the cloud.
\label{therm2}}
\end{figure}
\subsection{Isothermal emitting cloud}
If the cloud is isothermal, the Planck intensity $B(\tau')$ can be taken to have the $\tau'$-independent value $B$. Then we can use (\ref{gf4}) with (\ref{gf34}) to write the thermal intensity in the cloud as
\begin{eqnarray}
&&|T(\tau)\}= B\int_0^{\tau_c}d\tau'|G(\tau,\tau')\}\nonumber\\
&=&|B\}+\bigg[H(\tau-\tau')e^{-\hat\kappa_{\bf u}(\tau-\tau')}-
H(\tau'-\tau)e^{-\hat \kappa_{\bf d}(\tau-\tau')}+\mathcal{U}(\tau)\mathcal{I}^{-1} \mathcal{R}(\tau')\bigg]_{\tau'=0}^{\tau'=\tau_c}|B\}\nonumber\\
&=&\bigg[\hat 1 -e^{-\hat \kappa_{\bf d}(\tau-\tau_c)}+\mathcal{U}(\tau)\mathcal{I}^{-1} \mathcal{R}(\tau_c)-e^{-\hat\kappa_{\bf u}\tau}
-\mathcal{U}(\tau)\mathcal{I}^{-1} \mathcal{R}(0)\bigg]|B\}\nonumber\\
&=&\bigg[\hat 1 -\mathcal{U}(\tau)+\mathcal{U}(\tau)\mathcal{I}^{-1} \{\mathcal{R}(\tau_c)- \mathcal{R}(0)\}\bigg]|B\}.
\label{iec2}
\end{eqnarray}
From (\ref{gf24}), (\ref{sct12}) and (\ref{olp4}) we find

\begin{eqnarray}
\mathcal{R}(\tau_c)- \mathcal{R}(0)&=&-\mathcal{C}_{\bf d u}+\mathcal{C}_{\bf u d}
e^{\hat\kappa_{\bf d}\tau_c}+\mathcal{C}_{\bf d u}e^{-\hat\kappa_{\bf u}\tau_c}
-\mathcal{C}_{\bf u d}\nonumber\\
&=&\mathcal{I}-\hat 1.
\label{iec4}
\end{eqnarray}
Substituting (\ref{iec4}) into (\ref{iec2}) we find that the thermal radiation generated by the cloud is
\begin{equation}
|T(\tau)\}=
\bigg[\hat 1 -\mathcal{U}(\tau)\mathcal{I}^{-1}\bigg]|B\}.
\label{iec6}
\end{equation}
From (\ref{iec6}) we see that
the thermal outgoing radiation from the cloud is
\begin{eqnarray}
|T^{\{\rm out\}}\}&=&\mathcal{M}_{\bf d}|T(0)\}+\mathcal{M}_{\bf u}|T(\tau_c)\} \nonumber\\
&=&\left[\mathcal{M}_{\bf d}+\mathcal{M}_{\bf u}
-\{\mathcal{M}_{\bf d}\,\mathcal{U}(0)+\mathcal{M}_{\bf u}\,\mathcal{U}(\tau_c)\}\mathcal{I}^{-1}\right]|B\}\nonumber\\
&=&\left[\hat 1-\mathcal{O}\mathcal{I}^{-1}\right]|B\}\nonumber\\
&=&\mathcal{E}|B\}.
\label{iec8}
\end{eqnarray}
The outgoing matrix $\mathcal{O}$ was defined by (\ref{sct22}). According to (\ref{sct20}), the scattering matrix of the cloud is $\mathcal{S}=\mathcal{O}\mathcal{I}^{-1}$.
So from (\ref{iec12}) we see that the emissivity matrix is
\begin{equation}
\mathcal{E}=\hat 1 -\mathcal{O}\mathcal{I}^{-1}=\hat 1-\mathcal{S}.
\label{iec10}
\end{equation}
 Eq.  (\ref{iec10}) is Kirchhoff's law  (\ref{c24}).  
 As a consistency check, we use (\ref{iec6}) and (\ref{sct12}) to show that the thermal  radiation coming into the cloud is
\begin{eqnarray}
|T^{\{\rm in\}}\}&=&\mathcal{M}_{\bf u}|T(0)\}+\mathcal{M}_{\bf d}|T(\tau_c)\} \nonumber\\
&=&\left[\mathcal{M}_{\bf u}+\mathcal{M}_{\bf d}
-\{\mathcal{M}_{\bf u}\mathcal\,{U}(0)+\mathcal{M}_{\bf d}\,\mathcal{U}(\tau_c)\}\mathcal{I}^{-1}\right]|B\}\nonumber\\
&=&\left[\hat 1-\mathcal{I}\mathcal{I}^{-1}\right]|B\}\nonumber\\
&=&\hat 0.
\label{iec12}
\end{eqnarray}
\section{Summary}
Here we summarize the most important differences between the results of our $2n$-stream model and previous work. Additional interesting and useful results will be discussed in subsequent papers.

We consider axially symmetric variables, like the intensity $I(\mu,\tau)$, that depend on the direction cosine $\mu$ of the radiation, and on the optical depth $\tau$ above the bottom of a cloud.
Using Dirac notation, similar to that used for the analysis of  Schr\"odinger's equation in quantum mechanics, we write the equation of transfer in the formally simple, but completely general form (\ref{vet14}).  The propagation of radiation is controlled by the $2n\times 2n$ exponentiation-rate matrix $\hat\kappa$  of (\ref{vet12}) that is analogous to the quantum mechanical Hamiltonian operator $H$. The intensity is described by an abstract state vector $|I(\tau)\}$ that depends on the optical depth $\tau$ in a similar way to the  dependence of a spin wave function $|\psi(t)\rangle$ on the time $t$.

As exemplified in (\ref{sdbv10a}), 
we represent radiative variables as a superposition of $2n$ basis vectors, the right stream bases $|\mu_i)$  and the left stream bases $\lvec\mu_i|$ of (\ref{sdbv20}) and (\ref{sdbv22}). The stream indices are $i=1,2,3,\ldots, 2n$.  The streams have nominal zenith angles $\theta_i=\cos^{-1}\mu_i$.  According to (\ref{lg4}), the Gauss-Legendre direction cosine  $\mu_i$ is a root of the Legendre polynomial $P_{2n}$, that is, $P_{2n}(\mu_i)=0$. Representative projections $\langle\mu|\mu_i)$ of the stream bases  onto continuous direction-cosine states $\langle\mu|$ of (\ref{vc4}) are shown in Fig. \ref{mumuj}. 
The maximum values of  $|\langle \mu|\mu_i)|$ occur for $\mu\approx\mu_i$.  So the stream bases $|\mu_i)$ represent radiation propagating with directions close to that of the zenith angle $\theta_i=\cos^{-1}\mu_i$.

As in quantum mechanics, we think of bases as eigenvectors of ``observables." In (\ref{emu2}) we show that the stream bases $|\mu_i)$ are right eigenvectors of the $2n\times 2n$ direction-cosine matrix  $\hat\mu$ of (\ref{me3}), or of its inverse, the direction-secant matrix $\hat\varsigma$ of (\ref{me17}). The left eigenvectors of $\hat \mu$ are denoted by $\lvec\mu_i|$. The double left parenthesis is a reminder that the left eigenvectors are not Hermitian conjugates of the right eigenvectors. Many radiative-transfer variables in $2n$ space can be conveniently represented with non-Hermitian matrices. So it is not always possible to write left and right eigenvectors as Hermitian conjugate pairs.  

Radiative transfer is parameterized by the single scattering-albedo $\tilde\omega$ and by the scattering phase function $p(\mu,\mu')$. The fraction of photons scattered by a collision is $\tilde \omega$. The  fraction absorbed and converted to heat is $1-\tilde\omega$. The fraction of photons of initial direction cosine $\mu'$ that is scattered into direction cosines between $\mu$ and $\mu+d\mu$ is  $ p(\mu,\mu')\,d\mu/2$. 

We have introduced a family of phase functions, $p(\mu,\mu'=1) = \varpi^{\{p\}}(\mu)$ of (\ref{pfb2}) 
which are superpositions of the first $2p$ Legendre polynomials, and  which produce the maximum possible forward scattering.  For a $2n$-stream model one must have $p\le n$. The maximum-forward-scattering phase function $\varpi^{\{p\}}(\mu)$ satisfies the constraints (\ref{in14}) and (\ref{in16}). The forward-scattering values are  $\varpi^{\{p\}}(\mu =1) = p(p+1)$.  A representative forward-scattering phase function $\varpi^{\{3\}}(\mu,\mu')$ is shown in Fig. \ref{ph3}, and others are shown in Fig. \ref{maxphase}.

The homogeneous equation of transfer (\ref{sct2}) has $2n$ independent solutions, $|\lambda_i)e^{-\kappa_i\tau}$.  The penetration modes $|\lambda_i)$, defined by (\ref{plm4}), are eigenfunctions of the exponentiation-rate matrix $\hat \kappa$, or of its inverse, $\hat \lambda = \hat\kappa^{-1}$, the penetration-length matrix.  Representative projections $\langle\mu|\lambda_i)$ of the penetration modes   onto the direction-cosine states $\langle\mu|$ are shown in Fig. \ref{mulamj} for a weakly absorbing cloud with $1-\tilde \omega = 0.001$.  For negligible scattering, $\tilde \omega\to 0$, we noted in (\ref{lim2}) that the
penetration modes are identical to the stream bases, $|\lambda_i)= |\mu_i)$, and the penetration lengths are the same as the Gauss-Legendre cosines, $\lambda_i =\mu_i$. For any value of $\tilde\omega <0$, the 
 directional penetration modes $|\lambda_i)$, with $i=2,3,\dots,2n-1$, are qualitatively similar to the corresponding stream bases $|\mu_i)$.  But for weak absorption, the upward quasi-isotropic mode $|\lambda_{2n})$, shown in the left panel of Fig. \ref{mulamj},  has very little dependence on $\mu$, and it differs drastically from the stream basis $|\mu_{2n})$ on the left panel of Fig. \ref{mumuj}, which is strongly peaked near $\mu = 1$.  The downward quasi-isotropic mode $|\lambda_1)$ differs from $|\mu_1)$ in an analogous way. The quasi-isotropic modes represent the familiar, nearly uniform brightness seen from all directions  on a foggy day or from an aircraft flying through thick, sunlit clouds.

 Representative penetration lengths, $\lambda_i=1/\kappa_i$, the inverses of the exponentiation rates $\kappa_i$, are plotted in Fig. \ref{miom} as a function of the single-scattering albedo $\tilde \omega$. As $\tilde\omega \to 1$, the penetration lengths $\lambda_i$ separate into two groups. For the directional modes, with $i=2,3,\ldots, 2n-1$, the lengths  $|\lambda_i|$ are only  slightly larger than the corresponding Gauss-Legendre cosines. $|\mu_i|$.  The quasi-isotropic mode $|\lambda_{2n})$, and its mirror image, $|\lambda_1)$ have divergent penetration lengths,  as one can see by inspection of Fig. \ref{miom}. According to (\ref{lim8}), as $\tilde\omega \to 1$ the quasi-isotropic penetration lengths are very nearly given by $\lambda_{2n}=-\lambda_1\to \big[3(1-\tilde\omega p_1) (1-\tilde\omega)\big]^{1/2}$.

The optical-depth evolution operator $\mathcal{U}(\tau)$ of (\ref{sct6}) is the sum of exponentially growing or decaying terms of the form $e^{-\kappa_i(\tau-\tau_i)}|\lambda_i)\lvec\lambda_i|$. As indicated by (\ref{sct6a}), for radiative transfer  it is convenient to choose reference optical depths $\tau_i=0$, at the bottom of the cloud, for decaying modes with $\kappa_i>0$. For growing modes, with $\kappa_i<0$, the reference optical depths, $\tau_i=\tau_c$, are most conveniently chosen at the top of the cloud. According to  (\ref{sct16}), the intensity $|I(\tau)\}$ at an optical depth $\tau$ inside a cloud   is given in terms of the incident intensity $|I^{\{\rm in\}}\}$ at the top and bottom of the cloud by   $|I(\tau)\}=\mathcal{U}(\tau)\mathcal{I}^{-1}|I^{\{\rm in\}}\}$, where the incoming matrix $\mathcal{I}$ was given by (\ref{sct12}).

We represent the overall radiation transfer in non-emissive clouds by the scattering matrix $\mathcal{S}$ of (\ref{sct20}). According to (\ref{sct18}), the outgoing intensity, $|I^{\{\rm out\}}\}$, from the top and bottom of the cloud  is related to the incoming intensity $|I^{\{\rm in\}}\}$ by $|I^{\{\rm out\}}\}=\mathcal{S}|I^{\{\rm in\}}\}$.
The scattering matrices $\mathcal{S}^{(1)}$ and $\mathcal{S}^{(2)}$  of two clouds in series can be represented by a single scattering matrix  $\mathcal{S}$, given by the addition formula of (\ref{sad2}).

We  consider not only the scattering of external radiation by clouds but also the thermal emission of radiation by warm molecules or particulates inside clouds. The Green's function (or point-response function) $G(\tau,\tau')$  of (\ref{gf34}) can be used to calculate the thermally emitted intensity $|T(\tau)\}$ within a non-isothermal cloud with (\ref{gf4}). For the simple case of an isothermal cloud with a constant Planck intensity $|B\}$,  (\ref{iec8}) shows that the thermally generated intensity going out of the cloud is $|T^{\{\rm out\}}\}=\mathcal{E}|B\}$, where $\mathcal{E}$ is the emissivity matrix of the cloud. In agreement with Kirchhoff's law, (\ref{iec10}) shows that the sum of the isothermal emissivity matrix and the scattering matrix is the unit matrix, $\mathcal{E}+\mathcal{S}=\hat 1$.

Fig. \ref{Ch21} compares diffuse reflection and transmission  for pure isotropic scattering in clouds of various optical depths $\tau_c$. The top panels of Fig. \ref{Ch21} reproduce  Fig. 21 of Chandrasekhar\cite{Chandrasekhar}. The bottom panels are  the results of a 32-stream calculation with $\tilde \omega = 1-10^{-6}$ and with an input stream $ |I^{\{\rm in\}}\}=|\mu_{23})$, where $\mu_{23} = 0.5877$ is the Gauss-Legendre cosine that is closest to the input cosine $\mu = 0.6$ used by Chandrasekhar. In our Fig. \ref{Ch22} we show a similar comparison to Chandrasekhar's Fig. 22 where the single-scattering albedo was  $\tilde\omega = 0.9$.  Our results can hardly be distinguished from those of Chandrasekhar, who used ``$X$ and $Y$ functions'' of \S{\bf 61}(1) and \S{\bf 61}(2), the solutions of complicated integral equations. 

The $2n$ values of intensity, $I(\mu_i,\tau)=w_i^{-1}\lvec\mu_i|I(\tau)\}$, can be used to make physically instructive plots like Fig. \ref{fin2},  which shows how part the nearly vertical radiation  incident on the bottom of a cloud is diffusely reflected and part is transmitted. For optically thick clouds like that of Fig. \ref{fin2}, with $\tau_c = 8$, but with very small single-scattering absorption $1-\tilde \omega = 10^{-2}$, appreciable transmission (12\%) is possible because of multiple scattering. Multiple scattering nearly ``isotropizes'' the intensity near the center of the cloud. At the top of the cloud, there are upward transmitted streams, but no downward streams. At the bottom of the cloud there is a strong incident stream $|I^{\{\rm in\}}\}=|\mu_{2n})$, and relatively weak diffusely reflected streams $I(\mu_j,\tau =0)$ with $\mu_j<0$.

Fig. \ref{fin2a} shows what happens when the single-scattering absorption is reduced by a factor of 100 from that of Fig. \ref{fin2},  to $1-\tilde\omega = 10^{-4}$. For such small single-scattering absorption, the absorption in the cloud is also negligible and the flux is nearly constant from the bottom to the top of the cloud. The formalism outlined in this paper does not work if there is no absorption at all and $\tilde\omega = 1$. As we will show in a subsequent paper,  minor modifications in the computational steps outlined above allow us to calculate radiative tranfer for $\tilde \omega = 1$, Chandrasekhar's ``conservative scattering.''

Other interesting calculations are summarized in Fig. \ref{trans1}, which shows how the transmission of radiation from the bottom to the top of a cloud with negligibly small absorption $1-\tilde\omega \ll 1$ depends on the optical depth $\tau_c$ of the cloud and on the nature of the scattering phase function $p(\mu,\mu')$. Not suprisingly, the transmission is greatest for forward scattering and least for backward scattering.  

Fig. \ref{alb0}, for an optically thick cloud, shows the cloud albedo, $\langle\Omega\rangle$, of (\ref{alb14}). The cloud albedo is  the  fraction of the vertical flux reflected from the bottom of a cloud, as opposed to being absorbed. The cloud albedo depends on the single-scattering albedo $\tilde\omega$, on the nature of the  scattering phase function $p(\mu,\mu')$, and on the angular distribution of the incident intensity $|I^{\{\rm in\}}\}$. If  the cloud were not optically thick,  its albedo would also depend on the optical thickness $\tau_c$. 

Finally, Fig. \ref{therm2} shows how blackbody thermal flux emitted from a warm surface propagates into a cooler, isothermal cloud with single-scattering albedo $\tilde\omega = 0.9$, a forward scattering phase function, $\varpi^{\{3\}}(\mu)$, and an optical depth $\tau_c=8$.  Most of the surface radiation is absorbed and replaced by thermal flux emitted by the cloud.  A small  fraction of the radiation is diffusely reflected back into the surface, and a still smaller fraction is transmitted out of the top of the cloud, where most of the emergent radiation has been thermally generated  by cloud particulates.

It is easy to use modern mathematical software like Matlab with this new $2n$-stream formalism to calculate a variety of radiative-transfer quantities, including diffuse reflection or transmission, albedos, thermal emission, etc.   A few dozen lines of code is usually sufficient.  It is easy to go beyond  isotropic or Rayleigh-scattering phase functions of traditional radiative transfer calculations to much more realistic and complicated ones. The family of phase functions $\varpi^{\{p\}}(\mu)$ of (\ref{pfb2})  allows one to model radiation transfer with large forward-scattering amplitudes, like the Mie scattering of sunlight in Earth's clouds.  We believe that the $2n$-stream formalism discussed here will be a useful new tool for radiative-transfer calculations.
\appendix
\section{Phase Functions for Maximum Forward Scattering}
We turn now to the proof that the algebraic formula 
of (\ref{pfb2}) for $\varpi^{\{p\}}(\mu)$ gives the maximum possible forward scattering for a phase function constructed from the first $2p=2,4,6,\ldots$ Legendre polynomials.
We assume that the maximum forward-scattering phase function can be written as 
\begin{equation}
	\varpi^{\{p\}}(\mu)=(1+\mu)\phi^2(\mu). \label{pm2}
\end{equation}
Here $\phi(\mu)$ is a polynomial  of degree $p-1$ which we expand in Legendre polynomials as
\begin{eqnarray}
	\phi(\mu)
	&=&\sum_{l=0}^{p-1}\phi_{l}\, g_l\, P_l(\mu).
	\label{pm4}
\end{eqnarray}
The coefficient $g_l$ is a diagonal element of the 
statistical weight matrix $\hat g$, defined by (\ref{rpl4}).
To facilitate further discussion we introduce a statistical-weight column vector $|g\rangle$ and row vector $\langle g|$ with the elements
\begin{equation}
	|g\rangle=
	\left[\begin{array}{c}g_0\\g_1\\ g_2\\\vdots \\ g_{p-1}\end{array}\right],\quad\hbox{and}\quad\langle g| = [g_0,\, g_1,\, g_2,\,\cdots, g_{p-1}].
	\label{pm8}
\end{equation}
Using (\ref{pm4}) we write the forward-scattering phase (\ref{pm2}) as
\begin{eqnarray}
	\varpi^{\{p\}}(1)&=&2\phi^2(1)= 2\left(\sum_{l=0}^{p-1}\phi_lg_l\right)^2\nonumber\\
	&=&2\sum_{l l'}\phi_l q_{l l'}\phi_{l'}.
	\label{pm10}
\end{eqnarray}
In (\ref{pm10}) we have introduced the $p\times p$ symmetric matrix $\hat q$, defined by
\begin{equation}
	\hat q =|g\rangle\langle g|, \quad\hbox{or}\quad \lvec l|\hat q|l')=q_{l l'}=g_lg_{l'}.
	\label{pm12}
\end{equation}
As shown in (\ref{in16}), the area of the phase function  $\varpi^{\{p\}}(\mu)$ is constrained to be 
\begin{equation}
	\int_{-1}^1 d\mu\,\varpi^{\{p\}}(\mu)=\int_{-1}^1 d\mu\,(1+\mu)\sum_{l l'}g_l\phi_lP_l(\mu)g_{l'}\phi_{l'}P_{l'}(\mu)
	=2,
	\label{pm14}
\end{equation}
or 
\begin{equation}
	1=\sum_{l l'}\phi_l h_{l l'}\phi_{l'}.
	\label{pm16}
\end{equation}
Here the elements $h_{l l'}$ of the constraint matrix are
\begin{equation}
	h_{l l'}=g_l\delta_{l l'}+g_l\hat \mu_{l l'},\quad\hbox{or}\quad \hat h=\hat g+\hat g \hat\mu
	=\hat g(\hat 1+\hat \mu).
	\label{pm18}
\end{equation}
The elements of the direction-cosine matrix $\hat\mu$
were given by (\ref{mm18}) or (\ref{me3}), and they can be used to write
 elements of the constraint matrix $\hat h$ of (\ref{pm18}) as
\begin{equation}
	h_{ll'}=\lvec l|\hat h|l')
	=\left[\begin{array}{lllllll} 1&1&0&0&0&0&\cdots \\ 1&3&2&0&0&0&\cdots \\ 
		0&2&5&3&0&0&\cdots \\   0&0&3&7&4&0&\cdots \\ 
		0&0&0&4&9&5&\cdots \\ 0&0&0&0&5&11&\cdots \\ 
		\vdots&\vdots &\vdots&\vdots&\vdots&\vdots&\ddots\\ \end{array}\right]
	\label{pm19a}
\end{equation}

We use a Lagrangian multiplier $m$ to 
find the coefficients $\phi_l$ that maximize the forward-scattering phase $\varpi^{\{p\}}(1)$ of (\ref{pm10}), subject to the constraint (\ref{pm16}). Then we can vary the  amplitudes $\phi_k$ independently to find local maxima, minima or saddle points of the function
\begin{equation}
	M(\phi_0,\phi_1,\ldots,\phi_{n-1})=\varpi^{\{p\}}(1)-m\sum_{ll'}\phi_lh_{ll'}\phi_{l'}.
	\label{pm19b}
\end{equation}
For $\phi_k$'s at a local maximum of $M$,  we must have
\begin{eqnarray}
	0&=&\frac{\partial}{\partial \phi_k}M\nonumber\\
	&=&\frac{\partial}{\partial \phi_k}\sum_{l l'}\phi_l\left(2 q_{l l'}-m  h_{l l'}\right)\phi_{l'}\nonumber\\
	&=&2\sum_{l'}\left(2 q_{k l'}-m h_{k l'}\right)\phi_{l'},\quad\hbox{for}\quad k=0,1,2,\ldots,2p-1.
	\label{pm20}
\end{eqnarray}
In writing the last line of (\ref{pm20}) we made use of the symmetries
$q_{ll'}=q_{l'l}$ and $h_{l l'}=h_{l' l}$.
We can write the last line of (\ref{pm20}) as
\begin{equation}
	2\sum_{l'}q_{k l'}\phi_{l'}=m\sum_{l'} h_{k l'}\phi_{l'},\quad\hbox{or}\quad
	2|g\rangle\langle g|\phi)=m \hat h|\phi).
	\label{pm21}
\end{equation}
Here we have introduced the column vector of multipole amplitudes
\begin{equation}
	|\phi)=
	\left[\begin{array}{c}\phi_0\\\phi_1\\ \phi_2\\\vdots \\ \phi_{p-1}\end{array}\right].
	\label{pm21a}
\end{equation}
We can multiply  the left version of (\ref{pm21}) by $\phi_k$, sum over $k$, and use (\ref{pm10})  with (\ref{pm16})  to find
\begin{eqnarray}
	\varpi^{\{p\}}(1)&=&2\sum_{k l'}\phi_k  q_{k l'}\phi_{l'}\nonumber\\
	&=&m\sum_{k l'}\phi_k  h_{k l'}\phi_{l'}\nonumber\\
	&=&m.
	\label{pm21b}
\end{eqnarray}
The maximum value $\varpi^{\{p\}}(1)$ of the phase function is the same as the value of $m$, the Lagrange multiplier.
Multiplying both sides of the second, matrix equation of (\ref{pm21}) by $\hat h^{-1}$ we find
\begin{equation}
	\hat m |\phi\rangle=m |\phi\rangle,\quad\hbox{where}\quad \hat m =|f\rangle \langle g|.
	\label{pm26}
\end{equation}
The column vector $|f\rangle$ of  (\ref{pm26}) is
\begin{eqnarray}
	|f\rangle&=&2\hat h^{-1} | g\rangle.
	\label{pm28}
\end{eqnarray}

To facilitate further discussion 
we introduce a $p\times 1$ column vector, $|x\rangle$ of alternating 2's and 0's,  for which the last element is a 2.
\begin{equation}
	|x\rangle=
	\left[\begin{array}{c}\vdots\\0\\2 \\0 \\ 2 \end{array}\right].
	\label{vf4}
\end{equation}
From inspection of (\ref{vf4}) and (\ref{me3}) we see that 
\begin{equation}
	\hat \mu |x\rangle=
	\left[\begin{array}{c}\vdots\\2\\0 \\2 \\ 0 \end{array}\right].
	\label{vf5}
\end{equation}
\noindent Therefore,
\begin{equation}
	(\hat 1+ \hat \mu) |x\rangle=
	\left[\begin{array}{c}\vdots\\2\\2 \\2 \\ 2 \end{array}\right].
	\label{vf6}
\end{equation}
The inverse $\hat g^{-1}$ of the statistical-weight  matrix $\hat g $ of  (\ref{rpl4}) has the matrix elements
\begin{equation}
\lvec l|\hat g^{-1}|l')=g_l^{-1}\delta _{ll'}.
	\label{vf12}
\end{equation}
Using (\ref{vf12}) with (\ref{pm8})  and (\ref{vf6}) we find
\begin{equation}
	2\hat g^{-1} |g\rangle=
	\left[\begin{array}{c}\vdots\\2\\2 \\2 \\ 2 \end{array}\right]
	=(\hat 1+\hat\mu)|x\rangle.
	\label{vf7}
\end{equation}
Multiplying both sides of (\ref{vf7}) by $(\hat 1+\hat\mu)^{-1}$, we find
\begin{eqnarray}
|x\rangle &=&2(\hat 1+\hat\mu)^{-1}\hat g^{-1}|g\rangle\nonumber\\
&=&2[\hat g (\hat 1+\hat\mu)]^{-1}|g\rangle\nonumber\\
&=&2 \hat h^{-1}|g\rangle\nonumber\\
& =&|f\rangle. 
\label{vf16}
\end{eqnarray}
So the column  vector $|f\rangle$ of (\ref{pm28}) is the same as the column vector $|x\rangle$ of (\ref{vf4}), for which  the elements alternate between 2 and  0, with $x_{p-1}=2$.

In the eigenvalue equation of (\ref{pm26}), the matrix $\hat m=|f\rangle\langle g|$ is the outer product between the column vector $|f\rangle$ and row vector $\langle g|$. It is easy to find the eigenvectors $|\phi)$ and eigenvalues $m$ of outer products. There are $p-1$  degenerate eigenvalues, $m=0$, corresponding to any set of  $p-1$ independent vectors $|\phi)$ for which $\langle g|\phi)=0$. These solutions are of no interest to us.  The eigenvector corresponding to the single, potentially non-zero eigenvalue can be chosen to be
\begin{equation}
	|\phi\rangle =c|f\rangle.
	\label{pm30}
\end{equation}
We will determine the constant $c$ below.  Substituting (\ref{pm30}) into the eigenvalue equation of (\ref{pm26}) we find
\begin{equation}
	c|f\rangle\langle g|f\rangle=c\,m |f\rangle.
	\label{pm31}
\end{equation}
Eq. (\ref{pm31}) implies that
\begin{equation}
	\langle g|f\rangle =m.
	\label{pm32}
\end{equation}
Using (\ref{pm32})  with the explicit form of $|f\rangle=|x\rangle$ of (\ref{vf4}) and $\langle g|$ of (\ref{pm8}) we find that the value of the Lagrange multiplier $m$ must be
\begin{eqnarray}
	m&=& 2(g_{p-1}+g_{p-3}+g_{p-5}+\cdots)
	\nonumber\\
	&=&p(p+1).
	\label{pm34}
\end{eqnarray}
The sum on the first line of (\ref{pm34}) only includes statistical weights $g_l$ for which $l\ge 0$.
One can prove the last line of (\ref{pm34}) by mathematical induction.  Letting $m_p$ be the value of $m$ for the phase function (\ref{pm2}) that is expanded on $2p$ Legendre polynomials, we see that  (\ref{pm34}) is true for $p=1$ and $p=2$,
\begin{eqnarray}
	m_1&=&2( 1)\nonumber\\
	m_2&=&2(3).
	\label{pm36}
\end{eqnarray}
If $ m_p=p(p+1)$ for the integer $p$, then we can use (\ref{pm34}) to write
\begin{eqnarray}
	m_{p+2}&=&2g_{p+1}+p(p+1)\nonumber\\
	&=&2(2p+3)+p^2+p\nonumber\\
	&=&p^2+5p+6\nonumber\\
	&=&(p+2)(p+3).
	\label{pm38}
\end{eqnarray}
This completes the proof that (\ref{pm34}) is true for all positive integers $p$.

Substituting the coefficients $\phi_l = c f_l$ from (\ref{pm30}) into (\ref{pm4}) we find
\begin{eqnarray}
	\phi(\mu)&=&\sum_{l=0}^{p-1}\phi_{l}\, g_l\, P_l(\mu)\nonumber\\
	&=&2c\left[g_{p-1}P_{p-1}(\mu)+g_{p-3}P_{p-3}(\mu)+\cdots\right]\nonumber\\
	&=&2c\frac{dP_p}{d\mu}(\mu).
	\label{pm40}
\end{eqnarray}
An identity of Legendre polynomials \cite{Legendre} was used to go from the second to 
the third line of (\ref{pm40}). Substituting (\ref{pm40}) into (\ref{pm2}) we find
\begin{equation}
\varpi^{\{p\}}(\mu)=4 c^2(1+\mu)\left[\frac{dP_p}{d\mu}(\mu)\right]^2.\label{pm42}
\end{equation}
Evaluating (\ref{pm42}) for forward scattering ($\mu = 1$) and using
(\ref{pm21b}) with (\ref{pm34}) we find
\begin{equation}
\varpi^{\{p\}}(1)=8c^2\left[\frac{dP_p}{d\mu}(1)\right]^2=p(p+1).\label{pm43}
\end{equation}
The derivatives of the Legendre polynomials at the end point $\mu = 1$ have the values \cite{Legendre}
\begin{equation}
	\frac{dP_p}{d\mu}(1)=\frac{p(p+1)}{2}.\label{pm44}
\end{equation}
Using (\ref{pm44}) in (\ref{pm43}) we find that the squared coefficient $c^2$ of (\ref{pm42})  must be
\begin{equation}
	c^2=\frac{1}{2 p(p+1)}.\label{pm46}
\end{equation}
Substituting (\ref{pm46}) into (\ref{pm42}) gives (\ref{pfb2}) and completes the proof that $\varpi^{\{p\}}(\mu)$ is the phase function constructed from the first $2p=2,4,6,\ldots$ Legendre polynomials that has the maximum possible forward-scattering value. 

\section*{Acknowledgements}
 The Canadian Natural Science and Engineering Research  Council provided financial support of one of us.

\end{document}